\documentclass[twocolumn,a4paper]{sn-jnl}

\usepackage{epsfig}
\usepackage{hhline}
\usepackage{amsmath}
\usepackage{amssymb}
\usepackage{mathrsfs}
\usepackage{placeins}

\usepackage{ bbold }

\usepackage{color}
\usepackage{xspace}
\usepackage{paralist} 
\usepackage{caption}
\usepackage{bm} 
\usepackage{acronym}
\usepackage{adjustbox}

\usepackage{hyperref} 
\hypersetup{colorlinks=true, urlcolor=blue}
\usepackage{cite} 

\usepackage{booktabs} 
\usepackage{xcolor}
\usepackage{comment}

\usepackage{parskip}

\usepackage[utf8]{inputenc}
\newlength{\dinwidth}
\newlength{\dinmargin}
\renewcommand{\d}{\text{d}}
\newcommand{\pb}{\rm pb}

\begin{document}

\newcommand{\GeVsq}{\ensuremath{\mathrm{GeV}^2} }
\newcommand{\GeV}{\ensuremath{\mathrm{GeV}} }
\newcommand{\pt}{\ensuremath{P_{\text{T}}}}
\newcommand{\PP}{\ensuremath{\mathcal{P}}}
\newcommand{\Qsq}{\ensuremath{Q^{2}}\xspace}
\newcommand{\xbj}{\ensuremath{x_{\text{Bj}}}\xspace}
\newcommand{\chisq}{\ensuremath{\chi^{2}}}
\newcommand{\hchisq}{\ensuremath{\hat\chi^{2}}}
\newcommand{\chisqA}{\ensuremath{\chi_{\rm A}^{2}}}
\newcommand{\chisqL}{\ensuremath{\chi_{\rm L}^{2}}}
\newcommand{\ndf}{\ensuremath{n_{\rm dof}}}

\newcommand{\V}{\ensuremath{\mathbb{V}}}
\newcommand{\W}{\ensuremath{\mathbb{W}}}
\newcommand{\M}{\ensuremath{\tilde{M}}}
\newcommand{\Y}{\ensuremath{Y\M}}
\newcommand{\U}{\ensuremath{\Upsilon}}

\newcommand{\dt}{\ensuremath{\bm{d}}}
\newcommand{\mb}{\ensuremath{\bm{\bar{m}}}}
\newcommand{\mt}{\ensuremath{\bm{\tilde{m}}}}
\newcommand{\s}{\ensuremath{\bm{s}}}
\newcommand{\ahut}{\ensuremath{\bm{\hat{a}}}}

\newcommand{\ptjet}{\ensuremath{P_{\rm T}^{\rm jet}}}
\newcommand{\Mjj}{\ensuremath{m_{12}}}
\newcommand{\mz}{\ensuremath{m_{\rm Z}}\xspace}
\newcommand{\as}{\ensuremath{\alpha_{\rm s}}\xspace}
\newcommand{\asmz}{\ensuremath{\as(\mz)}\xspace}
\newcommand{\asmzPDF}{\ensuremath{\as^{\rm PDF}(\mz)}\xspace}
\newcommand{\asmzf}{\ensuremath{\as^{\Gamma}(\mz)}\xspace}
\newcommand{\asmur}{\ensuremath{\alpha_{\rm s}(\mur)}\xspace}
\newcommand{\chad}{\ensuremath{c_{\rm had}}\xspace}
\newcommand{\ord}{\ensuremath{\mathcal{O}}\xspace}
\newcommand{\PDFasResult}{0.1142}
\newcommand{\HonePDF}{H1PDF2017\,{\protect\scalebox{0.8}{[NNLO]}}}
\newcommand{\TODO}{\color{red} Todo.}
\newcommand{\NEW}{\color{blue}}

\newcommand{\tQ}{\ensuremath{\tau_{zQ}}\xspace}
\newcommand{\tb}{\ensuremath{\tau_1^b}\xspace}
\newcommand{\pzb}{\ensuremath{P_{z}^\text{Breit}}\xspace}
\newcommand{\pzbi}{\ensuremath{P_{z,i}^\text{Breit}}\xspace}


\newcommand{\redtext}[1]{\textcolor{red}{#1}}
\newcommand{\bluetext}[1]{\textcolor{blue}{#1}}
\definecolor{myGreen}{rgb}{0.0,0.6,0.1}
\newcommand{\greentext}[1]{\textcolor{myGreen}{#1}}

\newcommand{\commenting}[2]{\redtext{(#1: #2)}}
\newcommand{\reply}[2]{\greentext{(#1: #2)}}

\newcommand{\qsqr}{\ensuremath{Q^2}}
\newcommand{\sqrts}{\ensuremath{\sqrt{s}}}
\newcommand{\ep}{\ensuremath{ep}}
\newcommand{\invpb}{\ensuremath{\mathrm{pb}^{-1}}}

\hyphenation{mig-ra-tion}
\hyphenation{ge-ne-ra-tors}

\received{March 2024}
\printhistory
\motto{}{To be submitted to EPJC \hfill DESY-24-035}

\title{Measurement of the 1-jettiness event shape observable in 
  deep-inelastic electron-proton scattering at HERA}

\subtitle{H1 Collaboration}

\author[50]{\fnm{V.}\sur{Andreev}}
\author[34]{\fnm{M.}\sur{Arratia}}
\author[46]{\fnm{A.}\sur{Baghdasaryan}}
\author[9]{\fnm{A.}\sur{Baty}}
\author[40]{\fnm{K.}\sur{Begzsuren}}
\author[17]{\fnm{A.}\sur{Bolz}}
\author[30]{\fnm{V.}\sur{Boudry}}
\author[15]{\fnm{G.}\sur{Brandt}}
\author[26]{\fnm{D.}\sur{Britzger}}
\author[7]{\fnm{A.}\sur{Buniatyan}}
% ...
\affil[7]{\orgaddress{School of Physics and Astronomy, University of Birmingham, Birmingham, United Kingdom$^{f3}$}}
% ...
\affil[9]{\orgaddress{University of Illinois, Chicago, IL 60607, USA}}
% ...
\affil[15]{\orgaddress{II. Physikalisches Institut, Universität Göttingen, Göttingen, Germany}}
% ...
\affil[17]{\orgaddress{Deutsches Elektronen-Synchrotron DESY, Hamburg and Zeuthen, Germany}}
% ...
\affil[26]{\orgaddress{Max-Planck-Institut für Physik, München, Germany}}
% ...
\affil[30]{\orgaddress{LLR, Ecole Polytechnique, CNRS/IN2P3, Palaiseau, France}}
% ...
\affil[34]{\orgaddress{University of California, Riverside, CA 92521, USA}}
% ...
\affil[40]{\orgaddress{Institute of Physics and Technology of the Mongolian Academy of Sciences, Ulaanbaatar, Mongolia}}
% ...
\affil[46]{\orgaddress{Yerevan Physics Institute, Yerevan, Armenia}}
% ...
\affil[50]{\orgaddress{Affiliated with an institute covered by a current or former collaboration agreement with DESY}}

\abstract{\unboldmath%
The H1 Collaboration reports the first measurement of the 1-jettiness
event shape observable \tb in neutral-current deep-inelastic
electron-proton scattering (DIS). The observable \tb is equivalent to a thrust observable defined in the Breit frame.
The data sample was collected at the HERA $ep$ collider in the years 2003--2007 with center-of-mass energy of $\sqrts=319\,\GeV$, corresponding to an integrated luminosity of 351.1\,\invpb. 
Triple differential cross sections are provided as a function of \tb, event virtuality \Qsq, and inelasticity $y$, in the kinematic region $\Qsq>150\,\GeVsq$.
Single differential cross section are provided as a function of \tb\ in a limited kinematic range.
Double differential cross sections are measured, in contrast, integrated over \tb\ and represent the inclusive neutral-current DIS cross section measured as a function of \Qsq\ and $y$.
The data are compared to a variety of predictions and include
classical and modern Monte Carlo event generators, predictions in fixed-order
perturbative QCD where calculations up to $\mathcal{O}(\as^3)$ are
available for \tb\ or inclusive DIS, and resummed predictions at
next-to-leading logarithmic accuracy matched to fixed order
predictions at $\mathcal{O}(\as^2)$.
These comparisons reveal sensitivity of the 1-jettiness observable to QCD parton
shower and resummation effects, as well as the
modeling of hadronization and fragmentation.
Within their range of validity, the fixed-order predictions provide a good description of the data.
Monte Carlo event generators are predictive over the full measured range
and hence their underlying models and parameters can be constrained by comparing to the presented data.
}
\maketitle

\section{Introduction}
Measurements in high-energy lepton-proton deep-inelastic scattering
(DIS) have played an important role in understanding the structure of
Quantum Chromodynamics
(QCD)~\cite{Abramowicz:1998ii,Klein:2008di,Newman:2013ada,Gross:2022hyw}.  
Inclusive neutral-current DIS cross section measurements probe the distribution of partonic constituents of the proton, and test 
perturbative QCD (pQCD) over a wide range of energy scale.
Beyond those inclusive cross sections, dedicated measurements of the
shape and substructure of the hadronic final state (HFS) provide
rigorous tests of pQCD calculations.
Observables related to the HFS are among others the
properties of jets, heavy-quark production, or event
shape quantities.
They are sensitive to the strong coupling constant and the gluon content of the
proton. In addition, they can be used to test the modeling of
non-perturbative (NP) effects, particularly hadronization and
fragmentation.  
However, comprehensive measurements of HFS observables over the full
HFS phase space in neutral-current DIS were not performed in the past due to
experimental and theoretical limitations. 


Event shapes have been studied extensively in $e^+e^-$
collisions~\cite{ALEPH:1996sio,ALEPH:2003obs,L3:1995eyy,L3:1997bxr,L3:1998ubl,L3:2002oql,L3:2004cdh,DELPHI:1999vbd,DELPHI:2003yqh,DELPHI:2004omy,SLD:1994idb,MovillaFernandez:1997fr,Biebel:1999zt,JADE:1999zar,MovillaFernandez:2001ed,Pahl:2009zwz,CELLO:1989okb},
and in hadron-hadron
collisions~\cite{CDF:2011yfm,CMS:2011usu,ALICE:2012cor,ATLAS:2012tch,CMS:2013lua,CMS:2014tkl,ATLAS:2016hjr,CMS:2018svp,ATLAS:2020vup,Alvarez:2023fhi,ATLAS:2023tgo}.
Several event shape observables were also measured in neutral-current (NC)
DIS using data from the HERA-I data taking period (1992-2000)~\cite{H1:1997hbl,H1:1999wfh,ZEUS:2002tyf,H1:2005zsk,ZEUS:2006vwm},
which demonstrated sensitivity to the strong coupling constant \asmz,
as well as to hadronization and resummation effects.  
Nevertheless, event shapes have not been studied to date as extensively in DIS as in  $e^+e^-$ and hadronic collisions
due to the more limited precision in predicting event shapes in DIS as compared to $e^+e^-$ collisions~\cite{Kluth:2006bw,Becher:2008cf,Abbate:2010xh,LL2022}.
In this article, the H1 Collaboration reports the first measurement of the
1-jettiness event shape observable \tb\ in $ep$ collisions.
This variable has theoretical advantages over previously
studied event shape observables, since it is free of non-global
logarithms~\cite{Dasgupta:2001sh,Dasgupta:2003iq} and thus can be
calculated with high theoretical accuracy.
Furthermore, it is closely related  to event shapes in $e^+e^-$
collisions.

A traditional event shape observable is thrust $T$~\cite{Brandt:1964sa,Farhi:1977sg},
which quantifies the momentum distribution of the HFS along a defined axis.
There is freedom in choosing the projection axis, normalization and reference frames to analyzing the thrust observable in DIS.
A common choice of these conditions is given by the
Breit frame of reference~\cite{Streng:1979pv}, with the polar angle dividing the
event into two hemispheres. 
These are referred to as the \emph{current} (or \emph{jet})
hemisphere, $\mathcal{H_C}$, and the \emph{target fragmentation} (or
\emph{beam}) hemisphere, with polar angles smaller or larger than
$\tfrac{\pi}{2}$, respectively\,\footnote{This sign convention for the
$z$ axis in the Breit frame is opposite to that of the \emph{HERA
laboratory coordinate system}, where the proton moves along the
positive $z$ direction, while in the Breit frame as defined here it moves along the negative $z$ direction. The photon then moves in the
positive $z$ direction~\cite{Streng:1979pv}. Some HERA papers
define the $z$-axis in the Breit frame with opposite sign. }.  
%
In the Breit frame, the photon momentum\,\footnote{In NC DIS, the interaction is mediated by a photon, 
  $\gamma Z$ interference, or $Z$ exchange, which is denoted
  \emph{photon exchange} in the following. The photon four-momentum
  $q$ is determined from the incoming and 
  outgoing lepton four-vectors, $q=k-k^\prime$. The photon
  virtuality is $\Qsq=-q\cdot q$.} %
is aligned with the positive $z$ axis, i.e.\ $q_{b}=(0,0,Q;0)$.
A natural choice for the axis of Thrust is the photon
axis with normalization $\tfrac{Q}{2}$, and thus this variant of
thrust is computed as

\begin{equation}
  T_{zQ} = 2 \sum_{i\in\mathcal{H_C}} \frac{P_{z,i}}{Q}\,,
  \label{eq:Thrust}
\end{equation}

\noindent
where the sum runs over all HFS particles in the Breit frame current
hemisphere $\mathcal{H_C}$~\cite{Feynman:1973xc}.
This variant of Thrust is a scaling variable~\cite{Streng:1979pv}.
%
%

A generalized set of inclusive observables defined with respect to the Breit frame axis, named \emph{current jet thrust} observables, is discussed in Ref.~\cite{Bouchiat:1980nz}. Power corrections, next-to-leading order QCD corrections, and resummed predictions for such observables have been calculated~\cite{Webber:1995ka,Graudenz:1997sp,Dasgupta:1997ex,Dasgupta:1998xt,Antonelli:1999kx,Dasgupta:2002dc,Dasgupta:2003iq,Gehrmann:2019hwf}. A change in notation has likewise been introduced,
%
\begin{equation}
  \tQ = 1-T_{zQ}.
  \label{eq:tQ}
\end{equation}
\noindent
It was shown that \tQ\
is infrared and collinear safe, that it fulfills the
criteria required for analytic or automatized resummation, and that it is free of non-global logarithms~\cite{Streng:1979pv,Banfi:2003je,Banfi:2004yd,caesar}.

In the framework of soft-collinear effective theory (SCET), \tQ\ is one
variant of a more general class of global event-shape observables called 1-jettiness~\cite{Dasgupta:1998xt,Stewart:2010tn,Kang:2013nha,Kang:2014qba}, 
\begin{equation}
  \tQ = \tb = \frac{2}{\Qsq}\sum_{i\in X}\min\left(\xbj P\cdot p_i ,
  (q+\xbj P)\cdot p_i\right)\,,
  \label{eq:tau1b}
\end{equation}
\noindent
where in this case $i$ runs over {\it all} final state particles,
which renders the calculation of \tb\ free of non-global logarithms.
Here, \xbj{} is the Bjorken-$x$ scaling variable, $q$ is the exchanged photon four-momentum, and $P$ is the incoming proton four-momentum.
The observable \tb\ is a special case of a general class of $N$-jettiness observables~\cite{Stewart:2010tn,Jouttenus:2011wh,Kang:2012zr,Kang:2013lga,Cao:2024ota}. 
Using momentum conservation, \tb\ can be rewritten
\begin{equation}
  \tb
  = 1- 2\cdot\sum_{i\in X}\max\left( 0, \frac{q\cdot p_i  }{q\cdot q}\right) \\
  = 1- 2\cdot\sum_{i\in  \mathcal{H_C}} \frac{q\cdot p_i  }{q\cdot q}\,,
  \label{eq:tau1bH1}
\end{equation}
\noindent
where the sum runs over all HFS particles in the first expression, but
only over particles in the Breit frame current hemisphere $\mathcal{H_C}$
in the second expression.
Following Eq.~\eqref{eq:tau1bH1} the 1-jettiness is proportional to
the sum of particle 4-momenta radiated into $\mathcal{H}_C$, and
projected onto   the photon four-momentum.
%
%
It ranges from zero to unity, with $\tb\sim0$ indicating an
event structure with a single collimated jet emitted into
$\mathcal{H_C}$ along the photon direction.
There are DIS event configurations at low $\xbj$ where the
current hemisphere is empty, corresponding to
$\tb=1$~\cite{Streng:1979pv,Dasgupta:1997ex,Kang:2014qba}.  
When the full range of \tb\ is considered, $\tb\in[0,1]$, each event
in NC DIS has an associated value of \tb.


This article presents a first measurement of \tb, which constitutes the first triple-differential measurement of an hadronic event shape observable over the full phase space of selected NC DIS event kinematics. It is made possible by recent theoretical
developments and improved experimental reconstruction techniques and
is based on
data recorded with the H1 detector at the HERA collider for
\ep\ collisions at $\sqrts=319\,\GeV$ with an integrated luminosity of
$351.1\,\invpb$.
Differential cross sections as a function of {\tb}, as well as the
triple-differential cross section as a function of photon virtuality
$\Qsq$, event inelasticity $y$, and {\tb}, are reported over a large kinematic range.
Inclusive DIS cross sections as a function of
\Qsq\ and $y$, obtained by integrating the triple-differential
cross section, are also measured.
The data are compared to theoretical
calculations based on Monte Carlo (MC) event generators and pQCD, probing
their sensitivity to parton showering, resummation, \as,
and non-perturbative effects.
%

\section{Experimental setup}
The data were taken with the H1 detector in the years 2003 to 2007
using electron or positron\,\footnote{The term `electron' is used in the following to refer to both electrons and positrons.}
beams that were collided with a proton
beam at a center-of-mass energy of $\sqrt{s}=319\,\GeV$. 
The data correspond to an integrated luminosity of
$\mathcal{L}=351.1\,$pb$^{-1}$~\cite{H1:2012wor}.
The H1
experiment~\cite{H1:1993ids,H1CalorimeterGroup:1993boq,H1:1996prr,H1:1996jzy,H1SPACALGroup:1996ziw,Pitzl:2000wz}
is a general purpose particle detector with full 
azimuthal coverage around the electron--proton interaction region.
The H1 Collaboration uses a right handed coordinate system, where the
proton beam direction defines the positive $z$ axis. The nominal
interaction point is located at $z=0$.

The H1 detector consists of several subsystems.
The main subsystems used for this analysis are the tracking detectors,
the liquid argon (LAr) calorimeter, and the backward calorimeter
(SpaCal).
All these systems are situated inside a superconducting solenoid that provides a magnetic
field of 1.16\,T.
The central tracking system consists of drift and
proportional chambers, together with silicon-strip detectors close to
the interaction region and covers the polar angular range $15^\circ<\theta<165^\circ$.
The transverse momentum resolution of charged particles is
$\sigma_{p_\text{T}} /p_\text{T} = 0.2\,\%\,p_\text{T}/\GeV \oplus 1.5\,\%$.

The LAr sampling calorimeter consists of an electromagnetic section made of
lead absorbers and an hadronic section with steel absorbers, covering the polar angular range
$4^\circ <\theta < 154^\circ$. 
The energy resolution is $\sigma_E /E = 11\%/\sqrt{E/\GeV} \oplus 1\%$
for electrons and $\sigma_E /E \simeq 55\%/ \sqrt{E/\GeV} \oplus 3\%$ for
charged pions.
The LAr calorimeter is used for triggering and particle reconstruction in this analysis.

The SpaCal (`Spaghetti Calorimeter') is a lead-scintillating fiber
calorimeter with electromagnetic and hadronic sections, covering the
backward direction with polar angular range $153^\circ < \theta <
177^\circ$. The electromagnetic energy resolution $\sigma(E)/E$ is
 $7 \%/\sqrt{E/\GeV}\oplus 1\%$, and in the hadronic section it is $(56\pm13)\,\%$
for charged pions.

Online triggering and event selection follow the procedure described
in Ref.~\cite{H1:2014cbm}.
Events are triggered by a high-energy
cluster in the LAr calorimeter, with the  scattered electron identified using
isolation criteria.
Events are accepted if the scattered electron has
energy $E_{e}^{\prime}>11\,\GeV$ and is found in the high-efficient
regions of the LAr calorimeter trigger system, which corresponds to about 90\% of its
$\eta$--$\varphi$ coverage.
The trigger efficiency for inclusive DIS
events is greater than 99\,\%. 

In order to suppress non-collision backgrounds from cosmic muons,
beam-gas interactions, and high energetic muons produced off the proton
  beam in the HERA tunnel, a triggered event must fulfill
certain requirements~\cite{H1:2012qti,H1:2014cbm}.
Events with an energy deposition in the range $\theta>175^\circ$
are found to be sensitive to non-collision backgrounds and are removed. Events with a topology similar to QED Compton events are also removed~\cite{H1:2014cbm}.

Tracks of charged particles are
reconstructed from hits in the tracking detectors.
The energy
depositions of charged and neutral particles in the calorimeters are
clustered and calibrated.
Particle candidate four-vectors are then reconstructed offline using an
energy-flow algorithm which combines information from tracks with
that from
clusters~\cite{energyflowthesis,energyflowthesis2,energyflowthesis3}.
The energy of particle candidates is calibrated using a neural-network based
shower-classification algorithm and a dedicated jet-calibration
sample~\cite{Kogler:2011zz}.
The scattered electron candidate is identified as the electromagnetic
cluster in the LAr calorimeter
which has the highest energy in the event, and which satisfies isolation criteria and is  matched with a track~\cite{H1:2003xoe}.


Radiative photons may distort the kinematic reconstruction.
Isolated high-energy depositions in the central or backward part of
the electromagnetic calorimeters ($\theta>\frac{\pi}{3}$)
are found to have a good association with photons radiated off the incoming or scattered electron.
Such clusters are treated in the same manner as radiated photons at particle level.
If their angular distance to the scattered electron candidate is
smaller than the distance to the negative $z$-axis, they are recombined with
the scattered electron to form a dressed 
scattered electron with four-vector $p_e$; otherwise, their energy deposition is
removed from the event record.
This procedure suppresses photons from initial-state QED radiation,
which effectively reduce the energy of the incoming lepton.
In addition, it provides well-reconstructed
observables in the presence of final-state QED radiation.
%
%
All remaining particle candidates which are not classified as the
scattered electron, comprise the hadronic final state (HFS), with
their sum corresponding to the HFS four-vector, $p_h$. 
The 
incoming electron energy is reconstructed by the $\Sigma$ method~\cite{Bassler:1994uq}, which makes use of the energy and longitudinal momentum of $p_h+p_e$.
The photon four-vector $q$ is calculated using the incoming electron energy and $p_e$.
HFS particles satisfying $p_i\cdot q<0$ are associated to
the current hemisphere $\mathcal{H}_C$ in the Breit frame, with their
sum forming the current-hemisphere four-vector $p_c$. 
%
%

The DIS kinematic observables are
calculated from the four vectors $p_e$ with energy $E_e$ and transverse momentum $P_{\text{T},e}$, and $p_h$,
using the I$\Sigma$ method~\cite{Bassler:1994uq,Bassler:1997tv}, 
\begin{equation}
  \begin{array}{cc}
  y_\Sigma = \frac{\Sigma_h}{\Sigma_h + \Sigma_e}\,, &
  \Qsq_\Sigma = \frac{P_{\text{T},e}^2}{1-y_\Sigma}, \\
  \multicolumn{2}{c}{
 x_{\text{I}\Sigma} = \frac{E_{e}}{E_{p_0}}\frac{\cos^2\frac{\theta_{e}}{2}}{y_\Sigma}\,,}
  \end{array}
 \label{eq:DIS}
\end{equation}
\noindent
and for certain purposes the electron method is employed instead,
\begin{equation}
  y_e=1-\frac{\Sigma_e}{2E_{e_0}}
  ~~~\text{using}~~~
   \Qsq_e = \frac{P_{\text{T},e}^2}{1-y_e}\,,
 \label{eq:DISeMethod}
\end{equation}
where the variables $\Sigma_i$ denote $\Sigma_i = E_i-P_{z,i}$ for $i=e,h$.
The electron and proton beam energies are $E_{e_0}=27.6\,\GeV$ and
$E_{p_0}=920\,\GeV$, respectively. The I$\Sigma$ method achieves good
resolution in $y$ over the entire kinematic
range~\cite{Bassler:1994uq,Bassler:1997tv,Arratia:2021tsq}.
Inserting $-Q^2_\Sigma$ for $q\cdot q$ in Eq.~\eqref{eq:tau1bH1} provides best resolution
for the calculation of $\tb$~\cite{Arratia:2022wny}.
However, the electron method provides higher resolution in $Q^2$
and is therefore used for $Q^2$ dependent measurements, $\propto\frac{\text{d}\sigma}{\text{d}\Qsq}$.
The electron-beam energy does not enter the equations of the I$\Sigma$
method, which thereby is largely insensitive to initial state QED radiative
effects.
The hadronic angle is defined in this analysis as
$\gamma_h=2\arctan{\frac{\Sigma_h}{P_{\text{T},e}}}$.

To achieve high resolution and to reduce initial-state QED radiation
effects, events are selected as follows: 
the ratio of the transverse momentum of the HFS and scattered lepton
satisfies $0.6<P_{\text{T},h}/P_{\text{T},e}<1.6$; their difference 
satisfies $P_{\text{T},h} - P_{\text{T},e} < 5\,\GeV$; and the
longitudinal energy-momentum balance is in the range $45<\Sigma_h+\Sigma_e<62\,\GeV$.
%
%
The current hemisphere four-vector has polar angle $\theta_c$ and
pseudorapidity $\eta_c$, which are used to classify phase-space
regions 
with lower resolution or reconstruction performance, and to suppress 
QED radiative effects and contributions from non-collision
backgrounds.
The following types of events are rejected at detector-level:
\begin{compactitem}
\item Events with $\theta_c-\gamma_h>\frac{\pi}{2}$,
  or with $\theta_c >  3.05$,
  or with
  $\theta_c>2.6$ where $\beta$ of the boost vector $b=2\xbj P+q$ ($\beta=\tfrac{\vec{b}^2 }{b_E^2}$) fulfills
  $\beta>0.9$, are found to be events with an initial state radiated
  photon or where a radiated photon was converted to hadrons;
\item Events with $\theta_c<0.12$ cannot be reconstructed well due to limited acceptance in the forward direction;
\item Events at low \tb ($\tb<0.3$) are found to have poor resolution when $\theta_c/\gamma_h>1.3$;
\item Events at large \tb and low \Qsq ($\tb>0.65$ and $\Qsq<700\,\GeVsq$) have poor resolution when $\beta>0.9$ and  $P_{\text{T},c}/P_{\text{T},e}<0.35$;
\item Events at very high value \tb ($\tb>0.95$) have poor resolution when $\beta>0.97$;
\item Events at low \Qsq ($\Qsq<700\,\GeVsq$) have poor resolution for $\eta_c+\ln{(\tan{\frac{\gamma_h}{2}})}>0.3$.
\item Events with $y_e>0.94$ are removed in order to suppress background
contributions from photoproduction.

\end{compactitem}
%
%
After the application of all acceptance, background and cleaning cuts,
about 35~to 80\,\% of the events generated within the phase space of the
measurement are accepted at the detector-level.

\begin{figure}[h]
\centering
\includegraphics[height=0.48\textwidth,trim={0 0 0 0 },clip]{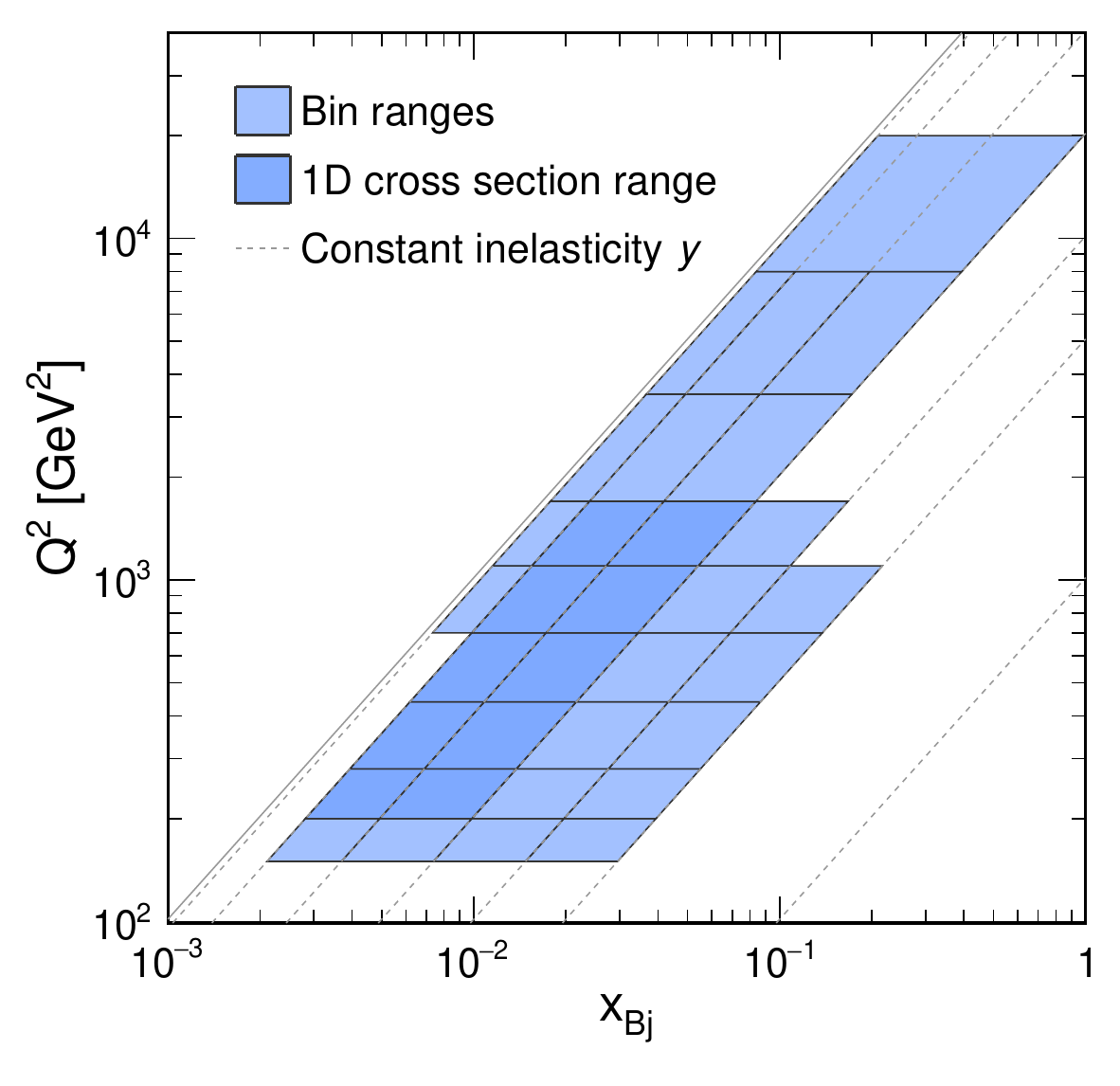}
\caption{
\label{fig:kineplane}
  DIS kinematic plane as a function of $\xbj$ and \Qsq. The
  kinematic range of the measurement is displayed as colored area.
  The dark shaded area indicates the kinematic range of the
  reported single-differential cross section
  ($200\leq\Qsq<1700\,\GeVsq$ and $0.2\leq y<0.7$).
}
\end{figure}

The final selected NC DIS kinematic range of the analysis is shown in
Figure~\ref{fig:kineplane} as a function of $\xbj$ and \Qsq.
The kinematic range is limited by acceptance, resolution, and trigger:
The polar-angle acceptance of the LAr calorimeter, $\theta\lesssim154^\circ$, defines the
\Qsq\ region of the measurement, $\Qsq>150\,\GeVsq$;
The electron energy acceptance corresponds to $y\lesssim0.7$
for $Q^2\lesssim1000\,\GeVsq$;
The inelasticity is required to be  $y>0.05$, since acceptance and
resolution deteriorate for lower $y$, where the hadronic angle 
$\gamma_h$ becomes small and  the Lorentz-factor $\beta$ of the boost
vector approaches unity.



The binning used in this analysis is also presented in
Table~\ref{tab:binnings}.
%
%
\begin{table*}[t]
  \footnotesize
  \begin{center}
    \begin{tabular}{lccc}
      \toprule
      Observable & Binning name &  Notation & Binning \\
      \midrule
      \tb   & \tb nominal          &  $b_{\tb}^{(11)}$  &  $[0,0.05,0.1,0.15,0.22,0.3,0.4,0.54,0.72,0.86,0.98,1.0+\epsilon]$ \\ 
      \tb   & \tb coarse           &  $b_{\tb}^{(8)}$   &  $[0,0.05,0.1,0.15,0.22,0.3,0.4,0.6,1.0+\epsilon]$  \\ 
      \tb   & \tb coarse low $y$   &  $b_{\tb}^{(7)}$   &  $[0,0.1,0.15,0.22,0.3,0.4,0.6,1.0+\epsilon]$ \\
      $y$   & $y$              &  $b_{y}$           &  $[0.05, 0.1, 0.2, 0.4, 0.7, 0.94]$ \\
      $\Qsq$& $\Qsq$           &  $b_{\Qsq}$        &  $[150, 200, 280, 440, 700, 1100, 1700, 3500, 8000, 20000[$ \GeVsq \\
          \bottomrule
    \end{tabular}
  \end{center}
    \caption{
    \label{tab:binnings}
      Binnings used in the analysis.
    }
\end{table*}
Over most of
the phase space the binning is dictated by the resolution of $\tb$. Coarser binning is chosen at the boundaries of the
kinematic range due to the following factors: 
\begin{compactitem}
\item low $y$ or large \Qsq: large values of $\tb$ have low statistics
or lower resolution; 
\item low $y$ and $\tb<0.05$: very low cross section which cannot be resolved. 
\end{compactitem}
Altogether 308 cross section values are measured.
The respective highest \tb\ bins ($0.98\leq\tb\leq1$ and
$0.6<\tb\leq1$) include events with $\tb=1$, in which the current
hemisphere is empty.
Such events are related to low values of
$x$~\cite{Streng:1979pv,Dasgupta:1997ex,Kang:2014qba}, 
and are present because \tb\ is defined in the Breit frame.
They are absent for event shape observables defined in the partonic
center-of-mass frame, like thrust in $e^+e^-$ collisions~\cite{Dasgupta:2003iq,Kang:2014qba}.
 

\section{Monte Carlo simulations and model predictions}
MC event generators are used to correct the data for
detector acceptance and resolution effects, and for 
contributions from $ep$ collisions outside the phase-space of this analysis.
The generated events are processed with a detailed
simulation of the H1 detector based on GEANT3~\cite{Brun:1987ma}, supplemented by fast shower simulations~\cite{Fesefeldt:1985yw,Grindhammer:1989zg,Gayler:1991cr,Kuhlen:1992ey,Grindhammer:1993kw,Glazov:2010zza}.
The simulated data are reconstructed and processed with the same
analysis algorithm as the real data~\cite{DPHEPStudyGroup:2009gfj,Steder:2011zz,South:2012vh,DPHEPStudyGroup:2012dsv,Britzger:2021xcx,DPHEP:2023blx,Brun:1997pa}.

Detector effects are corrected using \emph{regularized unfolding}, based on the simulation of NC DIS events using the MC event generators Djangoh~1.4~\cite{Charchula:1994kf} and
Rapgap~3.1~\cite{Jung:1993gf}.
Djangoh uses Born level matrix elements
for NC DIS and dijet production and applies the color dipole model from Ariadne~\cite{Lonnblad:1992tz} for higher-order emissions.
Rapgap implements Born level matrix elements for NC DIS and dijet production 
and uses the leading logarithmic approximation for parton shower emission.
Both generators are interfaced to Heracles~\cite{Kwiatkowski:1990es} for higher order QED effects at the lepton vertex. Both generators utilise the CTEQ6L parton distribution function (PDF) set~\cite{Pumplin:2002vw} and the Lund hadronization
model~\cite{Andersson:1983ia,Sjostrand:1994kzr}.
Hadronization model parameters were determined by the ALEPH Collaboration~\cite{ALEPH:1996oqp}.

Contributions to the event yield from processes other than NC DIS are
simulated with a variety of MC event generators.
Photoproduction events are simulated using
Pythia~6.2~\cite{Sjostrand:1993yb,Sjostrand:2001yu}.
Events with di-lepton production are generated using
Grape~\cite{Abe:2000cv}. A sample of QED Compton events is simulated using the
program Compton~\cite{Courau:1992ht}. Deeply virtual Compton
scattering (DVCS) is simulated with Milou~\cite{Perez:2004ig}. 
NC DIS events for lower values of \Qsq\ and for charged current DIS are
simulated with Djangoh.
Overall, more than $3.4\times10^{8}$ signal events were simulated with Djangoh and Rapgap,
and about $5.3\times10^{8}$ events were simulated in total. 



The fully-corrected \tb\ cross section measurements are
compared to a set of theoretical predictions.
The following classes of prediction are studied:
two common MC event generators as used at HERA; three modern
general-purpose event generators which are widely used in high-energy physics;
one dedicated event generator incorporating transverse-momentum dependent
effects; and
two variants of fixed-order predictions.
In order to explore the sensitivity of the data to various QCD effects, in each MC event generator one element is varied, such as the parton shower model, the fixed-order prediction and its
matching or merging procedure with the parton shower, the
hadronization model, or the PDF set.
The fixed-order predictions include variations of the renormalization and factorization
scales.
In addition, dedicated predictions for the inclusive NC DIS cross
sections are studied.
%
The following MC event generators from the HERA era are considered:
\begin{itemize}
  \item Djangoh~1.4 and Rapgap~3.1, similar to
    those described above, but with radiative effects switched off
    in Heracles.
\end{itemize}

The following modern MC event generators are studied:
\begin{itemize}
  \item Pythia~8.3~\cite{Sjostrand:2014zea,Bierlich:2022pfr}, where
    the impact of the parton-shower is studied by
    employing three different parton-shower models:
    i) `default' dipole-like $p_\perp$-ordered shower with a local dipole
recoil strategy (Pythia~8.310)~\cite{Cabouat:2017rzi},
    ii) $p_\perp$-ordered Vincia parton
    shower based on the antenna formalism
    at leading color (Pythia~8.307)~\cite{Giele:2007di,Giele:2011cb,Giele:2013ema,Fischer:2016vfv},
    and iii) the Dire~\cite{Hoche:2015sya,Hoche:2017iem,Hoche:2017hno} parton
    shower which is an improved dipole-shower with additional treatment
    of collinear enhancements (Pythia 8.307).
    All models use the Pythia~8.3 default Lund string model for
    hadronization~\cite{Bierlich:2022pfr} and the PDF4LHC21 PDF
    set~\cite{PDF4LHCWorkingGroup:2022cjn} for the hard PDFs.
    The Vincia and Dire parton shower use a value of 0.118 for the
    strong coupling at the mass of the $Z$ boson.
  \item Powheg Box plus Pythia (Powheg+Pythia)~\cite{Banfi:2023mhz} implements predictions in
    NLO QCD matched to parton showers using the Powheg
    method~\cite{Nason:2004rx,Frixione:2007vw}, where the radiation
    phase space is parameterised according to the  Frixione-Kunszt-Signer (FKS) subtraction
    technique~\cite{Frixione:1995ms}. Dedicated momentum mappings
    preserve the DIS kinematic variables. The NLO predictions (up to
    $\mathcal{O}(\as)$) are then interfaced to parton showers and
    hadronziation from Pythia~8.308.
  \item Herwig~7.2~\cite{Bellm:2015jjp}, where the impact of different
    modelling of the hard interaction and its merging or matching with
    the parton-shower model is studied in three variants:
    The default prediction implements leading-order
    matrix elements supplemented with an angular-ordered 
    parton shower~\cite{Gieseke:2003rz} and cluster hadronization
    model~\cite{Webber:1983if,Marchesini:1991ch}.
    The second variant utilises MC@NLO~\cite{Frixione:2002ik}  which implements NLO
    matrix element corrections. Matching with the default
    angular-ordered parton shower is performed~\cite{Platzer:2011bc}.
    The third variant also utilises NLO matrix elements, but with
    dipole merging and a dipole parton shower~\cite{Platzer:2011bc}.
    Herwig-generated events are further processed with Rivet~\cite{Bierlich:2019rhm}.
  \item Sherpa~2.2~\cite{Sherpa:2019gpd,Gleisberg:2008ta}, where the
    modelling of hadronization effects can be studied by using two
    different hadronization models.
    The default Sherpa~2.2 predictions are based on multi-leg
    tree-level matrix elements from Comix~\cite{Duhr:2006iq} that are
    combined in the CKKW merging formalism~\cite{Catani:2001cc} with
    dipole showers~\cite{Catani:1996vz,Schumann:2007mg} and
    supplemented with the  cluster hadronization as implemented in
    AHADIC++~\cite{Winter:2003tt}.
    As an alternative prediction, the parton-level calculation is supplemented
    with the Lund string fragmentation model~\cite{Sjostrand:2006za}.
  \item
     Predictions from the pre-release version
     of Sherpa~3.0~\cite{Sherpa3} are provided by the Sherpa authors
     featuring a new cluster hadronization model~\cite{Chahal:2022rid} and
     matrix element calculation at NLO QCD
     obtained from OpenLoops~\cite{Buccioni:2019sur} with the Sherpa dipole
     shower~\cite{Schumann:2007mg} based on the truncated shower
     method~\cite{Hoeche:2009rj,Hoeche:2012yf}.
     The predictions are associated with scale uncertainties from a
     7-point scale variation by factors of two.
\end{itemize}

The following dedicated MC event generators are studied:
\begin{itemize}
  \item Cascade~3~\cite{CASCADE:2021bxe} implements off-shell processes via the automated matrix
    element calculators  KaTie~\cite{vanHameren:2016kkz} and
    Pegasus~\cite{Lipatov:2019oxs}, parton shower and hadronization
    through Pythia~6. This utilises transverse momentum
    dependent (TMD) PDF sets in the parton branching (PB)
    methodology~\cite{Hautmann:2017xtx}.
    Two PB TMD PDF sets are studied, denoted as \emph{set 1} and \emph{set 2}~\cite{Jung:2021mox}. They differ primarily in the
    scale choice for QCD evolution.
\end{itemize}

The following exact QCD predictions are studied:
\begin{itemize}
  \item Next-to-next-to-leading order (NNLO) predictions in perturbative QCD
    up to third order in \as ($\mathcal{O}(\as^3)$) are obtained for the
    process $ep\to e+2\text{jets}+X$  with the program NNLOJET~\cite{Ridder:2016nkl,Currie:2017tpe,Currie:2016ytq,Gehrmann:2019hwf}. The factorization and renormalization scales are chosen to be
    $\mu=Q$. Scale uncertainties are determined by the largest
    difference in the 7-point scale-variation prescription, with scale
    factors 0.5 and 2.
    The PDF set PDF4LHC21~\cite{PDF4LHCWorkingGroup:2022cjn} is used.  
    Non-perturbative correction factors are applied to these
    parton-level predictions as multiplicative correction factors for
    hadronization effects.
    These NNLO predictions are valid only in the region where the
    $ep\to e+2\text{jets}+X$ process dominates and hadronization corrections are small.
    This corresponds to the region $\tb\gtrsim0.1$ and $\tb\neq1$.
    The NNLO predictions can formally be brought to
    next-to-next-to-next-to-leading (N3LO) order for NC DIS using the projection to
    Born method~\cite{Currie:2018fgr,Gehrmann:2019hwf}.
\item Analytic predictions are provided in NNLO$\otimes$NLL$^{\prime}\otimes$Had
    accuracy~\cite{Knobbe:2023ehi}, using NLO pQCD matrix elements for dijet
    production projected onto the NNLO NC DIS cross section~\cite{Hoche:2018gti} and
    supplemented with automatised next-to-leading logarithmic accuracy
    using the Caesar
    framework~\cite{caesar,Banfi:2003je,Banfi:2004yd} as implemented in the Sherpa framework~\cite{Gerwick:2014gya}.
    For these predictions, hadronization corrections are applied as a multiplicative
    correction matrix obtained from Sherpa~3.0~\cite{Reichelt:2021svh}.
\end{itemize}

Further predictions for \tb\ have been reported previously but are not
available here. These include NNLO
predictions supplemented with parton shower~\cite{Hoche:2018gti},
next-to-next-to-leading logarithmic and next-to-NLL predictions~\cite{Kang:2015swk},
analytic NLL predictions~\cite{Kang:2014qba}, and N3LO
predictions using the projection to Born method and a dispersive model
for hadronization effects~\cite{Currie:2018fgr,Gehrmann:2019hwf}.


The following predictions are compared to the inclusive NC DIS cross sections measured in bins of $\Qsq$ and $y$:
\begin{itemize}
\item Predictions in NNLO pQCD are obtained with the program
  Apfel++~\cite{Bertone:2013vaa,Bertone:2017gds}, using the three loop
  splitting and coefficient
  functions~\cite{Moch:2004pa,Vogt:2004mw,Moch:2004xu,Vermaseren:2005qc} 
  together with the PDF set NNPDF31\_nnlo\_as\_0118.
  The predictions are associated with PDF uncertainties and with
  scale uncertainties, where the latter are defined as 7-point scale
  variations in the coefficient functions.
  An alternative prediction is obtained with the H1PDF2017{\scriptsize NNLO} PDF set~\cite{H1:2017bml}.
\item Predictions in approximate next-to-next-to-next-to-leading order
  (aN3LO) are obtained for inclusive NC DIS cross sections
  using the program
  Apfel++~\cite{Bertone:2013vaa,Bertone:2017gds} and using the MSHT PDFs
  at a3NLO order~\cite{McGowan:2022nag} and four-loup splitting functions
  from Refs.~\cite{Moch:2017uml,Vogt:2018miu,Moch:2021qrk}.
\end{itemize}


\section{Data correction: regularized unfolding}

The data are corrected for detector effects, background processes,
and higher-order QED effects using regularized unfolding~\cite{Blobel:1984ku,Schmitt:2012kp}.
This section presents the unfolding procedure and the measurement of the fully-corrected cross section.
In addition, correction factors for hadronization and
electroweak effects are discussed.

\begin{figure*}[t]
\centering
\includegraphics[width=0.32\textwidth,trim={0 0 0 0 },clip]{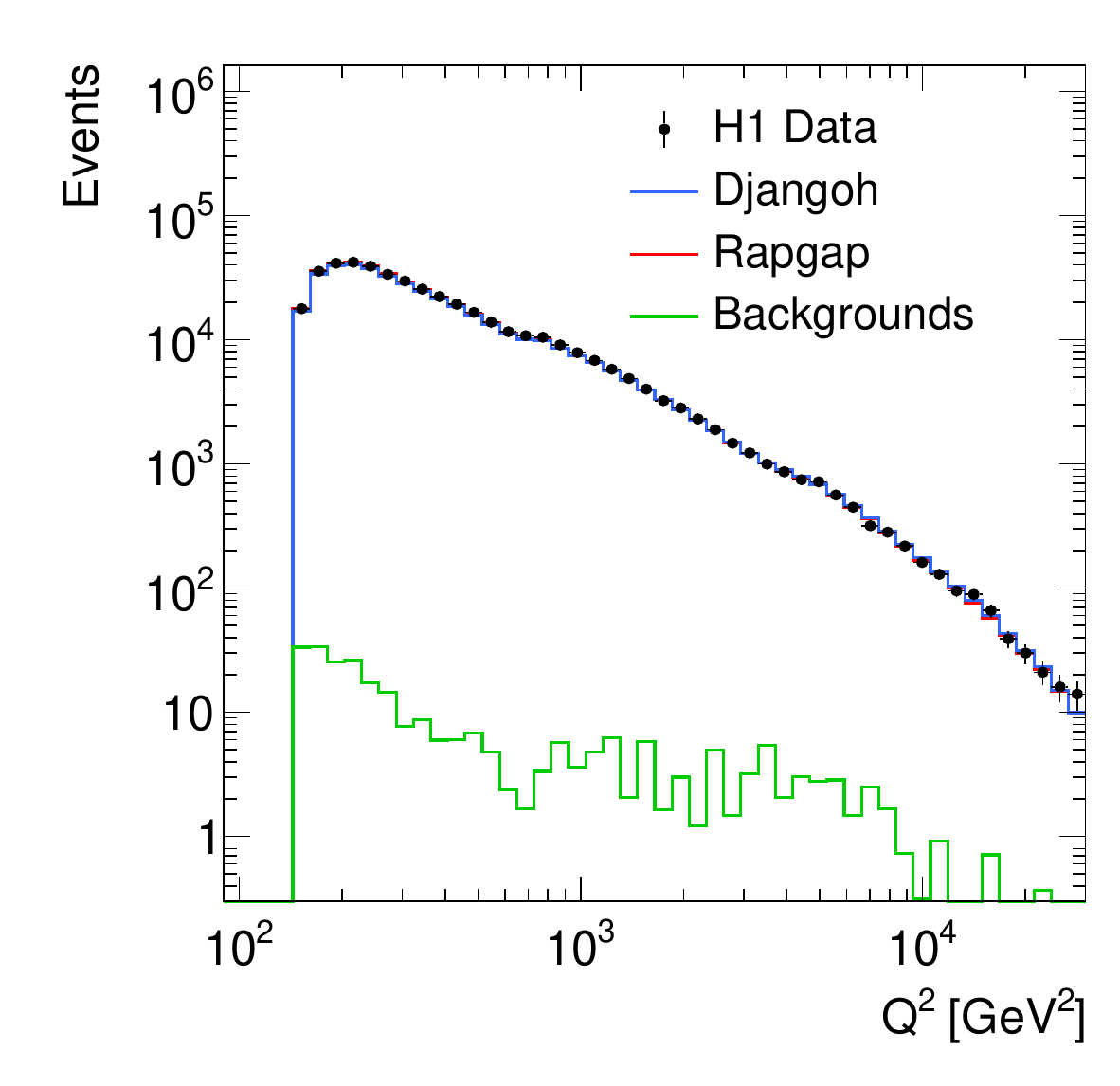}
\includegraphics[width=0.32\textwidth,trim={0 0 0 0 },clip]{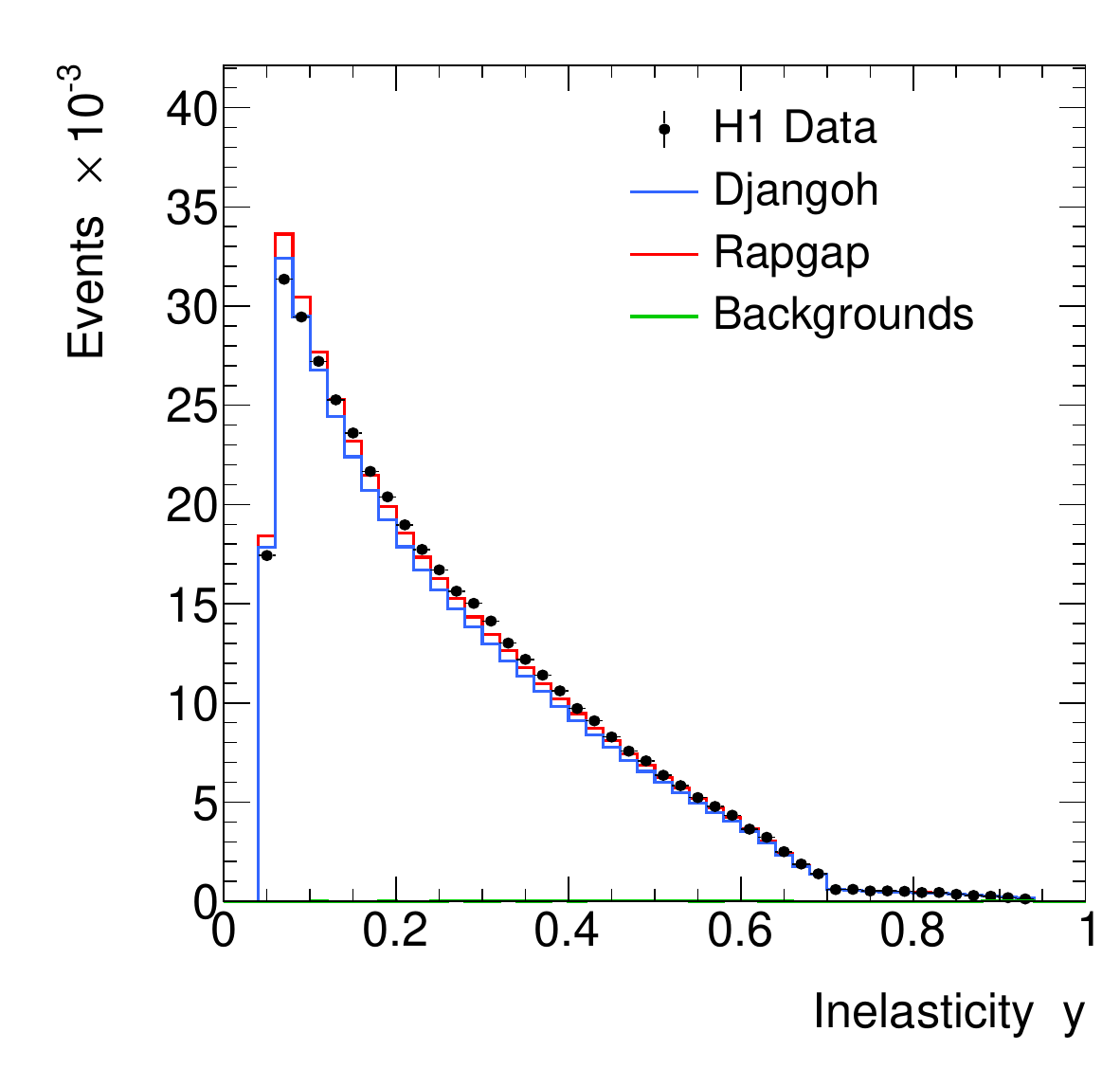}
\includegraphics[width=0.32\textwidth,trim={0 0 0 0 },clip]{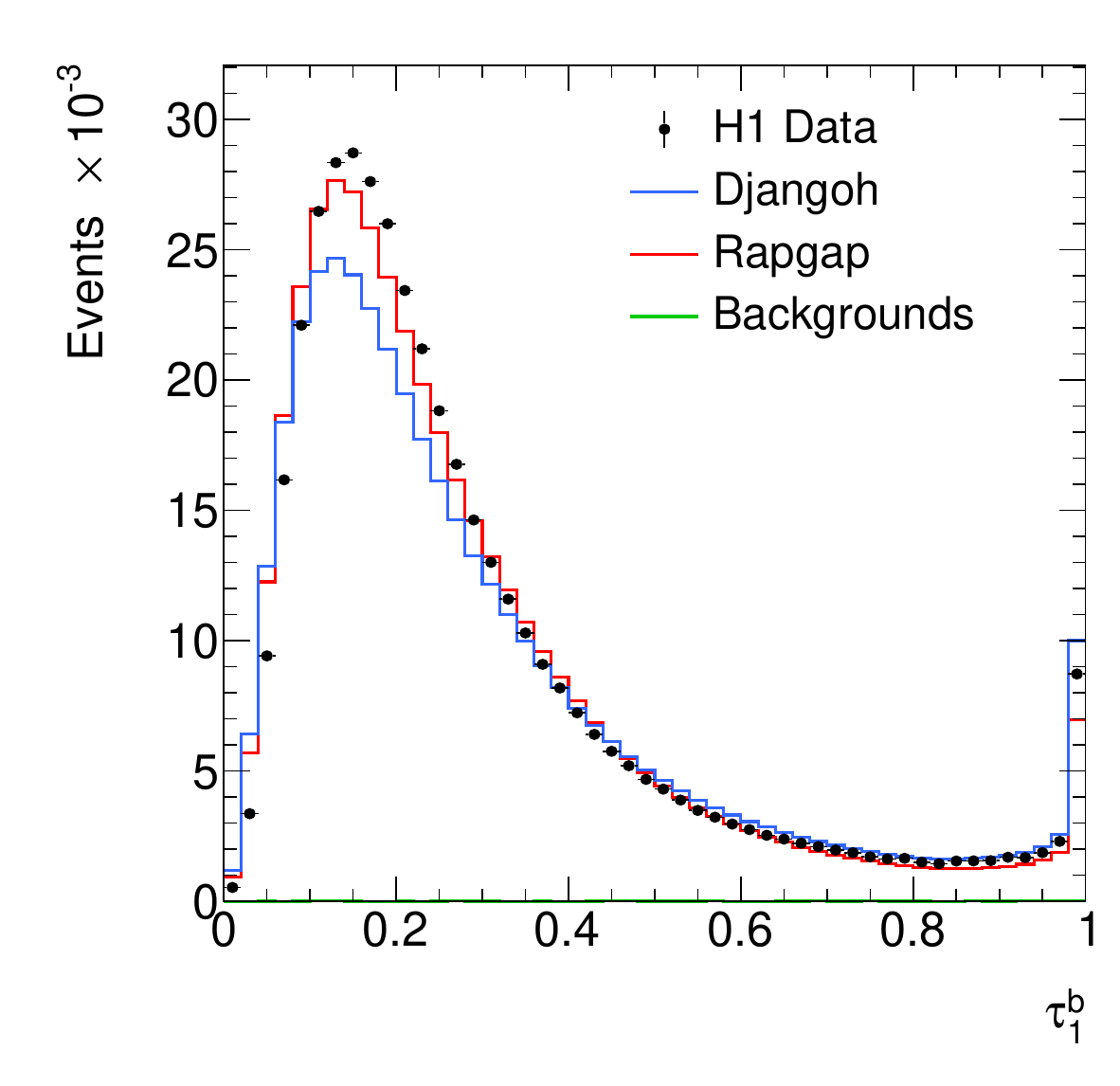}
\caption{
  \label{fig:ControlPlots}
  Detector level distributions of $\Qsq$ (left), $y$ (middle), and \tb~(right) of the
  selected data. The definition of these observables is
  given by Eq.~\eqref{eq:DIS} and~\eqref{eq:DISeMethod}. 
  The data are compared to the MC event simulations
  Djangoh and Rapgap.
  Both simulations include estimated background contributions from
  processes other than high-\Qsq\ NC DIS, as
  indicated by the green line (visible on logarithmic scale).
}
\end{figure*}

The signal Monte Carlo generators Djangoh and
Rapgap are used together with the detailed simulation of the H1
detector to generate synthetic detector-level events along with corresponding particle-level events (MC event simulation). 
Within the scope of the analysis, the detector response is found to
be well modelled in all studied aspects, 
which is essential for obtaining accurate unfolding results.
%
Figure~\ref{fig:ControlPlots} shows comparison of these simulations to data for the observables $\Qsq$, $y$, and $\tb$.
%
Both MC event simulations, although based on different physics models, are
found to describe the data well.
Comparisons to other observables, related closely to the detector performance, 
are discussed in the following.



\begin{figure*}[t]
\centering
\includegraphics[width=0.38\textwidth,trim={0 0 0 0 },clip]{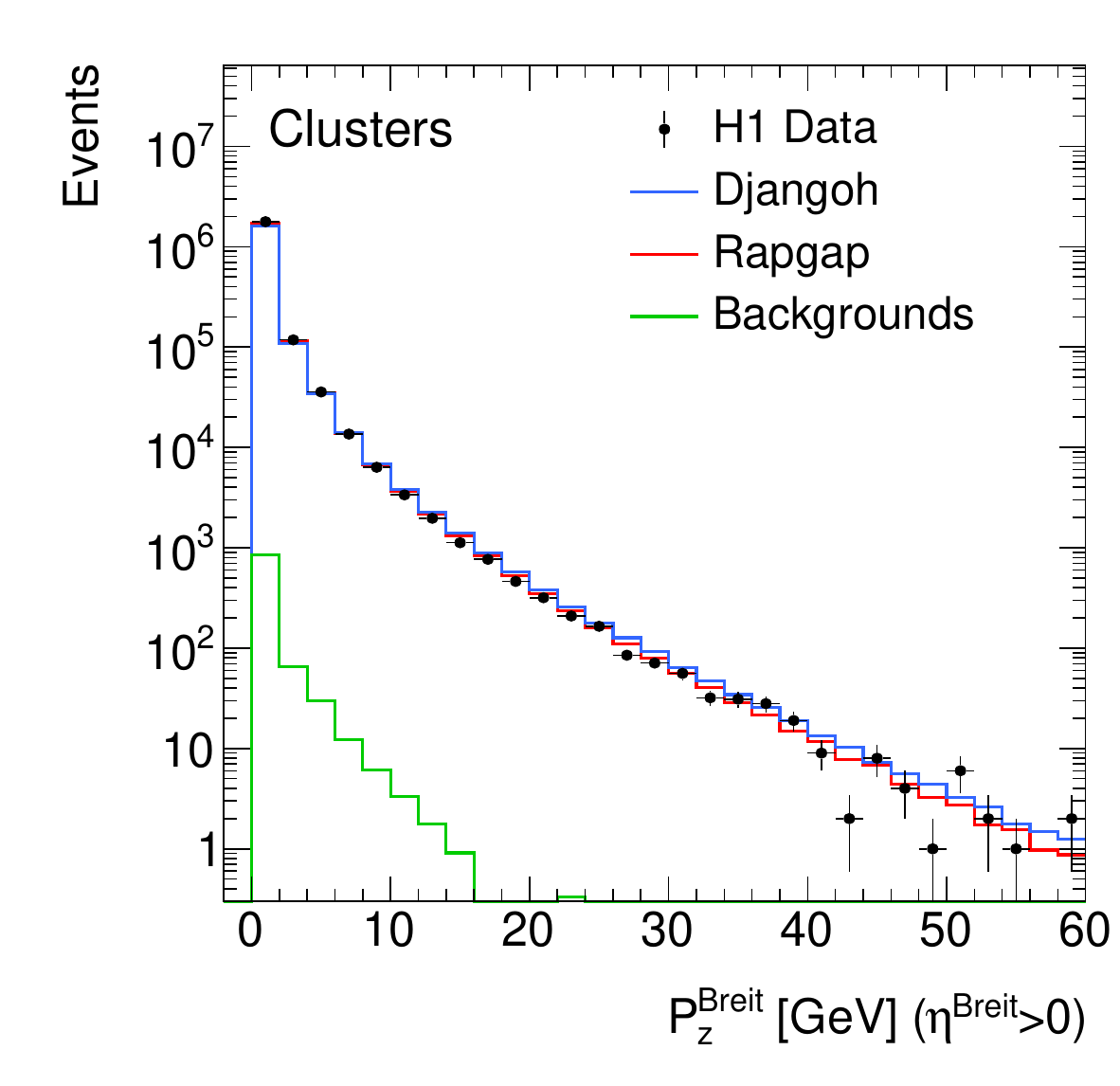}
\includegraphics[width=0.38\textwidth,trim={0 0 0 0 },clip]{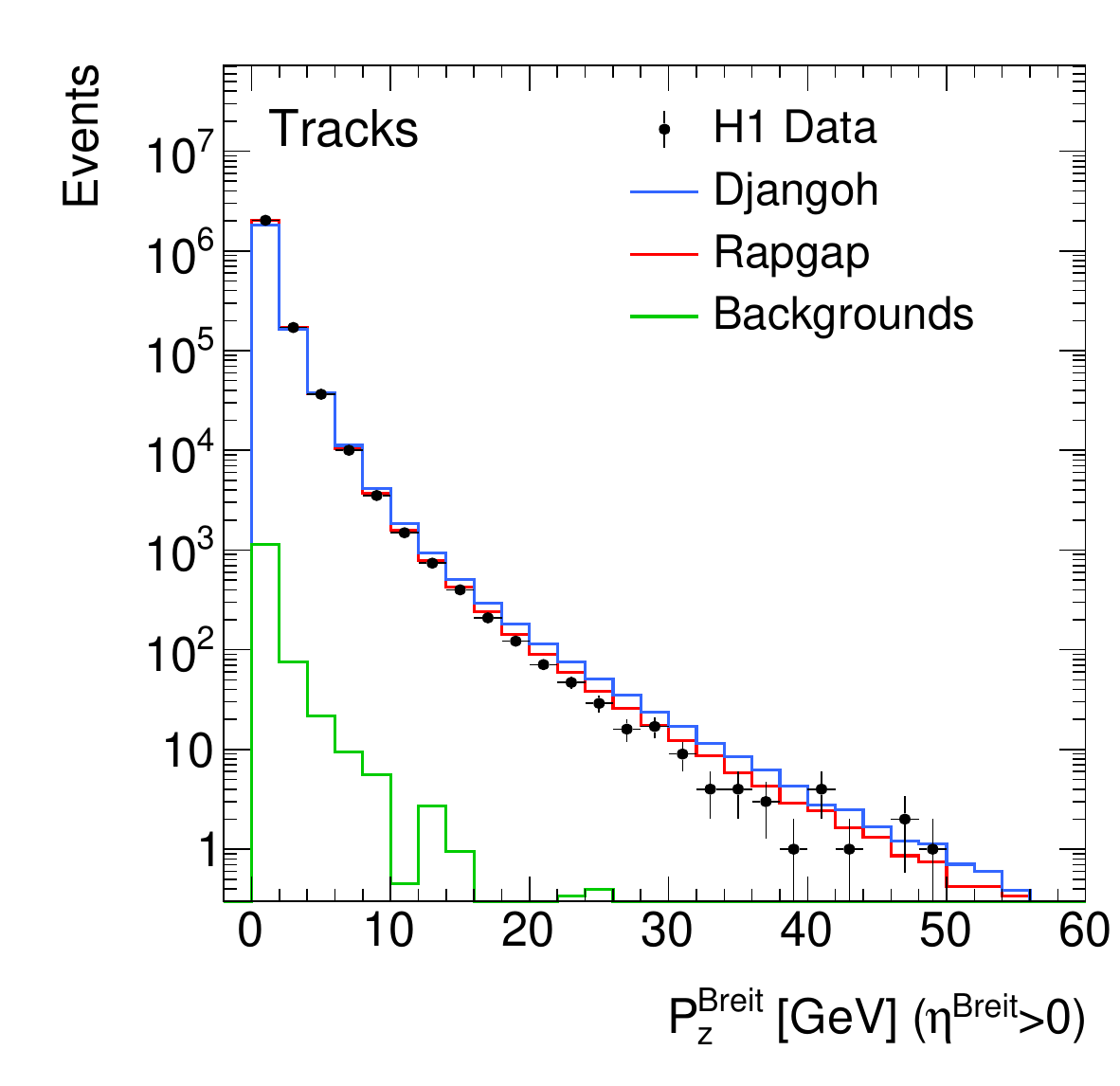}
\caption{
\label{fig:CurrentPz}
  The longitudinal momentum \pzb in the Breit frame of reconstructed particle
  candidates in the selected events.
  Only particle candidates with $\eta^\text{Breit}>0$ are displayed,
  since only those contribute to the actual calculation of \tb.
  Particle candidates are defined by an energy-flow algorithm taking
  clusters and tracks into account.
  Left: the \pzb\ distribution for clusters not associated with a track, right: the \pzb
  distribution for objects associated with a track.
  The detector-level data are compared to the MC event simulations Djangoh and
  Rapgap, which include further MC samples for background processes. 
  Further details are given in the Figure~\ref{fig:ControlPlots} caption.
}
\end{figure*}

\begin{figure*}[t]
\centering
\includegraphics[width=0.48\textwidth,trim={0 0 0 0 },clip]{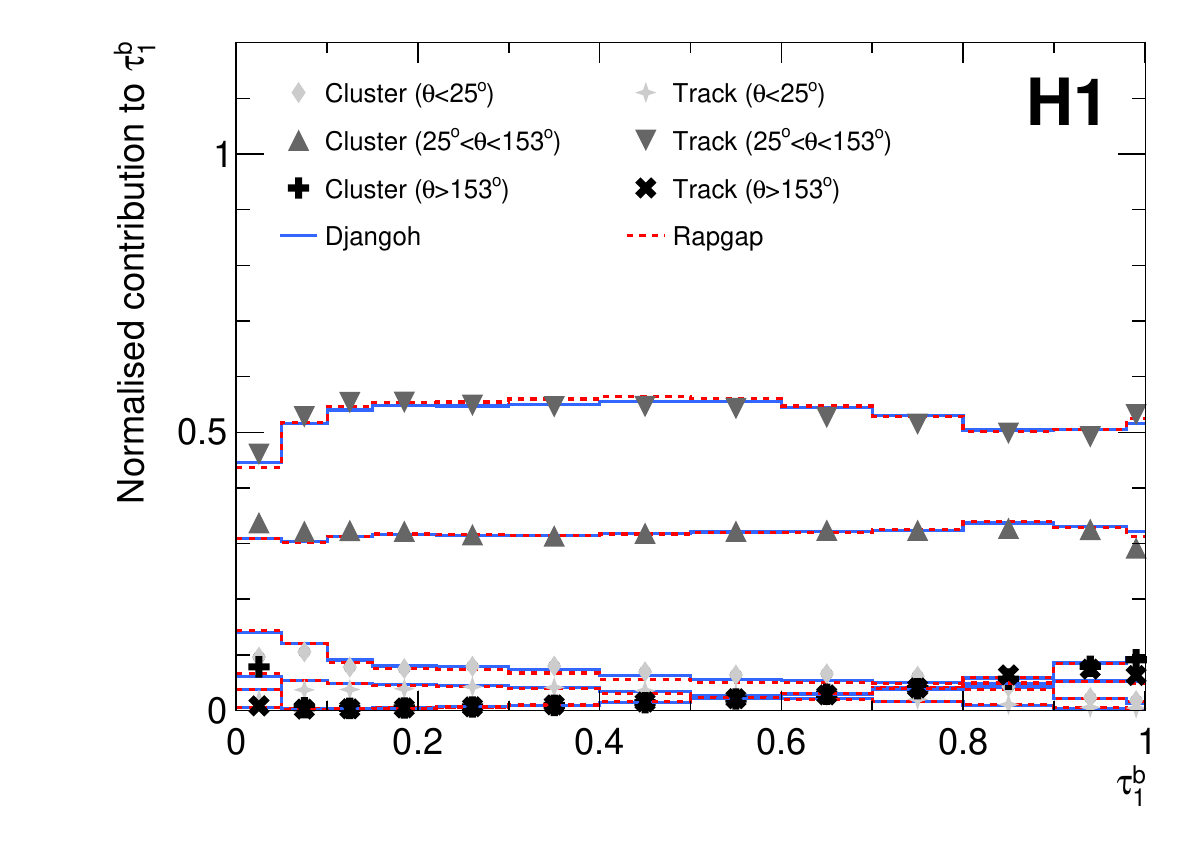}
\includegraphics[width=0.48\textwidth,trim={0 0 0 0 },clip]{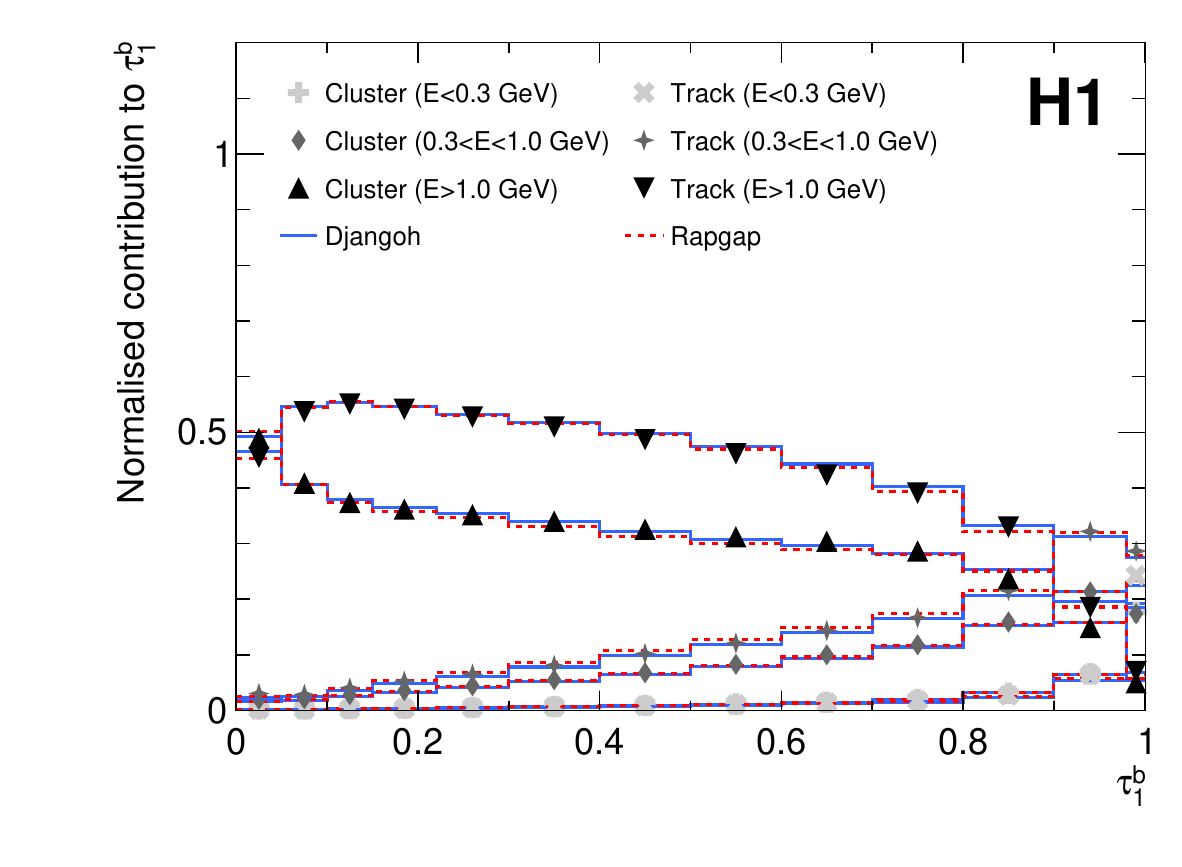}
\caption{
\label{fig:ContribTh}
  Left: The normalized contribution to \tb\ for differently reconstructed
  particle candidates (tracks or clusters as defined in Figure \ref{fig:CurrentPz}) in three distinct polar
  regions $\theta$.
  The regions are chosen according to $\theta$-ranges, where different
  components of the H1 detector are relevant for particle
  reconstruction~\cite{H1:1996prr,H1:1996jzy,H1SPACALGroup:1996ziw,Pitzl:2000wz}.
  Right: The normalised contribution to \tb\ for different ranges of particle
  energies $E$ and for differently reconstructed particle candidates.
  The data are compared to the MC event simulations Djangoh and Rapgap.
  The relative contributions are obtained by calculating a weight
  $w=\sum_\text{contrib}\pzb/\sum_\text{all}\pzb$
  for each event and each contribution.
  The resulting distributions are normalized to the \tb\ distribution.
}
\end{figure*}

Figure~\ref{fig:CurrentPz} shows the distribution of \pzb, the particle-candidate longitudinal momentum in the Breit frame, compared with simulations for reconstructed clusters and tracks. Good overall agreement between simulations and data is observed, for both clusters and tracks.
%
%
The measurement of \tb\ 
includes only particle candidates of the current
hemisphere. Figure~\ref{fig:ContribTh} shows relative contributions
from different polar angular regions $\theta$ and particle energies
$E$. The largest contributions are from objects in the central
detector ($25<\theta<153^\circ$), and from objects with large energy
($E>1.0\,\GeV$). Both types of object are measured well by the relevant H1
sub-detector components, and are modelled well by the
MC event simulations. 

At the particle level (also denoted as \emph{generator} or \emph{hadron level}), the
scattered electron is identified with the electron emitted at the DIS
vertex.
In the case of a photon radiated off the incoming or scattered lepton,
the photon is recombined with the scattered electron if the
angular distance is smaller than the angular distance to the $-z$-axis; otherwise it is
removed from the event record.

The HFS is formed from all remaining particles with proper lifetime
$c\tau>10\,\text{mm}$.
The DIS kinematic observables at the particle level are then computed
using equations~\eqref{eq:DIS} and~\eqref{eq:DISeMethod}.


The triple-differential cross section is reported as a
function of \tb in bins of $(\Qsq,y)$, and is obtained as
%
%
\begin{equation}
  \frac{d\sigma_i}{d\tb} (Q^2,y) = \frac{\left(A^+(\vec{n}_\text{data} - \vec{n}_\text{Bkg})\right)_i} {\mathcal{L} \cdot
    \Delta_{\tb,i}}\cdot c_{\text{QED},i}\,,
\end{equation}
\noindent
where $\vec{n}_\text{data}$ denotes a large data vector of event counts;
$\vec{n}_\text{Bkg}$ is a corresponding vector with the estimated number of events from processes other than high-\Qsq inclusive NC DIS; $\Delta_{\tb,i}$ is the
bin width of the $i$th bin of the respective \tb\ distribution; and $A^+$ denotes the
regularized inverse of the detector response matrix.
The parameter $c_\text{QED}$ is the QED correction factor, and $\mathcal{L}$ specifies the integrated luminosity of the measurement.
For this analysis,
$\mathcal{L}=351.1\,\text{pb}^{-1}$~\cite{H1:2012wor}, where the $e^+$
beam data contribute an integrated luminosity of 191.0\,pb$^{-1}$, and the
$e^-$ beam data have an integrated luminosity of 160.1\,pb$^{-1}$.

The detector response matrix $A$ is determined using both signal MC
event simulators. The TUnfold package~\cite{Schmitt:2012kp} is used to
calculate the regularised inverse $A^+$ and to propagate its
uncertainties. 
Finer binning is used at detector than at particle level; typically there are 12 bins in the
detector-level histogram. 
The vector $\vec{n}$ has 4331 entries in the
extended detector-level phase space, $0.01<y<0.94$ and
$Q^2>60\,\GeVsq$, 
and the vector $A^{+}\vec{n}$ has 572 entries at the particle level.
The extended detector-level phase space in $y$ and \Qsq\ accounts for
migrations into and out of the fiducial measurement phase space.
In order to obtain the final 324 cross section points from the 572 values,
two adjacent values are combined each, with exceptions made 
at lowest and largest \tb{}.
This procedure stabilizes the unfolding and reduces dependencies on
the MC models.
The regularization procedure employs the so-called \emph{curvature}
mode of TUnfold, which approximates second derivatives, and it acts on
the differences of the unfolded results and the MC predictions as a
bias distribution.  
The regularization is applied separately for each $\tb$ distribution in the individual ($Q^2,y$) intervals.
The regularization parameter is set to a small value
$\tau=10^{-3.2}$ as suggested by the analysis of Stein's Unbiased Risk
Estimator, SURE~\cite{Stein:1981}.
The regularization has a moderate impact on the result, and reduces
large fluctuations and sizable negative correlations between adjacent
data points in some regions of the analysis.

The background vector $\vec{n}_\text{Bkg}$ consists of events from photoproduction,
charged current DIS, di-lepton production, and DVCS processes.
A small number of simulated events which are outside the enlarged phase space at particle level (e.g.\ $y<0.01$, or $Q^2<120\,\GeVsq$) but migrate into the
enlarged detector-level phase space are treated as background, and are also
included in $\vec{n}_\text{Bkg}$ (``\emph{acceptance correction}'').

The multiplicative factor $c_\text{QED}$ corrects the data for
higher-order QED effects and for the positron charge.
It is defined as the ratio of the
cross section predicted without ($\sigma_\text{norad}^{e^-p}$) to the one with QED radiative effects ($\sigma_\text{rad}$).
The factors are determined using the Djangoh event
generator, where higher-order QED effects are implemented from
Heracles~\cite{Kwiatkowski:1990es}.
The factors are validated with Rapgap.
The denominator $\sigma_\text{rad}$ includes radiative cross sections for electron--proton and positron--proton
scattering with relative integrated luminosity weights as for real data. The
numerator $\sigma_\text{norad}^{e^-p}$ is defined for non-radiative electron--proton scattering ($e^-p$).
The resulting cross section measurements are therefore reported for $e^-p$
scattering.\,\footnote{At earlier HERA publications $e^+p$ cross sections were commonly reported.}
The effects of longitudinal polarization of the lepton
beam are negligible, due to a nearly balanced mixture of
running periods with opposite lepton beam helicities.
The QED correction $c_\text{QED}$ corrects for first-order real emission of
photons off the lepton line, including QED Compton enhancement,
photonic vertex correction, and
purely photonic self-energy contributions at the external lepton
lines.
This defines the \emph{non-radiative cross section level} ($\sigma_\text{norad}^{e^-p}$), which is reported
as the main result of the paper.
The data still include higher-order purely weak effects,
self-energy corrections of the exchanged bosons, second order
electroweak corrections, and photon-PDF induced contributions.

The QED correction factors $c_\text{QED}$ are largely independent of the kinematic variables $\tb$, \Qsq, and $y$
owing to the use of the  I$\Sigma$ method and the treatment of
radiated photons at the radiative particle level.
Effectively, $c_\text{QED}$ corrects mainly for the cut on the longitudinal energy-momentum balance
($(E-P_z)_{\text{gen}}>45\,\GeV$),
which has no effect at the non-radiative cross section level, but is
active at the radiative particle-level.
When the inverse of $c_\text{QED}$ is applied, the cross section at the radiative
level discussed above is obtained (see also
Ref.~\cite{Arratia:2022wny}), where the treatment of radiated photons is relevant.

To enable comparison with a range of theoretical calculations, additional correction factor are determined which optionally can be applied:
\begin{itemize}
\item The correction factor $c_\text{NoZ}$ corrects further for all leading and subleading electroweak corrections, which are $\gamma Z$ and pure $Z$-exchange, and first order purely weak vertex, self-energy and box corrections. Hence, only leptonic and hadronic corrections to the photon self-energy remain ($\sigma_{\gamma-\text{only}}$).
\item The correction factor $c_\text{Born}$ corrects further to the
so-called \emph{DIS Born level} ($\sigma_\text{Born}^{e^-p}$), where first order purely weak vertex, self-energy and box corrections are corrected. In addition, corrections for leptonic and hadronic contributions to the photon and $Z$ self-energy are applied.
\item The correction factor $c_{e^+p}$ corrects the reported $e^-p$
non-radiative cross sections to an $e^+p$ initial state ($\sigma_\text{norad}^{e^+p}$).
\end{itemize}
The QED correction factor and the optional correction factors can be summarized as:
\begin{equation}
  \begin{array}{cc}
  c_\text{QED}=\frac{\sigma_\text{norad}^{e^-p}}{\sigma_\text{rad}}
  \,, &
  c_\text{NoZ}=\frac{\sigma_{\gamma-\text{only}}}{\sigma_\text{norad}^{e^-p}}
  \,, \\
  c_\text{Born}=\frac{\sigma_\text{Born}^{e^-p}}{\sigma_\text{norad}^{e^-p}}
  \,, &
  c_{e^+p}    =\frac{\sigma_\text{norad}^{e^+p}}{\sigma_\text{norad}^{e^-p}}
  \,.
  \end{array}
\end{equation}

The single-differential cross section as a function of \tb\ 
in the NC DIS kinematic range, $200<\Qsq<1700\,\GeVsq$ and
$0.2<y<0.7$, is obtained by integrating the triple differential cross section for \tb\ over $(\Qsq,y)$.
The inclusive NC DIS cross section as a function of $\Qsq$ and $y$ is
obtained by integrating the \tb\ distribution over the $(\Qsq,y)$
range.
The single-differential and the inclusive NC DIS cross sections,
$\tfrac{d\sigma}{d\tb}$ and $\tfrac{d^2\sigma}{d\Qsq dy}$, are
reported separately for $e^-p$ and $e^+p$ scattering, where only data from the
respective running periods were analyzed.
Unfolding is carried out separately for $e^+p$ and $e^-p$ data in
these cases.

The result of integrating the \tb\ distribution, as described above, is consistent with the inclusive DIS cross section reported previously~\cite{H1:2012qti}.
However, the present measurement is not corrected to the bin center but is reported in intervals of
\Qsq\ and $y$, whereas previous inclusive DIS cross section measurements were reported as a function of
\Qsq\ and $\xbj$~\cite{H1:1999qgy,H1:2000muc,H1:2000olm,H1:2003xoe,H1:2010fzx,H1:2012qti}. 
This is the first inclusive NC DIS cross section measurement at HERA that is
corrected using regularised unfolding.


Fixed order calculations do not account for non-perturbative effects
due to hadronization of partons into stable hadrons. 
Such effects can approximately be accounted for by means of bin-wise
multiplicative correction factors $c_\text{Had}$, defined as
the ratio of calculated particle-level and parton-level cross sections.
Such hadronization corrections are determined using Djangoh and
Rapgap, which are found to be consistent with each other. 
These factors are also validated using Pythia 8.3.
The hadronization correction factors approximately decrease like $1/Q$, and are found to be similar in size to hadronization corrections reported for $e^+e^-$ at $\sqrt{s}\sim Q$ at PETRA~\cite{MovillaFernandez:2002zz,Pahl:2009aa,Pahl:2009zwz}.  
%


\section{Uncertainties}
The measurement is affected by a number of systematic
effects. The following sources of uncertainties are considered:
\begin{compactitem}
\item
  Statistical uncertainties of the data are propagated to the unfolded
  cross sections. This procedure results in bin-to-bin
  correlations, which are provided as
  supplementary material on the H1 webpage~\cite{h1pubs}.
\item
  The energy of the scattered lepton is measured with a precision of
  0.5\,\% in the central and backward region of the detector, and with
  $1\,\%$  precision in the forward region of the detector~\cite{H1:2012qti}.
  The efficiency of the electron identification algorithm is studied
  with an alternative track-based algorithm and the data and MC
  simulation are found to agree within 0.2\,\% at lower \Qsq\ and
  1\,\% for $\Qsq>3500\,\GeVsq$ ~\cite{Shushkevich:2012luo,H1:2012qti}.
\item The energies of all clusters and tracks receive scale 
  factors from a dedicated jet energy
  calibration~\cite{Kogler:2011zz,H1:2014cbm}.
  This calibration procedure results in two independent uncertainty
  contributions, whether clusters are contained in 
  jets or not. The two contributions are denoted as `jet energy scale uncertainty' (JES) and
  `remaining cluster energy scale uncertainty' (RCES), and 
  both uncertainties are determined by varying the energy of the
  respective HFS
  objects by $\pm1\,\%$.
 As compared to the JES, the RCES typically affects objects with lower transverse momenta.
\item The polar-angle position of the LAr calorimeter with respect to the Central Tracking
  Detector (CTD) is aligned with a precision of
  1\,mrad~\cite{H1:2012qti}.
  This uncertainty component is considered separately for the
  scattered electron and for the HFS objects.
  %
  %
\item The unfolding procedure is associated with several
  uncertainties.
  Differences in the migration matrix $A$ when determined from Djangoh
  or Rapgap are denoted as `model' uncertainty. Half of the
  difference in each element $A_{ij}$ is propagated to the
  unfolded cross section.
  This applies also to events that migrate from outside the particle-level
  phase space into detector-level phase space, and to events that are
  generated in the particle-level phase space but are not reconstructed at detector-level.
  %
  %
  Statistical uncertainties in $A$ from the limited event sample of
  the simulations are also considered.
  The size of the regularization parameter has an uncertainty of
  50\,\% and that variation is propagated to the resulting cross
  sections~\cite{Schmitt:2012kp}.
  The considered variation covers alternative procedures to determine
  the regularization strength, including the minimum global
  correlation coefficient or the $L$-curve scan~\cite{Schmitt:2012kp}.
\item Other uncertainty components are found to be negligible on
  their own, and a conservative bin-to-bin uncorrelated uncertainty of 0.5\,\% is
  introduced to cover the sum of these.
  An example of the uncertainties included here is the vertex and
  electron track reconstruction efficiency, which has an uncorrelated
  uncertainty of 0.2\,\%~\cite{Shushkevich:2012luo},
  The distance between the calorimeter
  cluster of the scattered electron and a vertex-associated track is
  not described perfectly by the simulation, and the corresponding efficiency
  correction introduces an uncorrelated 
  uncertainty of about 0.1 to 0.2\,\%. 
  The uncertainty associated with the choice of the regularization
  strength in the unfolding procedure is found to be below 0.1\,\%.
  The contributions from other processes than high-\Qsq\ NC DIS (\emph{backgrounds}) are estimated from MC
  simulations and a normalization uncertainty of 100\,\% is
  considered, which results in an uncertainty smaller than 0.1\,\%.
\item The normalization uncertainty is found to be 2.7\,\%, and
  is dominated by the uncertainty on the integrated luminosity~\cite{H1:2012wor}. Other
  normalization uncertainties are negligible in comparison. For example,
  the efficiency of the trigger is found to be higher than
  99.5\,\%~\cite{H1:2014cbm,H1:2012qti} and the related uncertainty is smaller than $0.5\%$. 
  Further normalization uncertainties are related to the electron identification, the noise
  suppression algorithm in the LAr, and to the track and vertex
  identification.
\item The QED correction factors were determined separately with
  Djangoh and Rapgap and very good agreement was found.
  While both MC generators implement QED corrections from Heracles,
  previous studies showed that these are consistent with the program
  Hector~\cite{Arbuzov:1995id} and EPRC~\cite{Spiesberger:1995pr}, and
  that the contribution from the two-photon exchange, which is not
  implemented in Heracles, is negligible~\cite{H1:2012qti,H1:2018mkk}.
  Consequently, the uncertainty of the QED correction factors
  represent only their statistical component, while other
  uncertainties would be of negligible size,
  when compared to other uncertainty sources.
\item Uncertainties of the hadronization correction factors, relevant for
  a subset of the comparisons only, are determined as half of the difference between the correction factor 
  obtained from Djangoh and Rapgap.
\end{compactitem}

\section{Results}
Results for the single-differential and triple-differential 1-jettiness cross
sections as well as for the double-differential inclusive NC DIS cross section
are presented.

\begin{figure*}[t]
\centering
\includegraphics[width=0.495\textwidth,trim={20 0 30 0 },clip]{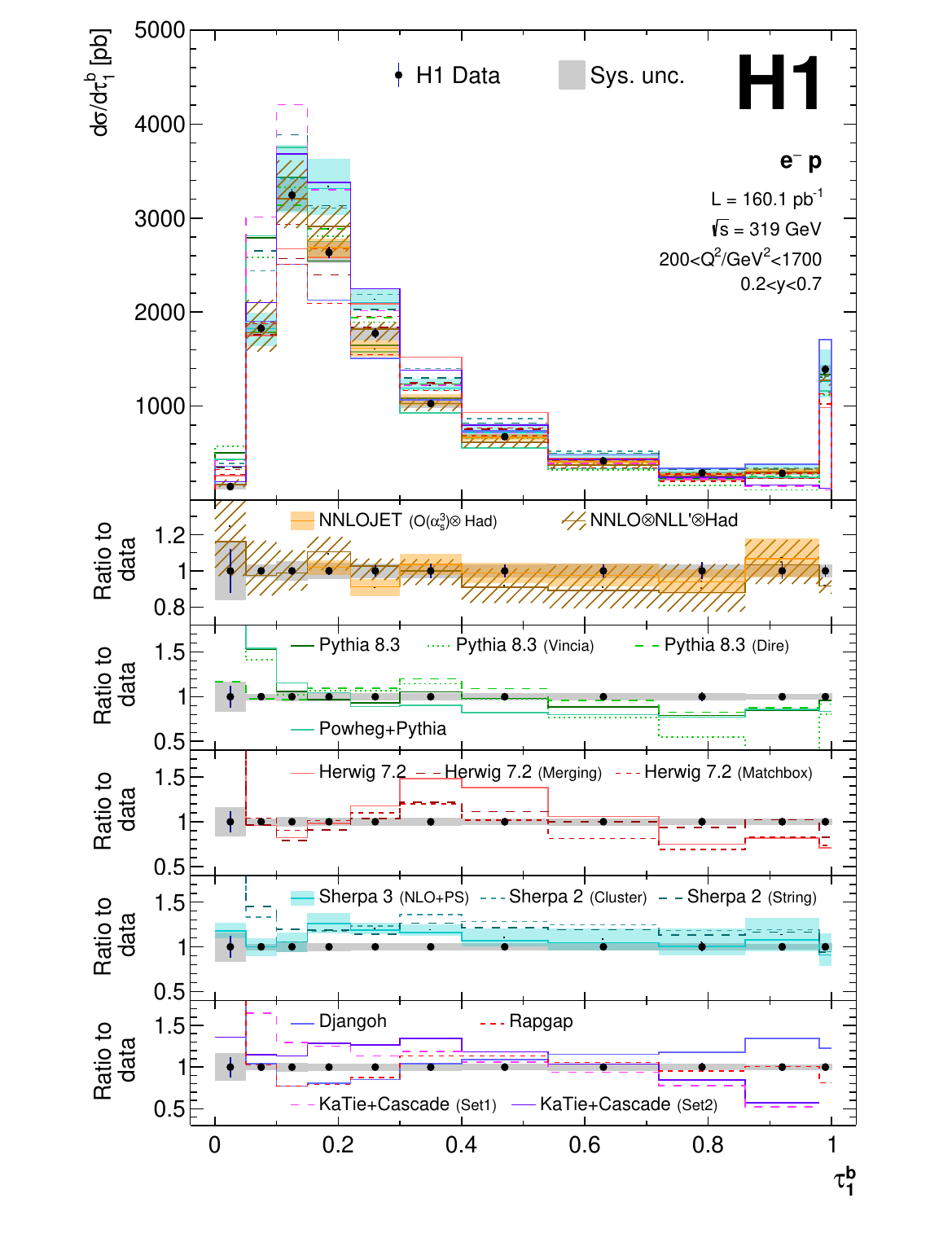}
\includegraphics[width=0.495\textwidth,trim={20 0 30 0 },clip]{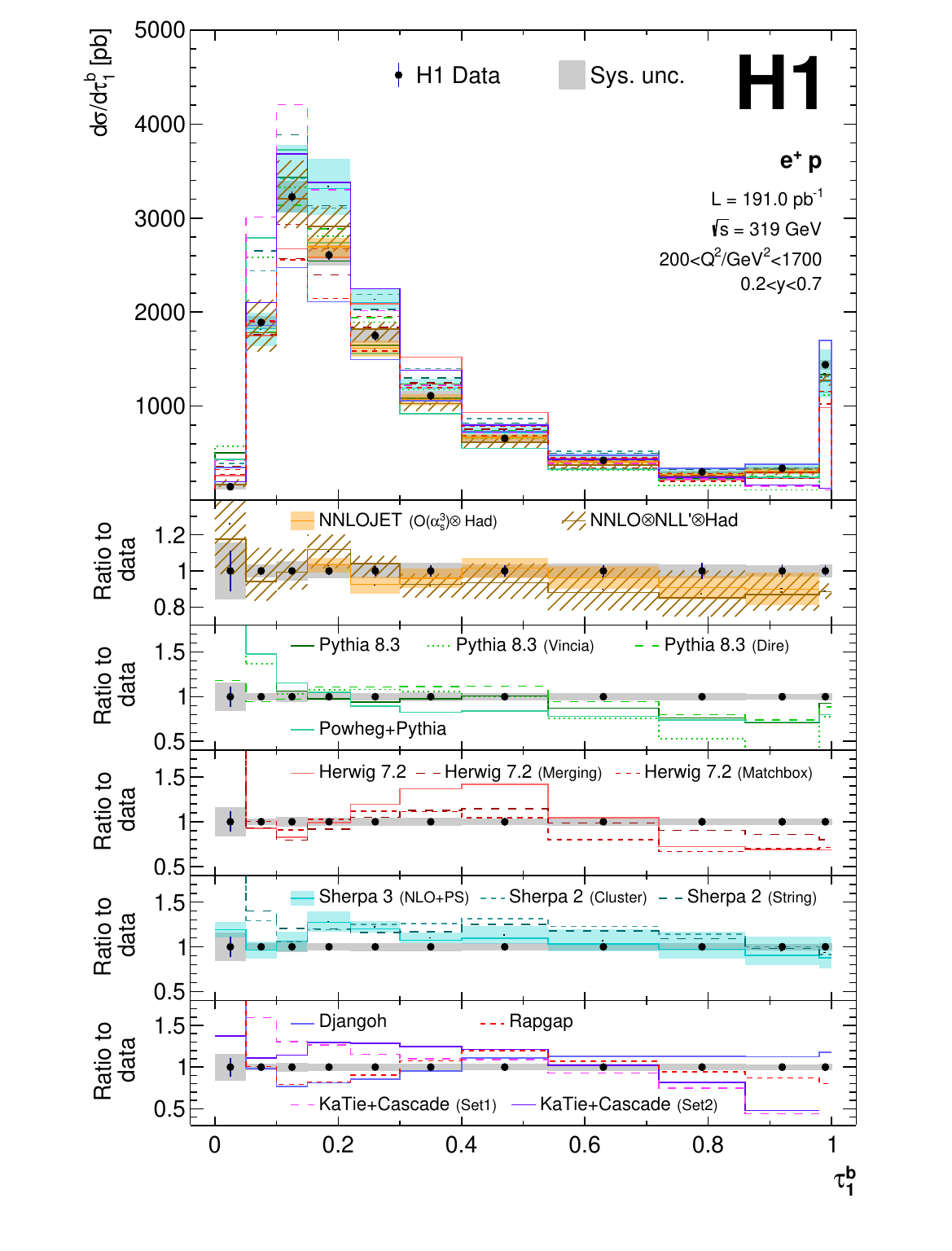}
\caption{
  The differential cross section $d\sigma/d\tb$ in the kinematic region
  $200\leq\Qsq/\,\GeVsq<1700$ and $0.2\leq y<0.7$ for $e^+p$ (left) and
  $e^-p$ scattering at $\sqrt{s}=319\,\GeV$.
  The data are corrected for detector effects (acceptance, resolution)
  and QED radiative effects.
  The statistical uncertainties are displayed as vertical error bars, 
  systematic uncertainties as shaded areas.
  The data are compared to the MC predictions from NNLOJET,
  NNLO$\otimes$NLL$^{\prime}\otimes$Had, Pythia~8.3, Powheg Box, Herwig~7.2, Sherpa~2.2,
  Sherpa~3.0, Djangoh, Rapgap, and KaTie+Cascade~3.
  Selected parameters are varied in the predictions as indicated in
  brackets. Colored areas indicate theoretical uncertainties
  associated with some of the predictions (see text).
  The lower panel displays the ratio of predictions to data.
}
\label{fig:1Da}
\end{figure*}

\begin{figure*}[t]
\centering
\includegraphics[width=0.65\textwidth,trim={0 0 0 0 },clip]{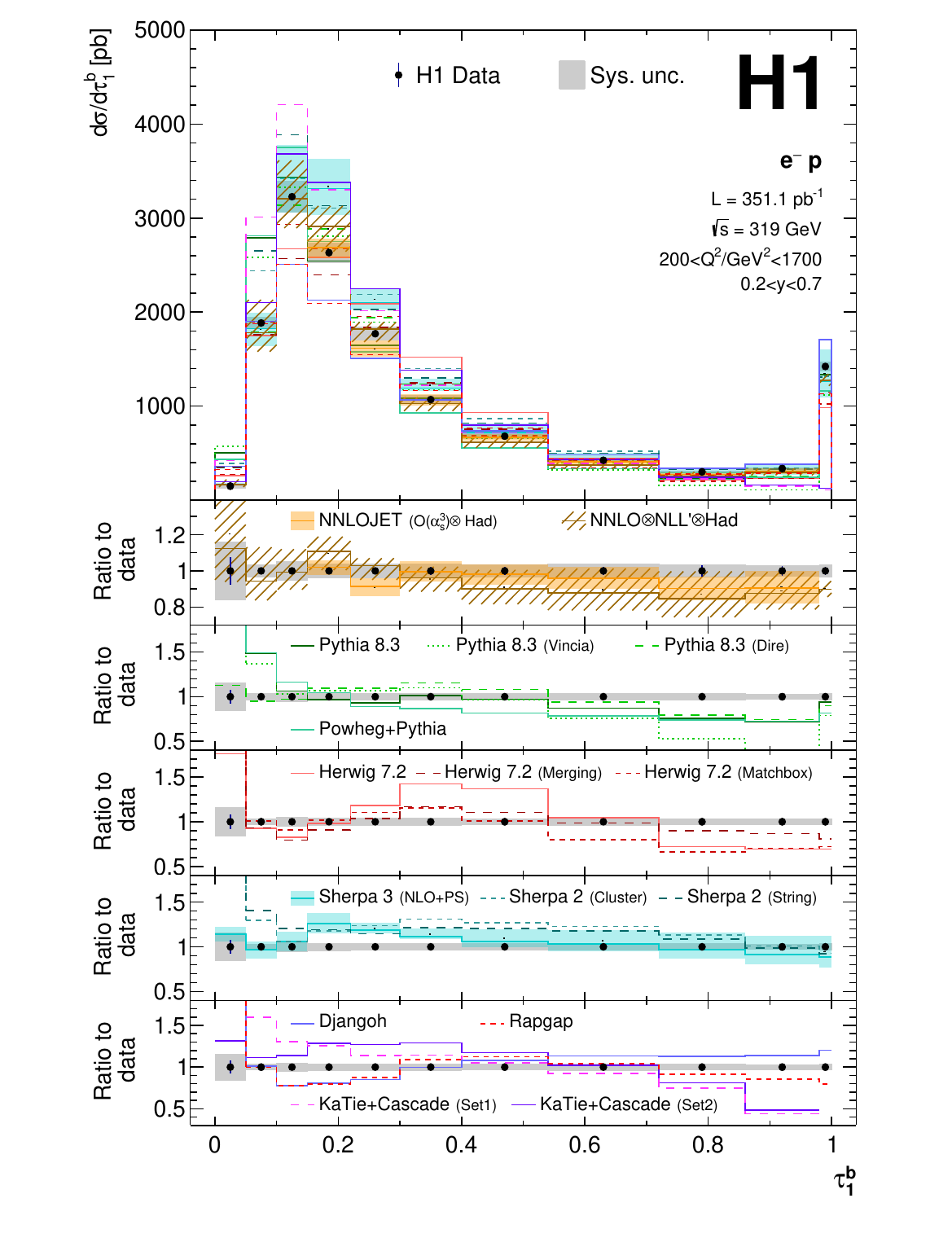}
\caption{
  The differential cross section $d\sigma/d\tb$ in the kinematic region
  $200\leq\Qsq/\,\GeVsq<1700$ and $0.2\leq y<0.7$ for $e^-p$ scattering
  at $\sqrt{s}=319\,\GeV$.
  The data are recorded with electron and positron beams, and
  corrected to $e^-p$ cross sections using Heracles.
  Further details are given in the caption of Figure~\ref{fig:1Da}.
}
\label{fig:1Db}
\end{figure*}

\paragraph{\boldmath Single-differential cross sections as a function of $\tau_1^b$}
The single-differential 1-jettiness cross sections
$\tfrac{d\sigma}{d\tb}$ in NC DIS $ep$
scattering in the kinematic range 
$200\leq\Qsq<1700$\,\GeVsq and $0.2\leq y<0.7$  are measured
for $e^+p$ and $e^-p$ collisions in Tables~\ref{tab:xs1D.em} and~\ref{tab:xs1D.ep} and displayed
in Figure~\ref{fig:1Da}.
As no significant differences are observed between the two cross sections, the measurement is repeated using the sum of both datasets, corrected to $e^-p$ scattering cross sections, in
Table~\ref{tab:xs1D.em351} and Figure~\ref{fig:1Db}. 
The statistical uncertainty in the data is typically around 2 to
4\,\% and the systematic uncertainty is of order 4\,\%. Larger
uncertainties are seen for the lowest \tb\ bin.
The differential \tb\ cross section exhibits a distinct peak at $\tb\sim0.13$
and a tail towards high values of \tb.
The distinct \emph{DIS peak} is populated by DIS Born-level kinematics with a single
hard parton, the position and shape of which are dominated by 
hadronization and resummation effects. 
The tail region is
populated by events with hard radiation, including 
two-jet topologies in the far tail.
The cross section at $\tb\simeq1$ has a sizeable value, as it includes
 events with empty current
hemisphere in the Breit frame. 
This event configuration can occur at low \xbj\ and is studied in a dedicated publication~\cite{EHEletter}.
The data are compared to predictions from NNLOJET up to
$\mathcal{O}(\as^3)$ in the strong coupling; resummed predictions at NLL accuracy
matched to fixed-order predictions at $\mathcal{O}(\as^2)$ and
corrected for hadronization; various predictions from the modern MC
event generators Pythia~8.3, Sherpa~2.2, Sherpa~3.0, Herwig~7.2, and
Powheg Box plus Pythia; as well as to the dedicated MC event generators for DIS, Djangoh,
Rapgap, and KaTie+Cascade.
The agreement between the predictions and data are very similar for $e^+p$ and
$e^-p$ cross sections.
Further details are discussed below.

\begin{figure*}[t]
\centering
\includegraphics[width=0.92\textwidth,trim={0 0 0 0 },clip]{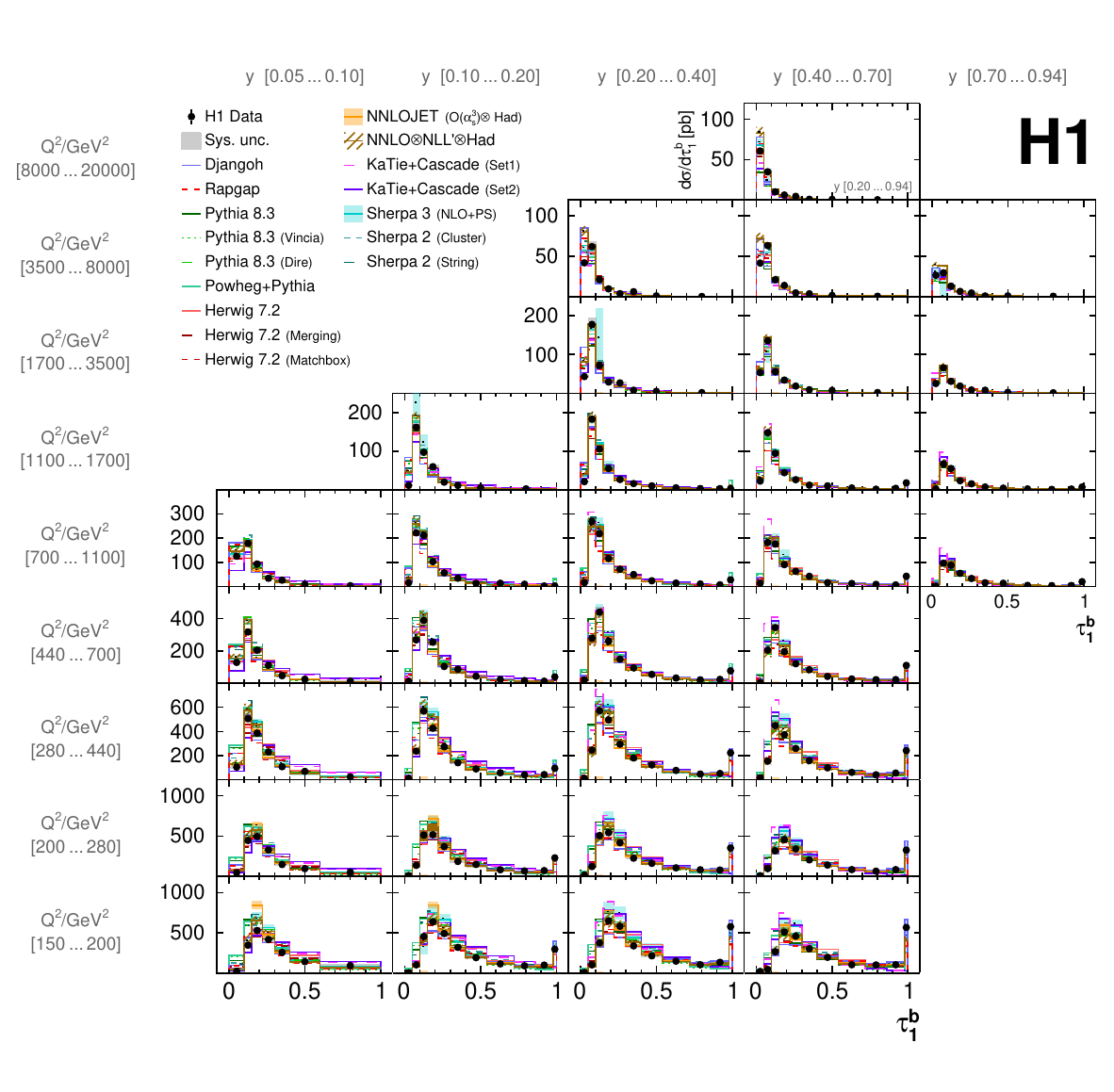}
\caption{
  The triple-differential cross section $\tfrac{d^3\sigma}{d\tb d\Qsq dy}$, presented as $\tfrac{d\sigma}{d\tb}$ 
  in regions of \Qsq\ and $y$.
  The \Qsq\ and $y$ ranges are displayed on the left and top,
  respectively. Each panel displays the differential cross
  section $d\sigma/d\tb$  in that phase space.
  The data are displayed as full circles, the vertical error
  bars indicate statistical uncertainties, while systematic uncertainties
  are displayed as shaded area.
  The data are compared to predictions from  NNLOJET,
  NNLO$\otimes$NLL$^{\prime}\otimes$Had, Pythia~8.3, Powheg Box, Herwig~7.2, Sherpa~2.2,
  Sherpa~3.0, Djangoh, Rapgap, and KaTie+Cascade~3. Selected
  parameters are varied in the predictions as indicated.
  The ratios of the data and predictions to the Sherpa~3.0 predictions
  are displayed in
  Figures~\ref{fig:xs3DRatioNNLO}--\ref{fig:xs3DRatioCascade}.
  At highest \Qsq ($8000<\Qsq/\GeVsq<20\,000\,$) only a single $y$
  interval is measured, $0.2<y<0.94$.
}
\label{fig:xs3Dall}
\end{figure*}

\begin{figure*}[t]
\centering
\includegraphics[width=0.92\textwidth,trim={0 0 0 0 },clip]{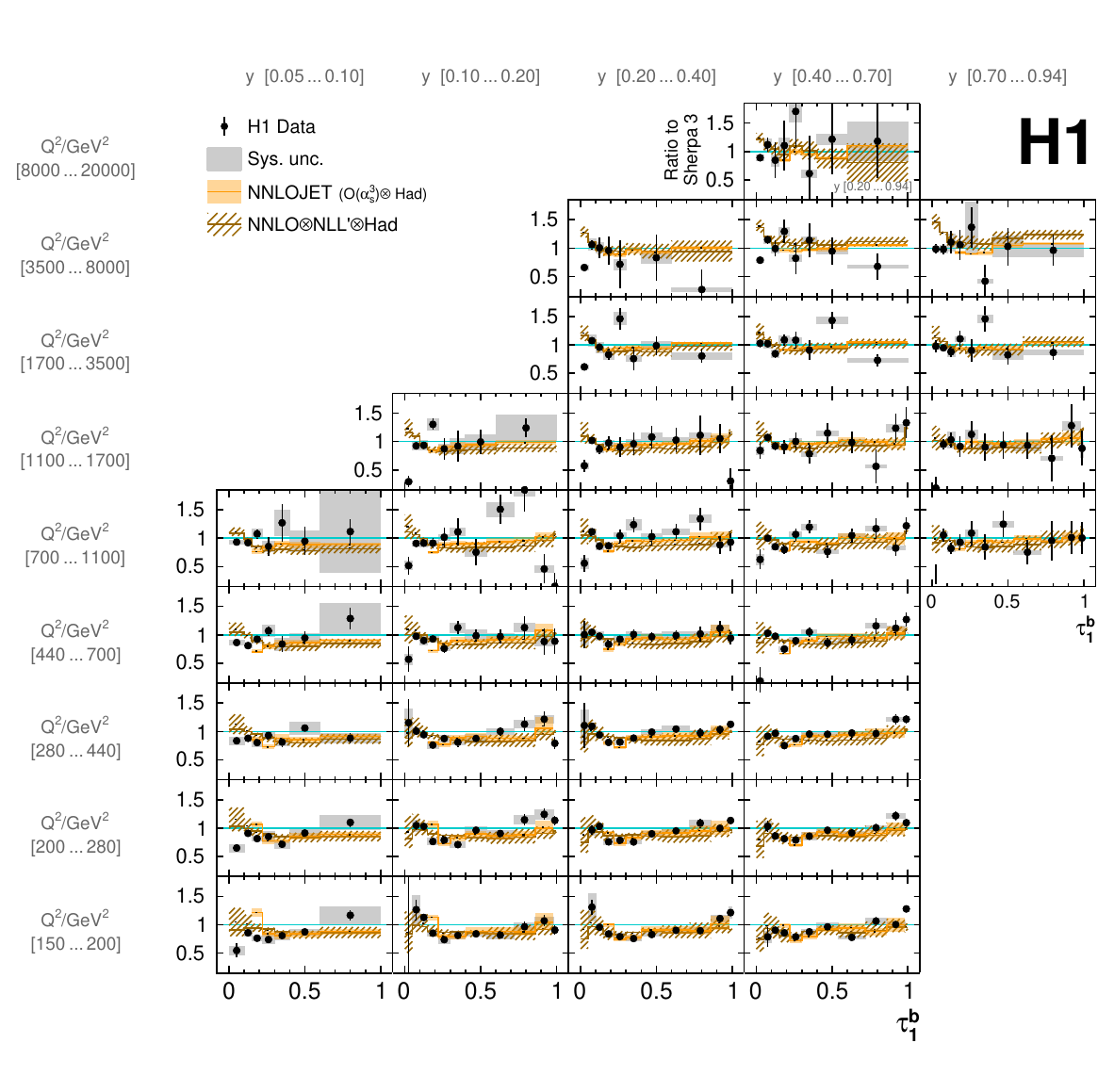}
\caption{
  The ratio of data and NNLOJET ($\mathcal{O}(\alpha_s^3)$)
  and NNLO$\otimes$NLL$^{\prime}\otimes$Had predictions to the Sherpa~3.0 predictions for the
  triple-differential cross section $\tfrac{d^3\sigma}{d\tb d\Qsq dy}$.
  See Figure~\ref{fig:xs3Dall} caption for further details.
  The data are displayed as full circles, the vertical error
  bars indicate statistical uncertainties, while systematic uncertainties
  are displayed as shaded area.
  The shaded coloured area indicates the theoretical uncertainty associated
  with the NNLOJET predictions.
  The hatched area indicates the theoretical uncertainty associated
  with the NNLO$\otimes$NLL$^{\prime}\otimes$Had predictions.
}
\label{fig:xs3DRatioNNLO}
\end{figure*}

\begin{figure*}[t]
\centering
\includegraphics[width=0.92\textwidth,trim={0 0 0 0 },clip]{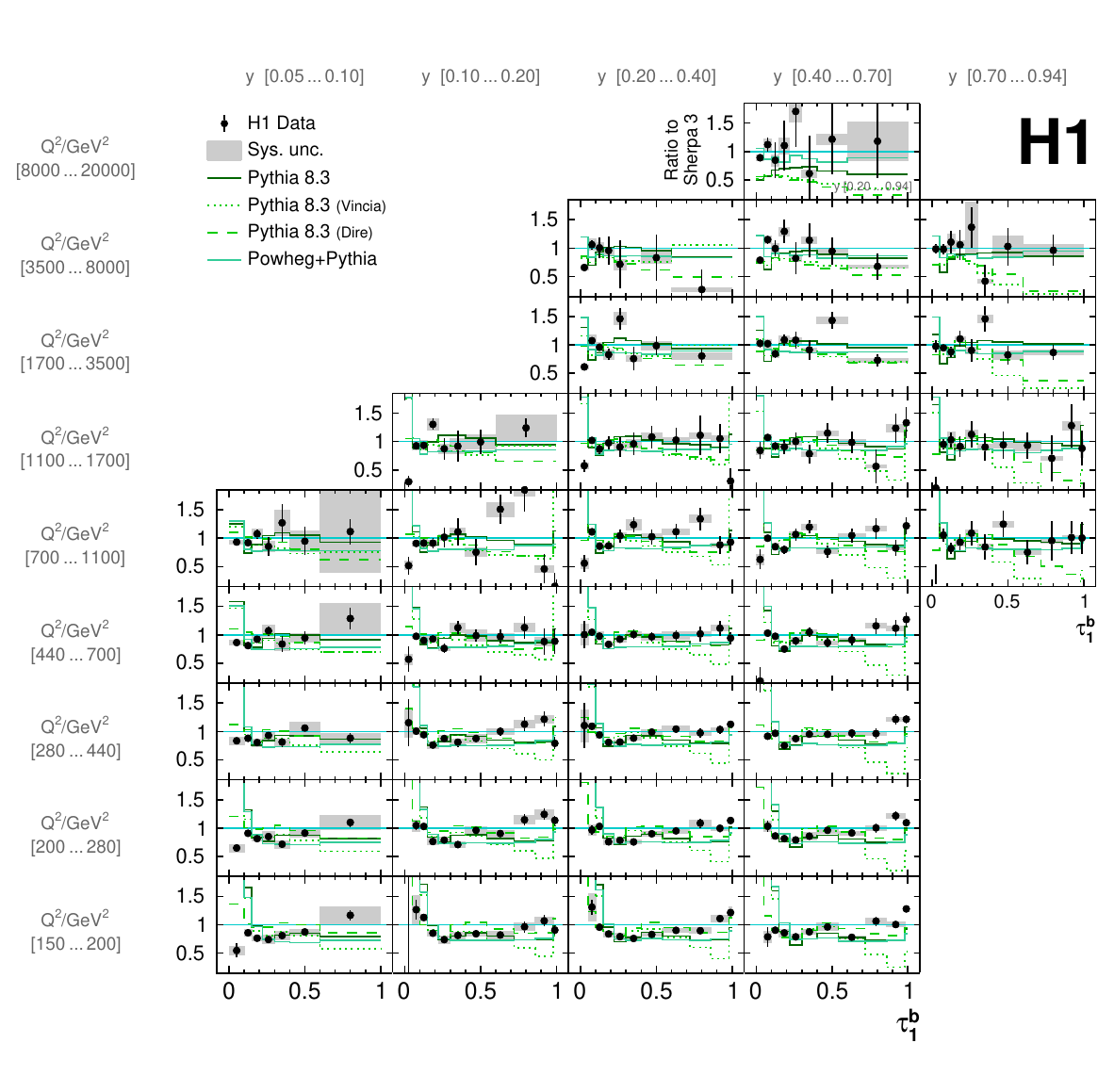}
\caption{
  The ratio of data and Pythia~8.3 predictions to the Sherpa~3.0 predictions for the
  triple-differential cross section $\tfrac{d^3\sigma}{d\tb d\Qsq dy}$.
  See Figure~\ref{fig:xs3Dall} caption for further details.
  The data are displayed as full circles, the vertical error
  bars indicate statistical uncertainties, while systematic uncertainties
  are displayed as shaded area.
  The full dark green line indicates predictions from Pythia~8.3
  employing the default parton shower adapted for DIS kinematics, the
  short dashed line indices predictions using the Vincia parton
  shower, and the long dashed line the Dire parton shower.
  The green full line shows predictions from Powheg Box interfaced to
  parton shower and hadronization from Pythia~8.3.
}
\label{fig:xs3DRatioPythia}
\end{figure*}

\begin{figure*}[t]
\centering
\includegraphics[width=0.92\textwidth,trim={0 0 0 0 },clip]{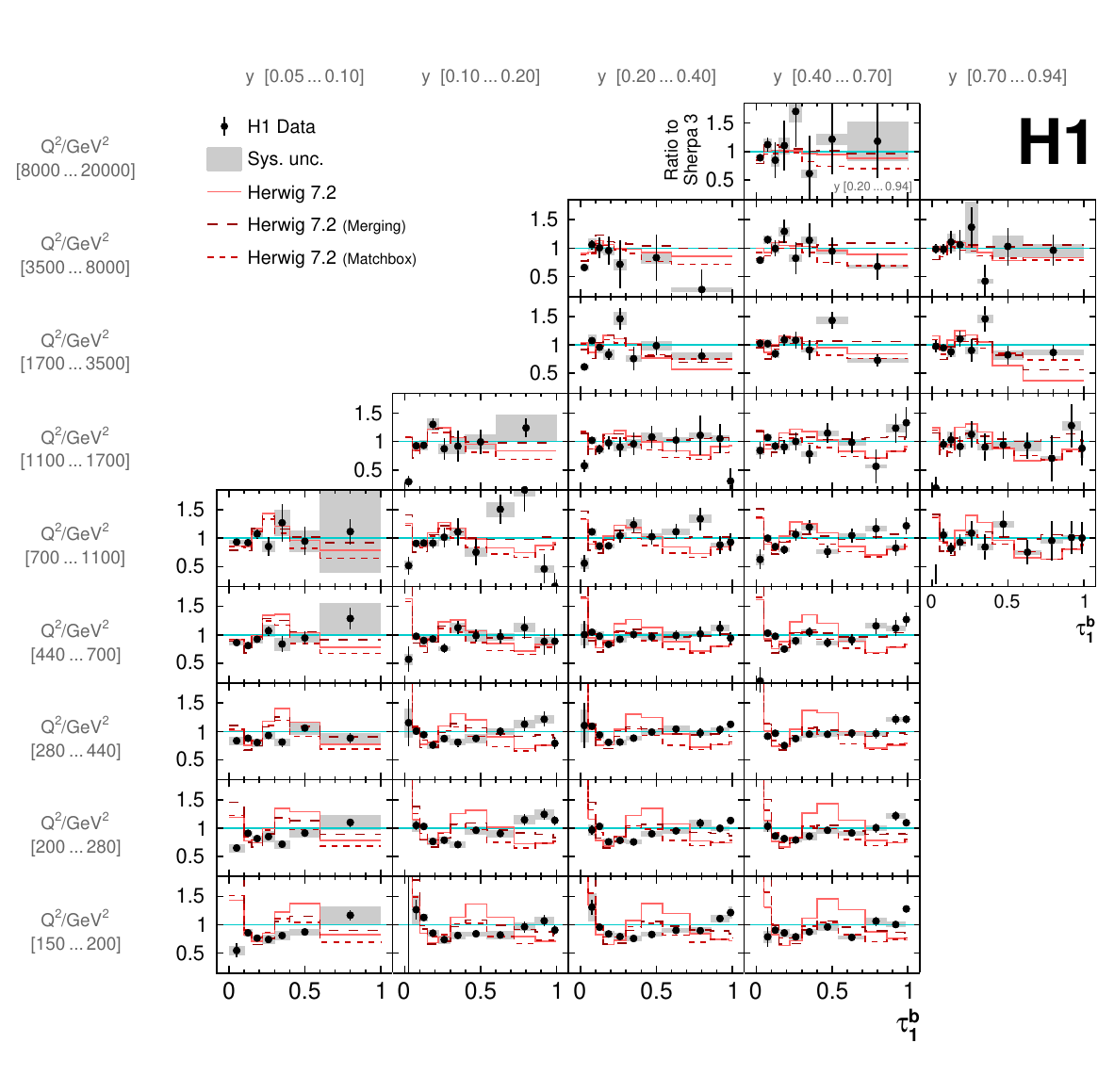}
\caption{
  The ratio of data and Herwig~7.2 predictions to the Sherpa~3.0 predictions for the
  triple-differential cross section $\tfrac{d^3\sigma}{d\tb d\Qsq dy}$.
  See Figure~\ref{fig:xs3Dall} caption for further details.
  The data are displayed as full circles, the vertical error
  bars indicate statistical uncertainties, while systematic uncertainties
  are displayed as shaded area.
  The full line indicates predictions from Sherpa~3.0, and the shaded
  coloured are associated theoretical uncertainties.
  The short and long dashed lines indicate predictions from Sherpa~2.2, where two
  hadronisation models are studies,  Cluster and String hadronisation
  model, respectively.
}
\label{fig:xs3DRatioHerwig}
\end{figure*}

\begin{figure*}[t]
\centering
\includegraphics[width=0.92\textwidth,trim={0 0 0 0 },clip]{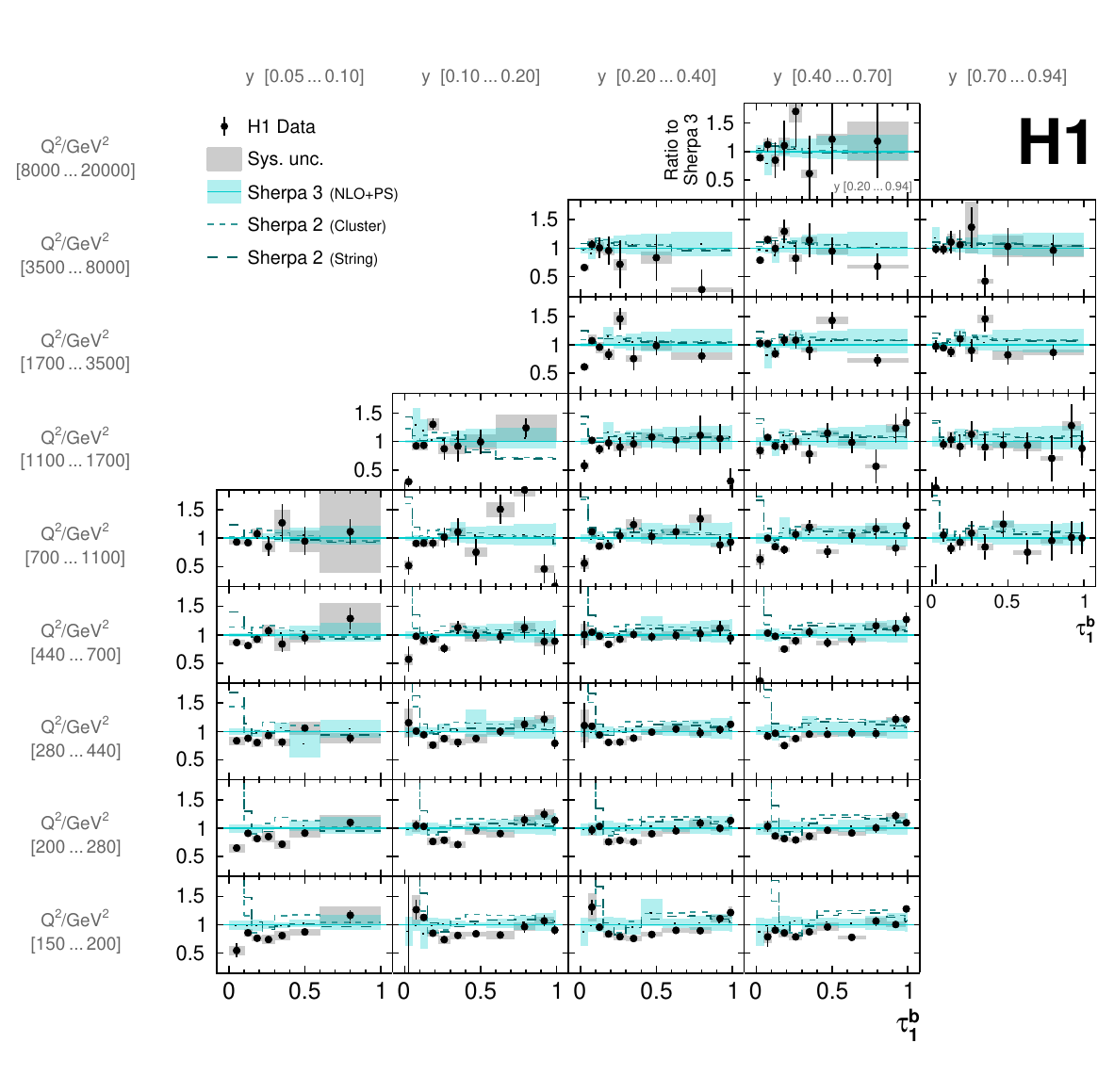}
\caption{
  The ratio of data and Sherpa 2 predictions to Sherpa~3.0 predictions for the
  triple-differential cross section $\tfrac{d^3\sigma}{d\tb d\Qsq dy}$.
  See Figure~\ref{fig:xs3Dall} caption for further details.
  The data are displayed as full circles, the vertical error
  bars indicate statistical uncertainties, while systematic uncertainties
  are displayed as shaded area.
  The full dark green line indicates predictions from Pythia~8.3
  employing the default parton shower adapted for DIS kinematics, the
  short dashed line indices predictions using the Vincia parton
  shower, and the long dashed line the Dire parton shower.
  The green full line shows predictions from Powheg Box interfaced to
  parton shower and hadronization from Pythia~8.3.
}
\label{fig:xs3DRatioSherpa}
\end{figure*}

\begin{figure*}[t]
\centering
\includegraphics[width=0.92\textwidth,trim={0 0 0 0 },clip]{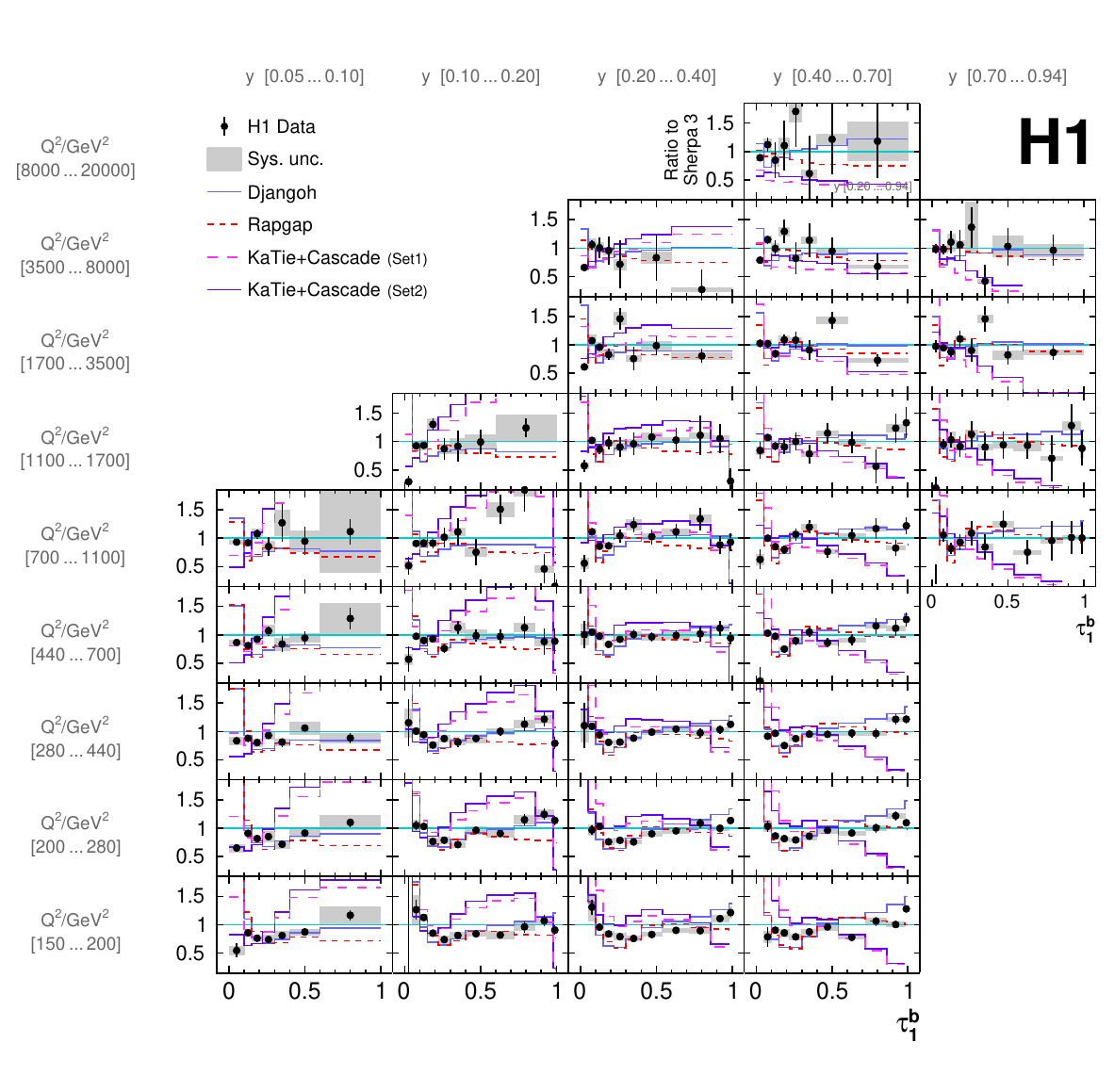}
\caption{
  The ratio of data and  predictions from dedicated DIS MC event
  generators to the Sherpa~3.0 predictions for the
  triple-differential cross section $\tfrac{d^3\sigma}{d\tb d\Qsq dy}$ for adjacent
  regions in \Qsq\ and $y$.
  See Figure~\ref{fig:xs3Dall} caption for further details.
  The data are displayed as full circles, the vertical error
  bars indicate statistical uncertainties, while systematic uncertainties
  are displayed as shaded area.
  The full blue line indicates predictions from Djangoh, the red
  dashed line from Rapgap, and the pink dashed and violet line line
  show predictions from KaTie+Cascade~3 for two different TMD PDFs and
  scale choices.
}
\label{fig:xs3DRatioCascade}
\end{figure*}

\paragraph{\boldmath Triple-differential cross sections as a function of $\tau_1^b$, \Qsq, and $y$}
Triple-differential cross sections are presented in Tables~\ref{tab:xstaub1b.em351.150_0.05}--~\ref{tab:xstaub1b.em351.8000_0.20} and are displayed in 
Figure~\ref{fig:xs3Dall}.
The ratio of the data and predictions to the predictions from
Sherpa~3.0 are displayed in
Figures~\ref{fig:xs3DRatioNNLO}--\ref{fig:xs3DRatioCascade}. 
The triple differential cross sections are presented in the kinematic range 
$150<\Qsq<20\,000\,\GeVsq$ and $0.05<y<0.94$ as a function of $\tb$.
Regions that are kinematically forbidden or experimentally
inaccessible are omitted.
At highest \Qsq\ only a single combined $y$ region ($0.2<y<0.94$) is presented due to
low event counts.
%
%

Events with a harder virtuality $Q$ produce more collimated particles, effectively shifting the DIS peak position in the $\tau_{1}^{b}$ distributions towards lower values. In addition, a reduced phase space for hard radiation at high $Q^2$ tends to lower the differential cross section in the larger $\tau_{1}^{b}$ region relative to the peak region. The relative contribution of topologies with $\tau_{1}^{b}$ close to unity increases with $y$ for fixed $Q$. Equivalently, that contribution increases as $x_{Bj}$ decreases.

%
%

%
\paragraph{Comparison to exact QCD predictions}
The fixed order predictions with NLL resummation of large logarithms
(NNLO$\otimes$NLL$^\prime\otimes$Had) provide an accurate description of the entire
\tb\ distribution within their uncertainties.
These predictions include resummation and are matched to the NNLO
inclusive DIS cross section ($\as^2$), such that they are valid over the entire \tb\ range.
Formally, they are one order lower in $\alpha_s$ than the NNLOJET predictions.

The NNLO dijet predictions from NNLOJET  ($\as^3$) provide a good description of the
data within their uncertainty over their entire range of validity ($0.22<\tb<0.98$).
Sizable hadronization corrections of up to $+40\,\%$ hinder quantitative comparisons
and an interpretation in terms of underlying parameters of the
theory; this point will have to be investigated in the future.
%


\paragraph{Comparison to modern MC event generators}
The recent MC event generators Pythia~8.3, Powheg+Pythia, Herwig~7.2,
Sherpa~2.2 and Sherpa~3.0 employ LO, multi-leg or NLO matrix elements
matched to parton shower and hadronization models. Those
generators generally provide a good description of the data.

The MC predictions from Pythia provide an overall reasonable
description of the data, with small but visible differences between the three
parton shower models studied.
The default Pythia and Pythia+Vincia predictions overestimate the data
at low $\tb<0.1$, whereas Pythia+Dire provides a good
description.
All three Pythia predictions are similar in the  parton-shower
region ($0.1<\tb<0.4$), but tend to underestimate the data in the tail region
($\tb>0.5$).
The latter could be related to the fact that matrix elements for
$eg\to eq\bar{q}$  enter at $\mathcal{O}(\as)$ only, and are not recovered
properly in a parton shower emerging from $eq\to eq$ alone.
The default Pythia predictions are a leading-order MC model and they
benefit from a large value of $\asmz$ for the parton shower, which
is default in Pythia, and thus raises the prediction at larger \tb
and brings them closer to the data. 
%
Looking only at the single-differential cross sections,
Pythia+Dire tends to overshoot the data in the parton-shower
region, but from the triple-differential data it is evident that the
discrepancy arises from events at lower values of \Qsq{}.
At very large values of $y$ ($y>0.7$), the Pythia+Dire predictions
fail to describe the data at large values of \tb.
%
%
%
The Pythia+Vincia prediction provides a good
description of the data in the range $0.1\lesssim\tb\lesssim0.5$. It
has difficulties at larger values of \tb, presumably related to the fact that
Vincia was developed for a symmetric collider setup and is not
yet fully validated for the asymmetric beams of $ep$ collisions.

The predictions from Powheg Box are lower than the default Pythia predictions
at medium to large \tb, and
overshoot the data at low \tb.
Larger values of \asmz\ in the simulation, or
corrections beyond NLO DIS, such as NLO dijet
matrix elements, may be required to improve the description of the data.
%

The Herwig variants are able to provide a good description of the data.
Although the default predictions from Herwig~7.2 provide the worst
description among all modern MC generators, a significant improvement
is obtained with NLO matrix elements.
The merging technique provides the best description among the Herwig
predictions, with benefits particularly at larger \tb.
All Herwig predictions fail to describe the lowest \tb\ cross section ($\tb<0.05$).
%
The default Herwig predictions exhibit a prominent structure at medium \tb.
%

The Sherpa~2.2 predictions provide a good description of the data, in
particular at larger \tb.
The two different hadronization models (String vs.\ Cluster) provide
very similar predictions, which do not differ by more than the data
uncertainties, albeit the string model is a bit closer to the data at
medium \tb.
Both Sherpa~2.2 predictions overshoot the data at lowest \tb ($\tb<0.1$),
and undershoot the data at $\tb\simeq1$. The underprediction at high \tb{} and overprediction at low \tb{} seems to be a common feature of NLO+PS models, c.f.\ Powheg Box and Herwig.

The Sherpa~3.0 predictions with NLO matrix elements, improved
parton shower and cluster hadronization model provide a better 
description of the data than those from Sherpa~2.2, and the
predictions provide an accurate description of the data within the
uncertainties. Only at lower \Qsq\ and values of $\tb\sim0.3$ the
Sherpa~3.0 predictions overestimate the data.
A noteworthy improvement over other MC predictions is observed in the
lowest \tb\ region, where also good agreement with data is observed.
This could be related to the modelling of an intrisic $k_t$ in
Sherpa~3.0.
%

\paragraph{Comparison to MC event generators from the HERA era and dedicated DIS models}
The DIS MC event generators Djangoh and Rapgap provide an
overall satisfactory description of the data, although both have
difficulties to describe the shape in the region $\tb<0.3$ accurately
and both models underestimate the data in the range $0.1<\tb<0.3$.
The high \tb region is well described. Djangoh is somewhat higher
than Rapgap in that region, which is consistent with the observation of a harder $P_{\text{T}}$-spectrum of
jets~\cite{H1:2016goa}.
The triple-differential cross sections are well described by Djangoh
and Rapgap. At low $y$, Rapgap underestimates the high \tb region more
than Djangoh. Both models fail to describe the region $\tb<0.05$.

The KaTie+Cascade~3 predictions, employing TMD-sensitive PDF sets with
two different choices for the QCD evolution scale, provide a reasonable
description of the data. TMD PDF Set~2 provides one of the best
descriptions at the low \tb region, whereas the predictions with TMD
PDF Set~1 are a bit higher.
However, at lowest \Qsq\ and large $y$, also predictions from
Set~2 overshoot the data significantly.
In the region of larger \tb, the KaTie+Cascade~3 predictions fail to describe the data, 
likely related to the absence of $eg\to eq\bar(q)$ processes in the hard matrix elements.
%
%

\begin{figure*}[t]
\centering
\includegraphics[width=0.48\textwidth,trim={0 0 0 0 },clip]{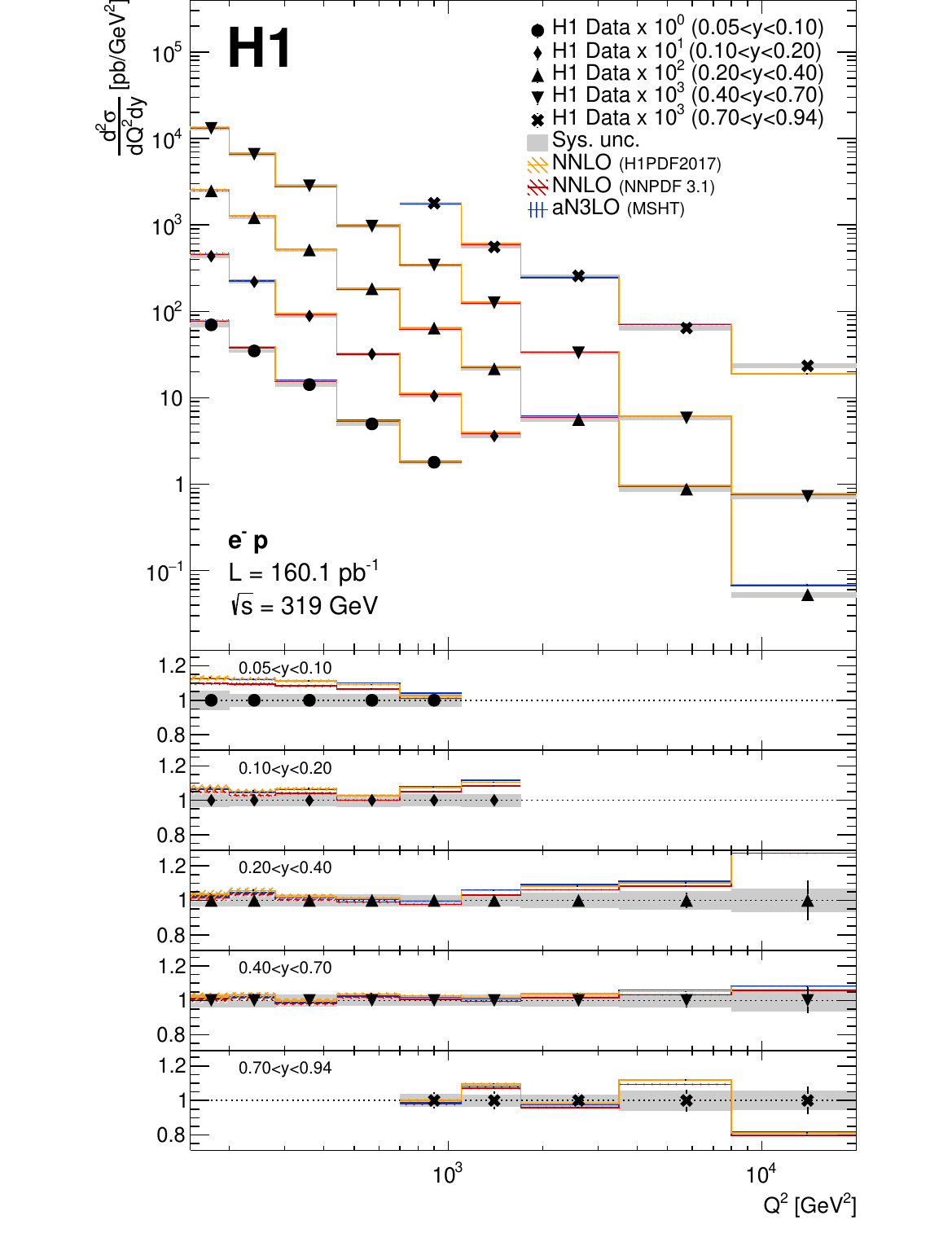}
\includegraphics[width=0.48\textwidth,trim={0 0 0 0 },clip]{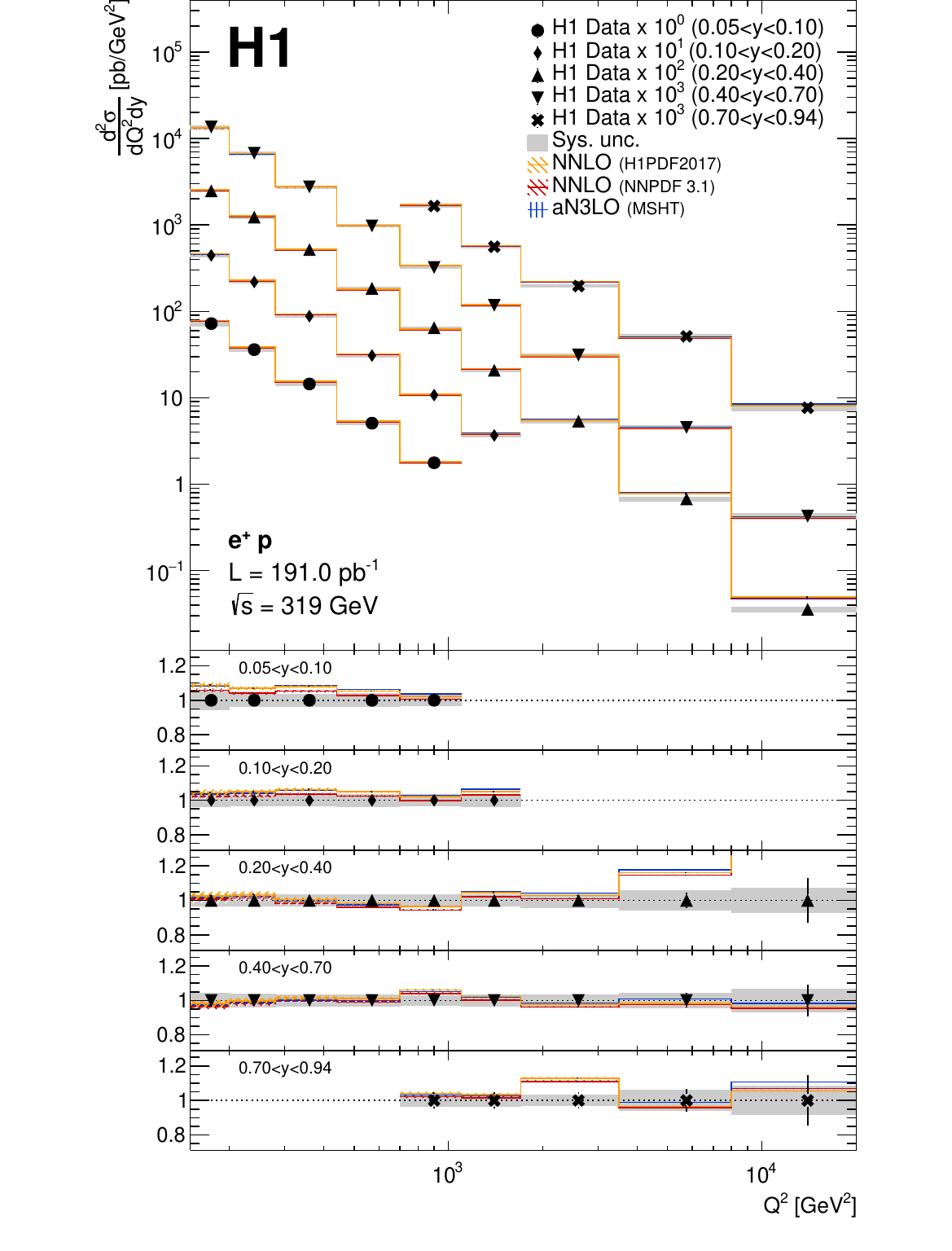}
\caption{
  The double-differential inclusive neutral current DIS cross section $\tfrac{d^{2}\sigma_\text{NC DIS}}{d\Qsq dy}$ for
  $e^-p$ (left) and $e^+p$ (right) scattering in the kinematic region
  $150\leq\Qsq<20\,000\,\GeVsq$ and $0.05\leq y<0.94$.
  The data are displayed as full circles, the vertical error
  bars indicate statistical uncertainties, while systematic uncertainties
  are displayed as shaded area.
  Different $y$ ranges are indicated by different markers, and are
  displaced vertically by a constant factor as indicated in the legend.
  The vertical error  bars indicate statistical uncertainties, whereas
  systematic experimental uncertainties are displayed as shaded area.
  The data are compared to NNLO and a3NLO predictions.
  The orange line indicates NNLO predictions using the
  H1PDF2017{\scriptsize NNLO} PDF set,
  the red line NNLO predictions using the
  NNPDF~3.1 PDF set, and the
  blue line indicates predictions in aN3LO accuracy using the MSHT PDF
  set.
  The hatched area indicates associated theoretical uncertainties from
  scale variations and PDF uncertainties.
}
\label{fig:NCDIS2D}
\end{figure*}

\paragraph{\boldmath Double-differential cross sections for inclusive
  neutral-current DIS as a function of \Qsq\ and $y$}
The inclusive NC DIS cross sections $\tfrac{d^{2}\sigma_\text{NC DIS}}{d\Qsq dy}$  are
presented in Tables~\ref{tab:NCDIS2D.em} and~\ref{tab:NCDIS2D.ep} and
are displayed in Figure~\ref{fig:NCDIS2D}, where NNLO and aN3LO
predictions for inclusive DIS are compared with the data.
For comparisons with the NC DIS predictions, the data are 
corrected to the DIS Born level by applying the factor $c_{\text  Born}$.
These cross sections are measured for $e^+p$ and $e^-p$ scattering 
by analyzing the data from the two lepton charges separately,
including dedicated unfolding matrices and QED correction factors.
The inclusive NC DIS cross sections $\tfrac{d^{2}\sigma_\text{NC DIS}}{d\Qsq dy}$  are 
obtained by integrating over \tb\ in every $(\Qsq,y)$ range.
Hence, kinematic migrations due to limited resolution of the
detector are corrected with high accuracy, since
different configurations of the hadronic final state, represented by \tb, are
considered in the unfolding.
The inclusive NC DIS cross sections can be compared with predictions
from structure function calculations, and serve as an important
validation of the absolute size of the \tb\ cross sections.
These data constitute the first bin integrated inclusive NC DIS cross section measurements at HERA in \Qsq\ and $y$, employing proper regularized unfolding of detector effects.
The results can be compared with predictions based on structure functions and serve as a crosscheck in the determination of the absolute normalization of the triple-differential cross section measurements.
The NC DIS measurements are also used together with the triple-differential cross sections to obtain normalized 1-jettiness event shape distributions~\cite{h1pubs}.
%
%

\paragraph{Predictions for inclusive NC DIS cross sections}
The double-differential inclusive DIS  $\tfrac{d^{2}\sigma_\text{NC DIS}}{d\Qsq
dy}$ are compared to predictions in NNLO and a3NLO accuracy.
It is observed that both the NNLO and the a3NLO predictions provide a
very good description of the $e^+p$ and $e^-p$ data in the entire
kinematic range, and small discrepancies are observed only at lowest
values of $y$ or largest values of \Qsq.
One can note that the H1 data~\cite{H1:2012qti,H1:2015ubc} are
already included in the PDF determination, which is the basis of the predictions.
Differences between the H1PDF2017{\scriptsize NNLO} PDF set and the
NNPDF3.1 PDF set are small.


\section{Summary}
The first measurement of the 1-jettiness event shape observable in
neutral-current deep-inelastic electron--proton scattering is
presented. 
The measured 1-jettiness observable \tb\ is equivalent to the classical event
shape observable Thrust normalized with $\frac{Q}{2}$, \tQ. It quantities the degree to which the hadronic final state in the current hemisphere is collimated along 
the exchanged bosons four-vector.

The data were recorded with the H1 experiment at the HERA
collider operating with a
center-of-mass energy of $\sqrt{s}=319\,\GeV$.
A single-differential cross section is measured as a function of \tb\ in the kinematic
region $200\leq\Qsq<1700\,\GeVsq$ and $0.2\leq y<0.7$.
A triple-differential cross section as a function of \tb, \Qsq,
and $y$ is reported in the kinematic range $0\leq\tb\leq1$,
$150\leq\Qsq<20\,000\,\GeVsq$, and
$0.05\leq y<0.94$.
The data are unfolded to the particle level and corrected for
higher-order QED radiative effects.


Given typical accuracies of $4$--$20\,\%$ (single differential) and $6$--$30\,\%$ (triple-differentials),
the data exhibit high sensitivity to the modelling of the hard
interaction, to the parton shower and hadronization models in MC
event generators.
The unfolded data are compared to a variety of predictions. Most of the 
models studied provide a satisfactory description of the data.
The classical MC event generators Djangoh and Rapgap, which were optimised for HERA
physics, provide an overall good description. 
The more modern general purpose MC event generators Herwig~7.2, Pythia~8.3,
Sherpa~2.2 and Sherpa~3.0 provide a good description of the
data, although some problems remain to be resolved in regions of the phase-space specific for each model.
When studying different parton shower or hadronization models,
the potential of these data for the optimization of event
generators is demonstrated.

Fixed order pQCD predictions provide a good description of the data,
but sizeable non-perturbative corrections for hadronization effects have to be applied. When incorporating NLL resummation, such predictions are seen to be in good agreement with the data over the full range of \tb{}.
Future pQCD analyses may be able to constrain parton distribution functions of the proton (PDFs) or the strong coupling constant from the data.

The \tb-integrated triple-differential cross
sections are found to be in agreement with the previously measured double-differential
inclusive NC DIS cross section.
This provides an important
consistency test of the data, but also a
consistency check with the HERA legacy measurements of inclusive NC DIS.
The inclusive NC DIS cross section is described by
structure function calculations which are available up to aN3LO and which are free of
hadronization effects.
The 1-jettiness measurement as a function of \tb, \Qsq, and $y$ can be viewed as a generalised NC DIS cross section,
and is the first triple-differential measurement of this kind.

In summary, it is believed that these data will be highly valuable to
improve MC event generators, which are of 
importance to achieve the physics goals of the HL-LHC physics
program~\cite{Dainese:2019rgk,ZurbanoFernandez:2020cco}. Further
improvements can be achieved when complemented with recent jet
substructure measurements~\cite{H1:2023fzk}.
With an improved understanding of soft and non-perturbative effects,
these data will become useful for the determination of PDFs or the
value of the strong coupling constant $\asmz$.
The presented measurement at HERA can be complemented in the future
with measurements in electron--proton collisions at lower
center-of-mass energies at  
the electron--ion collider in Brookhaven
(EIC)~\cite{AbdulKhalek:2021gbh} or at higher energies at the
LHeC or  FCC-eh at
CERN~\cite{FCC:2018byv,LHeC:2020van,AbdulKhalek:2021gbh}, or with
measurements in $pp$ at the LHC.

\section*{Acknowledgements}
We are grateful to the HERA machine group whose outstanding efforts
have made this experiment possible. We thank the engineers and
technicians for their work in constructing and maintaining the H1
detector, our funding agencies for financial support, the DESY
technical staff for continual assistance and the DESY directorate for
support and for the hospitality which they extend to the non-DESY members of the collaboration.
We express our thanks to all those involved in
securing not only the H1 data but also the software and working
environment for long term use,
allowing the unique H1 data set to continue to
be explored. The transfer from experiment specific to central resources with long term support,
including both storage and batch systems, has also been crucial to this enterprise.
We therefore also acknowledge the role played by DESY-IT and all
people involved during this transition and their
future role in the years to come.

We would like to thank
Valerio Bertone,
Silvia Ferrario Ravasio,
Ilkka Helenius,
Am{\'e}lie Henke,
Alexander Huss,
Hannes Jung,
Alexander Karlberg,
Christopher Lee,
Simon Plätzer,
Christian Preuss,
and
Hubert Spiesberger
for
many valuable comments and discussions, for providing us with theoretical
predictions, or for help with the predictions.
We thank Iain Stewart for motivating this analysis in the
`Future Physics with HERA data' workshop 2014~\cite{Stewart:2014}.

\par $^{f1}$ supported by the U.S. DOE Office of Science
\par $^{f2}$ supported by FNRS-FWO-Vlaanderen, IISN-IIKW and IWT and by Interuniversity Attraction Poles Programme, Belgian Science Policy
\par $^{f3}$ supported by the UK Science and Technology Facilities Council, and formerly by the UK Particle Physics and Astronomy Research Council
\par $^{f4}$ supported by the Romanian National Authority for Scientific Research under the contract PN 09370101
\par $^{f5}$ supported by the Bundesministerium für Bildung und Forschung, FRG, under contract numbers 05H09GUF, 05H09VHC, 05H09VHF, 05H16PEA
\par $^{f6}$ partially supported by Polish Ministry of Science and Higher Education, grant DPN/N168/DESY/2009
\par $^{f7}$ partially supported by Ministry of Science of Montenegro, no. 05-1/3-3352
\par $^{f8}$ supported by the Ministry of Education of the Czech Republic under the project INGO-LG14033
\par $^{f9}$ supported by CONACYT, México, grant 48778-F
\par $^{f10}$ supported by the Swiss National Science Foundation

\bibliography{desy24-035}



\FloatBarrier




\begin{table*}[t]
  \tiny\centering
    \begin{adjustbox}{width=\textwidth}
      \begin{tabular}{cc|c|cccccccc ccccc|ccccc c}
\toprule
\multicolumn{22}{c}{1D 1-jettiness cross sections  in $e^-p$ for  $200 < \Qsq < 1700\,\GeVsq$  and  $0.2 < y < 0.7$  --  $\d\sigma/\d\tb ~[\pb/\Delta\tb]$}\\
\midrule\multicolumn{2}{c|}{\tb range} & Results & \multicolumn{13}{c|}{Uncertainties} & \multicolumn{6}{c}{Correction factors}\\
\cmidrule(lr){1-2}\cmidrule(lr){3-3}\cmidrule(lr){4-16}\cmidrule(lr){17-22}
 $_\text{min}$ &  $_\text{max}$ & $\frac{\d\sigma}{\d\tb}$ & $_\text{stat}$ & $_\text{RCES}$ & $_\text{JES}$ & $_\text{HadTh}$ & $_\text{ElEn}$ & $_\text{ElTh}$ & $_\text{Model}$ & $_\text{MCstat}$ & $_\text{Unfold}$ &
 $_\text{ElecID}$ & $_\text{Lumi}$ & $_\text{Uncor}$ & $_\text{QED}$ & $c_\text{QED}$ & $c_\text{NoZ}$ & $c_\text{Born}$ & $c_\text{e+p}$ & $c_\text{Had}$ & $\delta_\text{HAD}$ \\
          &  &  $_{[\text{pb}]}$   & $_{[\%]}$  & $_{[\%]}$ & $_{[\%]}$ & $_{[\%]}$ & $_{[\%]}$ & $_{[\%]}$ & $_{[\%]}$ & $_{[\%]}$ & $_{[\%]}$ & $_{[\%]}$ & $_{[\%]}$ & $_{[\%]}$ & $_{[\%]}$ & & & & & & $_{[\%]}$\\
\midrule
   0.00 &   0.05 &      144.8 &     12.1 & $     -0.1$ & $     +1.6$ & $     +0.8$ & $     -0.6$ & $     +0.2$ & $    -15.6$ & $      0.5$ & $     -1.1$ &       0.2 &      2.7 &      0.5 &       0.3 &    1.080 &  0.975 &  0.912 &  0.960 &    0.102 &       1.1 \\
   0.05 &   0.10 &       1827 &      2.4 & $     +0.5$ & $     +1.0$ & $     +0.3$ & $     -0.9$ & $     +0.3$ & $     -0.3$ & $      0.5$ & $     +0.1$ &       0.2 &      2.7 &      0.5 &       0.1 &    1.070 &  0.981 &  0.909 &  0.974 &    0.769 &       1.3 \\
   0.10 &   0.15 &       3245 &      2.0 & $     -0.1$ & $     +0.6$ & $     +0.1$ & $     -0.4$ & $     +0.5$ & $     +4.0$ & $      0.6$ & $     +0.4$ &       0.2 &      2.7 &      0.5 &       0.1 &    1.066 &  0.988 &  0.912 &  0.984 &    1.342 &       6.7 \\
   0.15 &   0.22 &       2635 &      2.2 & $     +0.2$ & $     +0.1$ & $     -0.1$ & $     -1.0$ & $     +0.6$ & $     +2.8$ & $      0.6$ & $     -0.6$ &       0.2 &      2.7 &      0.5 &       0.1 &    1.065 &  0.989 &  0.913 &  0.990 &    1.397 &       1.1 \\
   0.22 &   0.30 &       1774 &      3.0 & $     +0.4$ & $     -0.3$ & $     -0.2$ & $     -0.7$ & $     +0.3$ & $     +2.5$ & $      0.6$ & $     -0.3$ &       0.2 &      2.7 &      0.5 &       0.1 &    1.071 &  0.988 &  0.911 &  0.989 &    1.307 &       1.7 \\
   0.30 &   0.40 &       1028 &      3.9 & $     +0.8$ & $     -0.7$ & $     -0.2$ & $     -1.0$ & $     +0.4$ & $     +2.1$ & $      0.4$ & $     +0.5$ &       0.2 &      2.7 &      0.5 &       0.1 &    1.071 &  0.987 &  0.912 &  0.989 &    1.311 &       4.5 \\
   0.40 &   0.54 &      675.7 &      3.4 & $     +1.0$ & $     -0.9$ & $     -0.2$ & $     -1.4$ & $     +0.5$ & $     +0.7$ & $      0.3$ & $     +0.1$ &       0.2 &      2.7 &      0.5 &       0.1 &    1.069 &  0.989 &  0.912 &  0.991 &    1.303 &       6.5 \\
   0.54 &   0.72 &      417.5 &      3.4 & $     +0.5$ & $     -0.9$ & $     -0.2$ & $     -1.3$ & $     +0.5$ & $     -1.6$ & $      0.2$ & $     +0.0$ &       0.2 &      2.7 &      0.5 &       0.1 &    1.076 &  0.987 &  0.909 &  0.991 &    1.268 &       5.1 \\
   0.72 &   0.86 &      290.1 &      4.7 & $     +0.0$ & $     -0.7$ & $     -0.1$ & $     -1.0$ & $     +0.5$ & $     -1.2$ & $      0.3$ & $     +0.1$ &       0.2 &      2.7 &      0.5 &       0.2 &    1.075 &  0.989 &  0.911 &  0.992 &    1.243 &       2.1 \\
   0.86 &   0.98 &      285.6 &      4.5 & $     +0.0$ & $     -0.9$ & $     +0.1$ & $     -0.5$ & $     +0.5$ & $     -0.4$ & $      0.3$ & $     +0.0$ &       0.2 &      2.7 &      0.5 &       0.2 &    1.073 &  0.992 &  0.913 &  0.994 &    1.340 &       1.6 \\
   0.98 &   1.00 &       1392 &      3.2 & $     -1.5$ & $     -0.2$ & $     -0.2$ & $     +0.0$ & $     +0.3$ & $     -1.0$ & $      0.2$ & $     -0.0$ &       0.2 &      2.7 &      0.5 &       0.2 &    1.076 &  0.988 &  0.913 &  0.991 &    0.763 &       1.1 \\
\bottomrule
\end{tabular}
\end{adjustbox}
\caption{ 
Single differential 1-jettiness cross sections for $e^-p$ scattering in the range $200 < \Qsq < 1700\,\GeVsq$  and  $0.2 < y < 0.7$,  $\d\sigma/\d\tb ~[\pb/\Delta\tb]$. The \tb{} ranges and the corresponding cross section measurements are indicated, together with the data statistical uncertainty (stat), and the systematic uncertainties: the hadronic (RCES and JES) and electron (ElEn) energy scale uncertainties, the hadron (HadTh) and electron (ElTh) polar angle uncertainties, the model uncertainty (Model), the simulation statistical uncertainty (MCstat), the unfolding uncertainty (Unfold), the electron identification uncertainty (ElecID), the luminosity uncertainty (Lumi), the residual uncorrelated uncertainty (Uncor), and the QED correction uncertainty (QED). The radiative correction factors are also shown.
}
\label{tab:xs1D.em} 
\end{table*}

\begin{table*}[t]
\tiny\centering\begin{adjustbox}{width=\textwidth}\begin{tabular}{cc|c|cccccccc ccccc|ccccc c}
\toprule
\multicolumn{22}{c}{1D 1-jettiness cross sections  in $e^+p$ for  $200 < \Qsq < 1700\,\GeVsq$  and  $0.2 < y < 0.7$  --  $\d\sigma/\d\tb ~[\pb/\Delta\tb]$}\\
\midrule\multicolumn{2}{c|}{\tb range} & Results & \multicolumn{13}{c|}{Uncertainties} & \multicolumn{6}{c}{Correction factors}\\
\cmidrule(lr){1-2}\cmidrule(lr){3-3}\cmidrule(lr){4-16}\cmidrule(lr){17-22}
 $_\text{min}$ &  $_\text{max}$ & $\frac{\d\sigma}{\d\tb}$ & $_\text{stat}$ & $_\text{RCES}$ & $_\text{JES}$ & $_\text{HadTh}$ & $_\text{ElEn}$ & $_\text{ElTh}$ & $_\text{Model}$ & $_\text{MCstat}$ & $_\text{Unfold}$ &
 $_\text{ElecID}$ & $_\text{Lumi}$ & $_\text{Uncor}$ & $_\text{QED}$ & $c_\text{QED}$ & $c_\text{NoZ}$ & $c_\text{Born}$ & $c_\text{e+p}$ & $c_\text{Had}$ & $\delta_\text{HAD}$ \\
          &  &  $_{[\text{pb}]}$   & $_{[\%]}$  & $_{[\%]}$ & $_{[\%]}$ & $_{[\%]}$ & $_{[\%]}$ & $_{[\%]}$ & $_{[\%]}$ & $_{[\%]}$ & $_{[\%]}$ & $_{[\%]}$ & $_{[\%]}$ & $_{[\%]}$ & $_{[\%]}$ & & & & & & $_{[\%]}$\\
\midrule
   0.00 &   0.05 &      137.5 &     11.4 & $     +0.2$ & $     +1.7$ & $     +0.9$ & $     -0.8$ & $     +0.2$ & $    -15.2$ & $      0.5$ & $     -0.7$ &       0.2 &      2.7 &      0.5 &       0.3 &    1.077 &  1.013 &  0.916 &  1.000 &    0.100 &       1.1 \\
   0.05 &   0.10 &       1841 &      2.1 & $     +0.8$ & $     +1.1$ & $     +0.4$ & $     -0.7$ & $     +0.5$ & $     +0.4$ & $      0.4$ & $     +0.7$ &       0.2 &      2.7 &      0.5 &       0.1 &    1.070 &  1.005 &  0.916 &  1.000 &    0.761 &       1.1 \\
   0.10 &   0.15 &       3174 &      1.8 & $     +0.2$ & $     +0.5$ & $     -0.1$ & $     -0.6$ & $     +0.3$ & $     +4.1$ & $      0.6$ & $     -1.0$ &       0.2 &      2.7 &      0.5 &       0.1 &    1.065 &  1.001 &  0.918 &  1.000 &    1.336 &       6.7 \\
   0.15 &   0.22 &       2580 &      1.9 & $     +0.0$ & $     -0.1$ & $     -0.1$ & $     -0.5$ & $     +0.4$ & $     +2.5$ & $      0.6$ & $     +0.7$ &       0.2 &      2.7 &      0.5 &       0.1 &    1.067 &  0.997 &  0.917 &  1.000 &    1.400 &       1.1 \\
   0.22 &   0.30 &       1730 &      2.7 & $     +0.4$ & $     -0.2$ & $     -0.1$ & $     -1.1$ & $     +0.6$ & $     +2.5$ & $      0.5$ & $     +0.4$ &       0.2 &      2.7 &      0.5 &       0.1 &    1.065 &  0.996 &  0.921 &  1.000 &    1.308 &       1.9 \\
   0.30 &   0.40 &       1099 &      3.3 & $     +0.8$ & $     -0.7$ & $     -0.2$ & $     -1.2$ & $     +0.4$ & $     +2.1$ & $      0.4$ & $     -1.5$ &       0.2 &      2.7 &      0.5 &       0.1 &    1.067 &  0.995 &  0.920 &  1.000 &    1.311 &       4.3 \\
   0.40 &   0.54 &      651.7 &      3.2 & $     +1.1$ & $     -0.9$ & $     -0.2$ & $     -1.5$ & $     +0.5$ & $     +0.5$ & $      0.3$ & $     +0.5$ &       0.2 &      2.7 &      0.5 &       0.1 &    1.068 &  0.996 &  0.919 &  1.000 &    1.302 &       6.7 \\
   0.54 &   0.72 &      419.5 &      3.1 & $     +0.6$ & $     -0.8$ & $     -0.2$ & $     -1.2$ & $     +0.5$ & $     -1.7$ & $      0.2$ & $     +0.4$ &       0.2 &      2.7 &      0.5 &       0.1 &    1.072 &  0.994 &  0.917 &  1.000 &    1.269 &       5.1 \\
   0.72 &   0.86 &      298.2 &      4.3 & $     -0.0$ & $     -0.8$ & $     -0.1$ & $     -1.0$ & $     +0.5$ & $     -1.8$ & $      0.3$ & $     +0.1$ &       0.2 &      2.7 &      0.5 &       0.2 &    1.072 &  0.994 &  0.917 &  1.000 &    1.243 &       2.2 \\
   0.86 &   0.98 &      337.4 &      3.7 & $     +0.1$ & $     -0.9$ & $     -0.0$ & $     -0.5$ & $     +0.4$ & $     -0.9$ & $      0.3$ & $     -0.0$ &       0.2 &      2.7 &      0.5 &       0.2 &    1.071 &  0.995 &  0.919 &  1.000 &    1.340 &       1.6 \\
   0.98 &   1.00 &       1428 &      2.9 & $     -1.4$ & $     -0.2$ & $     -0.1$ & $     +0.1$ & $     +0.4$ & $     -1.4$ & $      0.2$ & $     -0.0$ &       0.2 &      2.7 &      0.5 &       0.2 &    1.075 &  0.995 &  0.922 &  1.000 &    0.762 &       1.1 \\
\bottomrule
\end{tabular}\end{adjustbox}
\caption{ 
Single differential 1-jettiness cross sections for $e^+p$ scattering in the range $200 < \Qsq < 1700\,\GeVsq$  and  $0.2 < y < 0.7$,  $\d\sigma/\d\tb ~[\pb/\Delta\tb]$. Further details are given in the caption of Tab.~\ref{tab:xs1D.em}.}
\label{tab:xs1D.ep} 
\end{table*}

\begin{table*}[t]
\tiny\centering\begin{adjustbox}{width=\textwidth}\begin{tabular}{cc|c|cccccccc ccccc|ccccc c}
\toprule
\multicolumn{22}{c}{1D 1-jettiness cross sections  in $e^-p$ ($\mathcal{L}=351.1\,\pb^{-1}$) for  $200 < \Qsq < 1700\,\GeVsq$  and  $0.2 < y < 0.7$  --  $\d\sigma/\d\tb ~[\pb/\Delta\tb]$}\\
\midrule\multicolumn{2}{c|}{\tb range} & Results & \multicolumn{13}{c|}{Uncertainties} & \multicolumn{6}{c}{Correction factors}\\
\cmidrule(lr){1-2}\cmidrule(lr){3-3}\cmidrule(lr){4-16}\cmidrule(lr){17-22}
 $_\text{min}$ &  $_\text{max}$ & $\frac{\d\sigma}{\d\tb}$ & $_\text{stat}$ & $_\text{RCES}$ & $_\text{JES}$ & $_\text{HadTh}$ & $_\text{ElEn}$ & $_\text{ElTh}$ & $_\text{Model}$ & $_\text{MCstat}$ & $_\text{Unfold}$ &
 $_\text{ElecID}$ & $_\text{Lumi}$ & $_\text{Uncor}$ & $_\text{QED}$ & $c_\text{QED}$ & $c_\text{NoZ}$ & $c_\text{Born}$ & $c_\text{e+p}$ & $c_\text{Had}$ & $\delta_\text{HAD}$ \\
          &  &  $_{[\text{pb}]}$   & $_{[\%]}$  & $_{[\%]}$ & $_{[\%]}$ & $_{[\%]}$ & $_{[\%]}$ & $_{[\%]}$ & $_{[\%]}$ & $_{[\%]}$ & $_{[\%]}$ & $_{[\%]}$ & $_{[\%]}$ & $_{[\%]}$ & $_{[\%]}$ & & & & & & $_{[\%]}$\\
\midrule
   0.00 &   0.05 &      149.8 &      7.8 & $     +0.0$ & $     +1.7$ & $     +0.9$ & $     -0.6$ & $     +0.2$ & $    -15.4$ & $      0.4$ & $     -1.2$ &       0.2 &      2.7 &      0.5 &       0.2 &    1.101 &  0.975 &  0.912 &  0.960 &    0.102 &       1.1 \\
   0.05 &   0.10 &       1885 &      1.5 & $     +0.7$ & $     +1.1$ & $     +0.3$ & $     -0.8$ & $     +0.4$ & $     +0.0$ & $      0.3$ & $     +0.7$ &       0.2 &      2.7 &      0.5 &       0.1 &    1.084 &  0.981 &  0.909 &  0.974 &    0.769 &       1.3 \\
   0.10 &   0.15 &       3227 &      1.2 & $     +0.1$ & $     +0.5$ & $     -0.0$ & $     -0.6$ & $     +0.4$ & $     +4.0$ & $      0.4$ & $     -0.8$ &       0.2 &      2.7 &      0.5 &       0.1 &    1.073 &  0.988 &  0.912 &  0.984 &    1.342 &       6.7 \\
   0.15 &   0.22 &       2632 &      1.2 & $     +0.1$ & $     -0.0$ & $     -0.1$ & $     -0.8$ & $     +0.4$ & $     +2.7$ & $      0.4$ & $     +0.5$ &       0.2 &      2.7 &      0.5 &       0.1 &    1.070 &  0.989 &  0.913 &  0.990 &    1.397 &       1.1 \\
   0.22 &   0.30 &       1770 &      1.7 & $     +0.4$ & $     -0.2$ & $     -0.1$ & $     -0.9$ & $     +0.5$ & $     +2.5$ & $      0.3$ & $     -0.1$ &       0.2 &      2.7 &      0.5 &       0.1 &    1.072 &  0.988 &  0.911 &  0.989 &    1.307 &       1.7 \\
   0.30 &   0.40 &       1070 &      2.3 & $     +0.8$ & $     -0.7$ & $     -0.2$ & $     -1.1$ & $     +0.5$ & $     +2.2$ & $      0.3$ & $     -0.6$ &       0.2 &      2.7 &      0.5 &       0.1 &    1.073 &  0.987 &  0.912 &  0.989 &    1.311 &       4.5 \\
   0.40 &   0.54 &      681.3 &      2.2 & $     +1.1$ & $     -0.9$ & $     -0.2$ & $     -1.5$ & $     +0.6$ & $     +0.6$ & $      0.2$ & $     +0.3$ &       0.2 &      2.7 &      0.5 &       0.1 &    1.072 &  0.989 &  0.912 &  0.991 &    1.303 &       6.5 \\
   0.54 &   0.72 &      423.8 &      2.3 & $     +0.5$ & $     -0.9$ & $     -0.2$ & $     -1.3$ & $     +0.5$ & $     -1.6$ & $      0.2$ & $     +0.2$ &       0.2 &      2.7 &      0.5 &       0.1 &    1.077 &  0.987 &  0.909 &  0.991 &    1.268 &       5.1 \\
   0.72 &   0.86 &        302 &      3.1 & $     +0.0$ & $     -0.8$ & $     -0.2$ & $     -1.0$ & $     +0.5$ & $     -1.4$ & $      0.2$ & $     +0.1$ &       0.2 &      2.7 &      0.5 &       0.1 &    1.076 &  0.989 &  0.911 &  0.992 &    1.243 &       2.1 \\
   0.86 &   0.98 &      336.7 &      2.7 & $     +0.1$ & $     -0.9$ & $     +0.0$ & $     -0.5$ & $     +0.5$ & $     -0.5$ & $      0.2$ & $     -0.0$ &       0.2 &      2.7 &      0.5 &       0.1 &    1.074 &  0.992 &  0.913 &  0.994 &    1.340 &       1.6 \\
   0.98 &   1.00 &       1422 &      2.2 & $     -1.4$ & $     -0.2$ & $     -0.2$ & $     +0.1$ & $     +0.3$ & $     -1.1$ & $      0.1$ & $     -0.0$ &       0.2 &      2.7 &      0.5 &       0.1 &    1.079 &  0.988 &  0.913 &  0.991 &    0.763 &       1.1 \\
\bottomrule
\end{tabular}\end{adjustbox}
\caption{ 
Single differential 1-jettiness cross sections for $e^-p$ ($\mathcal{L}=351.1\,\pb^{-1}$) scattering in the range $200 < \Qsq < 1700\,\GeVsq$  and  $0.2 < y < 0.7$,  $\d\sigma/\d\tb ~[\pb/\Delta\tb]$. Further details are given in the caption of Tab.~\ref{tab:xs1D.em}.
}
\label{tab:xs1D.em351} 
\end{table*}



\begin{table*}[t]
\tiny\centering\begin{adjustbox}{width=\textwidth}
\begin{tabular}{cc|c|cccccccc ccccc|ccccc c}
\toprule
\multicolumn{21}{c}{Cross sections in $e^-p$ ($\mathcal{L}=351.1\,\pb^{-1}$) for $150 < \Qsq < 200$\,\GeVsq  and  $0.05 < y < 0.10$  --  $\d\sigma/\d\tb ~[\pb/\Delta\tb]$} \\
\midrule\multicolumn{2}{c|}{\tb range} & Results & \multicolumn{13}{c|}{Uncertainties} & \multicolumn{6}{c}{Correction factors}\\
\cmidrule(lr){1-2}\cmidrule(lr){3-3}\cmidrule(lr){4-16}\cmidrule(lr){17-22}
 $_\text{min}$ &  $_\text{max}$ & $\frac{\d\sigma}{\d\tb}$ & $_\text{stat}$ & $_\text{RCES}$ & $_\text{JES}$ & $_\text{HadTh}$ & $_\text{ElEn}$ & $_\text{ElTh}$ & $_\text{Model}$ & $_\text{MCstat}$ & $_\text{Unfold}$ &
 $_\text{ElecID}$ & $_\text{Lumi}$ & $_\text{Uncor}$ & $_\text{QED}$ & $c_\text{QED}$ & $c_\text{NoZ}$ & $c_\text{Born}$ & $c_\text{e+p}$ & $c_\text{Had}$ & $\delta_\text{HAD}$ \\
          &  &  $_{[\text{pb}]}$   & $_{[\%]}$  & $_{[\%]}$ & $_{[\%]}$ & $_{[\%]}$ & $_{[\%]}$ & $_{[\%]}$ & $_{[\%]}$ & $_{[\%]}$ & $_{[\%]}$ & $_{[\%]}$ & $_{[\%]}$ & $_{[\%]}$ & $_{[\%]}$ & & & & & & $_{[\%]}$\\
\midrule
   0.00 &   0.10 &      22.81 &     22.8 & $     +2.7$ & $     +1.8$ & $     +2.2$ & $     +4.4$ & $     +0.8$ & $    -11.0$ & $      2.3$ & $     +1.5$ &       0.2 &      2.7 &      0.5 &       0.8 &    1.063 &  1.010 &  0.918 &  0.992 &    0.144 &       1.6 \\
   0.10 &   0.15 &      345.4 &      4.0 & $     +2.7$ & $     +1.1$ & $     +1.1$ & $     +3.7$ & $     +0.9$ & $     -2.9$ & $      1.3$ & $     +0.4$ &       0.2 &      2.7 &      0.5 &       0.5 &    1.072 &  0.996 &  0.925 &  1.001 &    1.068 &       4.5 \\
   0.15 &   0.22 &      528.3 &      2.5 & $     +1.6$ & $     +0.4$ & $     +0.2$ & $     +3.0$ & $     +0.9$ & $     +4.0$ & $      0.9$ & $     +0.7$ &       0.2 &      2.7 &      0.5 &       0.4 &    1.065 &  1.001 &  0.922 &  1.002 &    2.035 &      31.9 \\
\bottomrule
\end{tabular}\end{adjustbox}
\caption{ 
Differential 1-jettiness cross sections for $e^-p$ ($\mathcal{L}=351.1\,\pb^{-1}$) scattering in the range  $150 < \Qsq < 200$\,\GeVsq  and  $0.05 < y < 0.10$,  $\d\sigma/\d\tb ~[\pb/\Delta\tb]$. Further details are given in the caption of Tab.~\ref{tab:xs1D.em}.
}
\label{tab:xstaub1b.em351.150_0.05}
\end{table*}

\begin{table*}[t]
\tiny\centering\begin{adjustbox}{width=\textwidth}\begin{tabular}{cc|c|cccccccc ccccc|ccccc c}
\toprule
\multicolumn{21}{c}{Cross sections in $e^-p$ ($\mathcal{L}=351.1\,\pb^{-1}$) for $150 < \Qsq < 200$\,\GeVsq  and  $0.10 < y < 0.20$  --  $\d\sigma/\d\tb ~[\pb/\Delta\tb]$} \\
\midrule\multicolumn{2}{c|}{\tb range} & Results & \multicolumn{13}{c|}{Uncertainties} & \multicolumn{6}{c}{Correction factors}\\
\cmidrule(lr){1-2}\cmidrule(lr){3-3}\cmidrule(lr){4-16}\cmidrule(lr){17-22}
 $_\text{min}$ &  $_\text{max}$ & $\frac{\d\sigma}{\d\tb}$ & $_\text{stat}$ & $_\text{RCES}$ & $_\text{JES}$ & $_\text{HadTh}$ & $_\text{ElEn}$ & $_\text{ElTh}$ & $_\text{Model}$ & $_\text{MCstat}$ & $_\text{Unfold}$ &
 $_\text{ElecID}$ & $_\text{Lumi}$ & $_\text{Uncor}$ & $_\text{QED}$ & $c_\text{QED}$ & $c_\text{NoZ}$ & $c_\text{Born}$ & $c_\text{e+p}$ & $c_\text{Had}$ & $\delta_\text{HAD}$ \\
          &  &  $_{[\text{pb}]}$   & $_{[\%]}$  & $_{[\%]}$ & $_{[\%]}$ & $_{[\%]}$ & $_{[\%]}$ & $_{[\%]}$ & $_{[\%]}$ & $_{[\%]}$ & $_{[\%]}$ & $_{[\%]}$ & $_{[\%]}$ & $_{[\%]}$ & $_{[\%]}$ & & & & & & $_{[\%]}$\\
\midrule
   0.00 &   0.05 &      3.299 &    153.8 & $     -3.4$ & $     +2.9$ & $     +1.9$ & $     +0.1$ & $     +0.9$ & $    -38.8$ & $     10.6$ & $     -2.5$ &       0.2 &      2.7 &      0.5 &       2.3 &    1.107 &  1.010 &  0.967 &  0.990 &    0.019 &       0.2 \\
   0.05 &   0.10 &      102.2 &     14.1 & $     +1.6$ & $     +1.5$ & $     +1.4$ & $     +4.3$ & $     -1.1$ & $    -18.5$ & $      2.6$ & $     +1.1$ &       0.2 &      2.7 &      0.5 &       0.7 &    1.079 &  0.994 &  0.904 &  0.985 &    0.248 &       5.2 \\
   0.10 &   0.15 &      454.1 &      4.6 & $     +2.4$ & $     +1.6$ & $     +0.7$ & $     +0.3$ & $     +1.0$ & $     +0.0$ & $      2.0$ & $     -0.8$ &       0.2 &      2.7 &      0.5 &       0.4 &    1.077 &  0.986 &  0.914 &  1.000 &    0.869 &       3.0 \\
\bottomrule
\end{tabular}\end{adjustbox}
\caption{ 
Differential 1-jettiness cross sections for $e^-p$ ($\mathcal{L}=351.1\,\pb^{-1}$) scattering in the range  $150 < \Qsq < 200$\,\GeVsq  and  $0.10 < y < 0.20$,  $\d\sigma/\d\tb ~[\pb/\Delta\tb]$. Further details are given in the caption of Tab.~\ref{tab:xs1D.em}.
}
\label{tab:xstaub1b.em351.150_0.10}
\end{table*}

\begin{table*}[t]
\tiny\centering\begin{adjustbox}{width=\textwidth}\begin{tabular}{cc|c|cccccccc ccccc|ccccc c}
\toprule
\multicolumn{21}{c}{Cross sections in $e^-p$ ($\mathcal{L}=351.1\,\pb^{-1}$) for $150 < \Qsq < 200$\,\GeVsq  and  $0.20 < y < 0.40$  --  $\d\sigma/\d\tb ~[\pb/\Delta\tb]$} \\
\midrule\multicolumn{2}{c|}{\tb range} & Results & \multicolumn{13}{c|}{Uncertainties} & \multicolumn{6}{c}{Correction factors}\\
\cmidrule(lr){1-2}\cmidrule(lr){3-3}\cmidrule(lr){4-16}\cmidrule(lr){17-22}
 $_\text{min}$ &  $_\text{max}$ & $\frac{\d\sigma}{\d\tb}$ & $_\text{stat}$ & $_\text{RCES}$ & $_\text{JES}$ & $_\text{HadTh}$ & $_\text{ElEn}$ & $_\text{ElTh}$ & $_\text{Model}$ & $_\text{MCstat}$ & $_\text{Unfold}$ &
 $_\text{ElecID}$ & $_\text{Lumi}$ & $_\text{Uncor}$ & $_\text{QED}$ & $c_\text{QED}$ & $c_\text{NoZ}$ & $c_\text{Born}$ & $c_\text{e+p}$ & $c_\text{Had}$ & $\delta_\text{HAD}$ \\
          &  &  $_{[\text{pb}]}$   & $_{[\%]}$  & $_{[\%]}$ & $_{[\%]}$ & $_{[\%]}$ & $_{[\%]}$ & $_{[\%]}$ & $_{[\%]}$ & $_{[\%]}$ & $_{[\%]}$ & $_{[\%]}$ & $_{[\%]}$ & $_{[\%]}$ & $_{[\%]}$ & & & & & & $_{[\%]}$\\
\midrule
   0.00 &   0.05 &      7.584 &     48.1 & $     -0.6$ & $     +1.1$ & $     +1.0$ & $     +1.3$ & $     +0.6$ & $    -34.2$ & $      7.2$ & $     +6.3$ &       0.2 &      2.7 &      0.5 &       1.9 &    1.047 &  1.015 &  0.954 &  0.979 &    0.018 &       0.2 \\
   0.05 &   0.10 &      101.9 &     10.4 & $     +1.3$ & $     +1.6$ & $     +0.2$ & $     -1.7$ & $     -0.2$ & $    -18.1$ & $      2.4$ & $     +0.5$ &       0.2 &      2.7 &      0.5 &       0.6 &    1.085 &  0.994 &  0.911 &  0.980 &    0.209 &       3.4 \\
   0.10 &   0.15 &      376.1 &      4.5 & $     +1.3$ & $     +2.3$ & $     +1.0$ & $     +0.5$ & $     +0.7$ & $     +0.2$ & $      1.7$ & $     -0.5$ &       0.2 &      2.7 &      0.5 &       0.3 &    1.067 &  1.002 &  0.926 &  0.998 &    0.714 &       1.6 \\
\bottomrule
\end{tabular}\end{adjustbox}
\caption{ 
Differential 1-jettiness cross sections for $e^-p$ ($\mathcal{L}=351.1\,\pb^{-1}$) scattering in the range  $150 < \Qsq < 200$\,\GeVsq  and  $0.20 < y < 0.40$,  $\d\sigma/\d\tb ~[\pb/\Delta\tb]$. Further details are given in the caption of Tab.~\ref{tab:xs1D.em}.
}
\label{tab:xstaub1b.em351.150_0.20}
\end{table*}

\begin{table*}[t]
\tiny\centering\begin{adjustbox}{width=\textwidth}\begin{tabular}{cc|c|cccccccc ccccc|ccccc c}
\toprule
\multicolumn{21}{c}{Cross sections in $e^-p$ ($\mathcal{L}=351.1\,\pb^{-1}$) for $150 < \Qsq < 200$\,\GeVsq  and  $0.40 < y < 0.70$  --  $\d\sigma/\d\tb ~[\pb/\Delta\tb]$} \\
\midrule\multicolumn{2}{c|}{\tb range} & Results & \multicolumn{13}{c|}{Uncertainties} & \multicolumn{6}{c}{Correction factors}\\
\cmidrule(lr){1-2}\cmidrule(lr){3-3}\cmidrule(lr){4-16}\cmidrule(lr){17-22}
 $_\text{min}$ &  $_\text{max}$ & $\frac{\d\sigma}{\d\tb}$ & $_\text{stat}$ & $_\text{RCES}$ & $_\text{JES}$ & $_\text{HadTh}$ & $_\text{ElEn}$ & $_\text{ElTh}$ & $_\text{Model}$ & $_\text{MCstat}$ & $_\text{Unfold}$ &
 $_\text{ElecID}$ & $_\text{Lumi}$ & $_\text{Uncor}$ & $_\text{QED}$ & $c_\text{QED}$ & $c_\text{NoZ}$ & $c_\text{Born}$ & $c_\text{e+p}$ & $c_\text{Had}$ & $\delta_\text{HAD}$ \\
          &  &  $_{[\text{pb}]}$   & $_{[\%]}$  & $_{[\%]}$ & $_{[\%]}$ & $_{[\%]}$ & $_{[\%]}$ & $_{[\%]}$ & $_{[\%]}$ & $_{[\%]}$ & $_{[\%]}$ & $_{[\%]}$ & $_{[\%]}$ & $_{[\%]}$ & $_{[\%]}$ & & & & & & $_{[\%]}$\\
\midrule
   0.00 &   0.05 &      19.92 &     23.3 & $     +0.4$ & $     +2.1$ & $     +1.9$ & $     +0.5$ & $     +0.8$ & $    -23.5$ & $      8.7$ & $     -7.8$ &       0.2 &      2.7 &      0.5 &       1.8 &    1.111 &  0.970 &  0.850 &  0.933 &    0.017 &       0.2 \\
   0.05 &   0.10 &      45.17 &     21.7 & $     -2.7$ & $     -0.4$ & $     +0.3$ & $     +1.3$ & $     +0.5$ & $    -11.3$ & $      3.0$ & $     -1.6$ &       0.2 &      2.7 &      0.5 &       0.5 &    1.054 &  1.012 &  0.918 &  1.006 &    0.180 &       2.1 \\
   0.10 &   0.15 &      265.3 &      6.0 & $     -0.1$ & $     +1.2$ & $     -0.3$ & $     +0.1$ & $     -0.2$ & $     +2.0$ & $      1.9$ & $     +2.4$ &       0.2 &      2.7 &      0.5 &       0.3 &    1.066 &  0.992 &  0.914 &  0.994 &    0.630 &       1.4 \\
\bottomrule
\end{tabular}\end{adjustbox}
\caption{ 
Differential 1-jettiness cross sections for $e^-p$ ($\mathcal{L}=351.1\,\pb^{-1}$) scattering in the range  $150 < \Qsq < 200$\,\GeVsq  and  $0.40 < y < 0.70$,  $\d\sigma/\d\tb ~[\pb/\Delta\tb]$. Further details are given in the caption of Tab.~\ref{tab:xs1D.em}.
}
\label{tab:xstaub1b.em351.150_0.40}
\end{table*}

\begin{table*}[t]
\tiny\centering\begin{adjustbox}{width=\textwidth}\begin{tabular}{cc|c|cccccccc ccccc|ccccc c}
\toprule
\multicolumn{21}{c}{Cross sections in $e^-p$ ($\mathcal{L}=351.1\,\pb^{-1}$) for $200 < \Qsq < 280$\,\GeVsq  and  $0.05 < y < 0.10$  --  $\d\sigma/\d\tb ~[\pb/\Delta\tb]$} \\
\midrule\multicolumn{2}{c|}{\tb range} & Results & \multicolumn{13}{c|}{Uncertainties} & \multicolumn{6}{c}{Correction factors}\\
\cmidrule(lr){1-2}\cmidrule(lr){3-3}\cmidrule(lr){4-16}\cmidrule(lr){17-22}
 $_\text{min}$ &  $_\text{max}$ & $\frac{\d\sigma}{\d\tb}$ & $_\text{stat}$ & $_\text{RCES}$ & $_\text{JES}$ & $_\text{HadTh}$ & $_\text{ElEn}$ & $_\text{ElTh}$ & $_\text{Model}$ & $_\text{MCstat}$ & $_\text{Unfold}$ &
 $_\text{ElecID}$ & $_\text{Lumi}$ & $_\text{Uncor}$ & $_\text{QED}$ & $c_\text{QED}$ & $c_\text{NoZ}$ & $c_\text{Born}$ & $c_\text{e+p}$ & $c_\text{Had}$ & $\delta_\text{HAD}$ \\
          &  &  $_{[\text{pb}]}$   & $_{[\%]}$  & $_{[\%]}$ & $_{[\%]}$ & $_{[\%]}$ & $_{[\%]}$ & $_{[\%]}$ & $_{[\%]}$ & $_{[\%]}$ & $_{[\%]}$ & $_{[\%]}$ & $_{[\%]}$ & $_{[\%]}$ & $_{[\%]}$ & & & & & & $_{[\%]}$\\
\midrule
   0.00 &   0.10 &      45.03 &      8.0 & $     +1.8$ & $     +1.4$ & $     +1.1$ & $     -1.4$ & $     +0.5$ & $     -8.9$ & $      0.9$ & $     +2.3$ &       0.2 &      2.7 &      0.5 &       0.6 &    1.062 &  0.993 &  0.929 &  0.997 &    0.220 &       2.2 \\
   0.10 &   0.15 &      445.6 &      2.1 & $     +2.2$ & $     +0.7$ & $     +0.6$ & $     -1.6$ & $     +0.6$ & $     -0.4$ & $      0.8$ & $     -0.6$ &       0.2 &      2.7 &      0.5 &       0.4 &    1.083 &  0.989 &  0.907 &  0.992 &    1.767 &      24.8 \\
   0.15 &   0.22 &      495.6 &      1.7 & $     +0.6$ & $     +0.1$ & $     -0.2$ & $     -0.9$ & $     +0.5$ & $     +4.7$ & $      0.8$ & $     -0.8$ &       0.2 &      2.7 &      0.5 &       0.4 &    1.068 &  0.991 &  0.909 &  0.998 &    2.198 &      28.3 \\
\bottomrule
\end{tabular}\end{adjustbox}
\caption{ 
Differential 1-jettiness cross sections for $e^-p$ ($\mathcal{L}=351.1\,\pb^{-1}$) scattering in the range  $200 < \Qsq < 280$\,\GeVsq  and  $0.05 < y < 0.10$,  $\d\sigma/\d\tb ~[\pb/\Delta\tb]$. Further details are given in the caption of Tab.~\ref{tab:xs1D.em}.
}
\label{tab:xstaub1b.em351.200_0.05}
\end{table*}

\begin{table*}[t]
\tiny\centering\begin{adjustbox}{width=\textwidth}\begin{tabular}{cc|c|cccccccc ccccc|ccccc c}
\toprule
\multicolumn{21}{c}{Cross sections in $e^-p$ ($\mathcal{L}=351.1\,\pb^{-1}$) for $200 < \Qsq < 280$\,\GeVsq  and  $0.10 < y < 0.20$  --  $\d\sigma/\d\tb ~[\pb/\Delta\tb]$} \\
\midrule\multicolumn{2}{c|}{\tb range} & Results & \multicolumn{13}{c|}{Uncertainties} & \multicolumn{6}{c}{Correction factors}\\
\cmidrule(lr){1-2}\cmidrule(lr){3-3}\cmidrule(lr){4-16}\cmidrule(lr){17-22}
 $_\text{min}$ &  $_\text{max}$ & $\frac{\d\sigma}{\d\tb}$ & $_\text{stat}$ & $_\text{RCES}$ & $_\text{JES}$ & $_\text{HadTh}$ & $_\text{ElEn}$ & $_\text{ElTh}$ & $_\text{Model}$ & $_\text{MCstat}$ & $_\text{Unfold}$ &
 $_\text{ElecID}$ & $_\text{Lumi}$ & $_\text{Uncor}$ & $_\text{QED}$ & $c_\text{QED}$ & $c_\text{NoZ}$ & $c_\text{Born}$ & $c_\text{e+p}$ & $c_\text{Had}$ & $\delta_\text{HAD}$ \\
          &  &  $_{[\text{pb}]}$   & $_{[\%]}$  & $_{[\%]}$ & $_{[\%]}$ & $_{[\%]}$ & $_{[\%]}$ & $_{[\%]}$ & $_{[\%]}$ & $_{[\%]}$ & $_{[\%]}$ & $_{[\%]}$ & $_{[\%]}$ & $_{[\%]}$ & $_{[\%]}$ & & & & & & $_{[\%]}$\\
\midrule
   0.00 &   0.05 &      11.36 &     35.5 & $     +0.2$ & $     +1.2$ & $     +1.8$ & $     -0.6$ & $     +0.9$ & $    -26.9$ & $      4.7$ & $     -2.2$ &       0.2 &      2.7 &      0.5 &       1.6 &    1.097 &  0.961 &  0.918 &  0.999 &    0.031 &       0.4 \\
   0.05 &   0.10 &      137.6 &      8.2 & $     +1.3$ & $     +1.9$ & $     +1.0$ & $     -2.1$ & $     +0.7$ & $     -8.2$ & $      1.5$ & $     +1.1$ &       0.2 &      2.7 &      0.5 &       0.5 &    1.089 &  0.996 &  0.910 &  1.000 &    0.402 &       7.2 \\
   0.10 &   0.15 &      513.3 &      3.3 & $     +1.7$ & $     +0.8$ & $     +0.6$ & $     -0.3$ & $     +0.6$ & $     -0.6$ & $      1.5$ & $     -1.7$ &       0.2 &      2.7 &      0.5 &       0.3 &    1.077 &  0.988 &  0.908 &  0.995 &    1.398 &      18.7 \\
\bottomrule
\end{tabular}\end{adjustbox}
\caption{ 
Differential 1-jettiness cross sections for $e^-p$ ($\mathcal{L}=351.1\,\pb^{-1}$) scattering in the range  $200 < \Qsq < 280$\,\GeVsq  and  $0.10 < y < 0.20$,  $\d\sigma/\d\tb ~[\pb/\Delta\tb]$. Further details are given in the caption of Tab.~\ref{tab:xs1D.em}.
}
\label{tab:xstaub1b.em351.200_0.10}
\end{table*}

\begin{table*}[t]
\tiny\centering\begin{adjustbox}{width=\textwidth}\begin{tabular}{cc|c|cccccccc ccccc|ccccc c}
\toprule
\multicolumn{21}{c}{Cross sections in $e^-p$ ($\mathcal{L}=351.1\,\pb^{-1}$) for $200 < \Qsq < 280$\,\GeVsq  and  $0.20 < y < 0.40$  --  $\d\sigma/\d\tb ~[\pb/\Delta\tb]$} \\
\midrule\multicolumn{2}{c|}{\tb range} & Results & \multicolumn{13}{c|}{Uncertainties} & \multicolumn{6}{c}{Correction factors}\\
\cmidrule(lr){1-2}\cmidrule(lr){3-3}\cmidrule(lr){4-16}\cmidrule(lr){17-22}
 $_\text{min}$ &  $_\text{max}$ & $\frac{\d\sigma}{\d\tb}$ & $_\text{stat}$ & $_\text{RCES}$ & $_\text{JES}$ & $_\text{HadTh}$ & $_\text{ElEn}$ & $_\text{ElTh}$ & $_\text{Model}$ & $_\text{MCstat}$ & $_\text{Unfold}$ &
 $_\text{ElecID}$ & $_\text{Lumi}$ & $_\text{Uncor}$ & $_\text{QED}$ & $c_\text{QED}$ & $c_\text{NoZ}$ & $c_\text{Born}$ & $c_\text{e+p}$ & $c_\text{Had}$ & $\delta_\text{HAD}$ \\
          &  &  $_{[\text{pb}]}$   & $_{[\%]}$  & $_{[\%]}$ & $_{[\%]}$ & $_{[\%]}$ & $_{[\%]}$ & $_{[\%]}$ & $_{[\%]}$ & $_{[\%]}$ & $_{[\%]}$ & $_{[\%]}$ & $_{[\%]}$ & $_{[\%]}$ & $_{[\%]}$ & & & & & & $_{[\%]}$\\
\midrule
   0.00 &   0.05 &      16.71 &     23.2 & $     +0.3$ & $     +1.7$ & $     +1.8$ & $     -1.5$ & $     +0.3$ & $    -30.4$ & $      3.9$ & $     -9.2$ &       0.2 &      2.7 &      0.5 &       1.3 &    1.044 &  1.002 &  0.940 &  1.023 &    0.027 &       0.3 \\
   0.05 &   0.10 &      122.6 &      8.4 & $     +1.2$ & $     +1.6$ & $     +1.0$ & $     -0.9$ & $     +0.3$ & $     -8.8$ & $      1.5$ & $     +2.4$ &       0.2 &      2.7 &      0.5 &       0.4 &    1.068 &  0.997 &  0.917 &  0.992 &    0.332 &       4.7 \\
   0.10 &   0.15 &      507.5 &      3.1 & $     -0.0$ & $     +1.1$ & $     -0.1$ & $     -0.4$ & $     +0.3$ & $     +2.4$ & $      1.2$ & $     -2.3$ &       0.2 &      2.7 &      0.5 &       0.3 &    1.070 &  0.996 &  0.921 &  0.992 &    1.145 &      12.5 \\
\bottomrule
\end{tabular}\end{adjustbox}
\caption{ 
Differential 1-jettiness cross sections for $e^-p$ ($\mathcal{L}=351.1\,\pb^{-1}$) scattering in the range  $200 < \Qsq < 280$\,\GeVsq  and  $0.20 < y < 0.40$,  $\d\sigma/\d\tb ~[\pb/\Delta\tb]$. Further details are given in the caption of Tab.~\ref{tab:xs1D.em}.
}
\label{tab:xstaub1b.em351.200_0.20}
\end{table*}

\begin{table*}[t]
\tiny\centering\begin{adjustbox}{width=\textwidth}\begin{tabular}{cc|c|cccccccc ccccc|ccccc c}
\toprule
\multicolumn{21}{c}{Cross sections in $e^-p$ ($\mathcal{L}=351.1\,\pb^{-1}$) for $200 < \Qsq < 280$\,\GeVsq  and  $0.40 < y < 0.70$  --  $\d\sigma/\d\tb ~[\pb/\Delta\tb]$} \\
\midrule\multicolumn{2}{c|}{\tb range} & Results & \multicolumn{13}{c|}{Uncertainties} & \multicolumn{6}{c}{Correction factors}\\
\cmidrule(lr){1-2}\cmidrule(lr){3-3}\cmidrule(lr){4-16}\cmidrule(lr){17-22}
 $_\text{min}$ &  $_\text{max}$ & $\frac{\d\sigma}{\d\tb}$ & $_\text{stat}$ & $_\text{RCES}$ & $_\text{JES}$ & $_\text{HadTh}$ & $_\text{ElEn}$ & $_\text{ElTh}$ & $_\text{Model}$ & $_\text{MCstat}$ & $_\text{Unfold}$ &
 $_\text{ElecID}$ & $_\text{Lumi}$ & $_\text{Uncor}$ & $_\text{QED}$ & $c_\text{QED}$ & $c_\text{NoZ}$ & $c_\text{Born}$ & $c_\text{e+p}$ & $c_\text{Had}$ & $\delta_\text{HAD}$ \\
          &  &  $_{[\text{pb}]}$   & $_{[\%]}$  & $_{[\%]}$ & $_{[\%]}$ & $_{[\%]}$ & $_{[\%]}$ & $_{[\%]}$ & $_{[\%]}$ & $_{[\%]}$ & $_{[\%]}$ & $_{[\%]}$ & $_{[\%]}$ & $_{[\%]}$ & $_{[\%]}$ & & & & & & $_{[\%]}$\\
\midrule
   0.00 &   0.05 &      9.507 &     39.2 & $     -0.9$ & $     +2.6$ & $     +0.6$ & $     +1.1$ & $     -0.0$ & $    -24.6$ & $      3.8$ & $     -2.2$ &       0.2 &      2.7 &      0.5 &       1.3 &    1.076 &  1.015 &  0.912 &  0.989 &    0.026 &       0.3 \\
   0.05 &   0.10 &      96.74 &     10.0 & $     -0.7$ & $     +2.4$ & $     +0.6$ & $     +1.3$ & $     +0.2$ & $     -6.6$ & $      1.8$ & $     -2.0$ &       0.2 &      2.7 &      0.5 &       0.4 &    1.069 &  0.999 &  0.920 &  0.994 &    0.287 &       3.3 \\
   0.10 &   0.15 &      319.3 &      4.7 & $     -1.0$ & $     +0.9$ & $     +0.1$ & $     -0.1$ & $     +0.7$ & $     +3.7$ & $      1.3$ & $     +2.2$ &       0.2 &      2.7 &      0.5 &       0.2 &    1.076 &  0.988 &  0.911 &  0.982 &    0.985 &       8.4 \\
\bottomrule
\end{tabular}\end{adjustbox}
\caption{ 
Differential 1-jettiness cross sections for $e^-p$ ($\mathcal{L}=351.1\,\pb^{-1}$) scattering in the range  $200 < \Qsq < 280$\,\GeVsq  and  $0.40 < y < 0.70$,  $\d\sigma/\d\tb ~[\pb/\Delta\tb]$. Further details are given in the caption of Tab.~\ref{tab:xs1D.em}.
}
\label{tab:xstaub1b.em351.200_0.40}
\end{table*}

\begin{table*}[t]
\tiny\centering\begin{adjustbox}{width=\textwidth}\begin{tabular}{cc|c|cccccccc ccccc|ccccc c}
\toprule
\multicolumn{21}{c}{Cross sections in $e^-p$ ($\mathcal{L}=351.1\,\pb^{-1}$) for $280 < \Qsq < 440$\,\GeVsq  and  $0.05 < y < 0.10$  --  $\d\sigma/\d\tb ~[\pb/\Delta\tb]$} \\
\midrule\multicolumn{2}{c|}{\tb range} & Results & \multicolumn{13}{c|}{Uncertainties} & \multicolumn{6}{c}{Correction factors}\\
\cmidrule(lr){1-2}\cmidrule(lr){3-3}\cmidrule(lr){4-16}\cmidrule(lr){17-22}
 $_\text{min}$ &  $_\text{max}$ & $\frac{\d\sigma}{\d\tb}$ & $_\text{stat}$ & $_\text{RCES}$ & $_\text{JES}$ & $_\text{HadTh}$ & $_\text{ElEn}$ & $_\text{ElTh}$ & $_\text{Model}$ & $_\text{MCstat}$ & $_\text{Unfold}$ &
 $_\text{ElecID}$ & $_\text{Lumi}$ & $_\text{Uncor}$ & $_\text{QED}$ & $c_\text{QED}$ & $c_\text{NoZ}$ & $c_\text{Born}$ & $c_\text{e+p}$ & $c_\text{Had}$ & $\delta_\text{HAD}$ \\
          &  &  $_{[\text{pb}]}$   & $_{[\%]}$  & $_{[\%]}$ & $_{[\%]}$ & $_{[\%]}$ & $_{[\%]}$ & $_{[\%]}$ & $_{[\%]}$ & $_{[\%]}$ & $_{[\%]}$ & $_{[\%]}$ & $_{[\%]}$ & $_{[\%]}$ & $_{[\%]}$ & & & & & & $_{[\%]}$\\
\midrule
   0.00 &   0.10 &      104.9 &      3.9 & $     +1.7$ & $     +1.1$ & $     +0.8$ & $     -1.3$ & $     +0.4$ & $     -5.8$ & $      0.5$ & $     +2.6$ &       0.2 &      2.7 &      0.5 &       0.4 &    1.071 &  0.985 &  0.911 &  0.992 &    0.367 &       1.9 \\
   0.10 &   0.15 &        506 &      1.8 & $     +1.5$ & $     +0.3$ & $     +0.1$ & $     -1.4$ & $     +0.6$ & $     +3.3$ & $      0.7$ & $     -1.2$ &       0.2 &      2.7 &      0.5 &       0.4 &    1.066 &  1.002 &  0.914 &  0.997 &    2.570 &      56.1 \\
   0.15 &   0.22 &      386.6 &      2.1 & $     +0.2$ & $     +0.0$ & $     -0.3$ & $     -1.3$ & $     +0.7$ & $     +6.6$ & $      1.0$ & $     -3.4$ &       0.2 &      2.7 &      0.5 &       0.4 &    1.060 &  0.997 &  0.916 &  0.993 &    1.878 &       4.7 \\
\bottomrule
\end{tabular}\end{adjustbox}
\caption{ 
Differential 1-jettiness cross sections for $e^-p$ ($\mathcal{L}=351.1\,\pb^{-1}$) scattering in the range  $280 < \Qsq < 440$\,\GeVsq  and  $0.05 < y < 0.10$,  $\d\sigma/\d\tb ~[\pb/\Delta\tb]$. Further details are given in the caption of Tab.~\ref{tab:xs1D.em}.
}
\label{tab:xstaub1b.em351.280_0.05}
\end{table*}

\begin{table*}[t]
\tiny\centering\begin{adjustbox}{width=\textwidth}\begin{tabular}{cc|c|cccccccc ccccc|ccccc c}
\toprule
\multicolumn{21}{c}{Cross sections in $e^-p$ ($\mathcal{L}=351.1\,\pb^{-1}$) for $280 < \Qsq < 440$\,\GeVsq  and  $0.10 < y < 0.20$  --  $\d\sigma/\d\tb ~[\pb/\Delta\tb]$} \\
\midrule\multicolumn{2}{c|}{\tb range} & Results & \multicolumn{13}{c|}{Uncertainties} & \multicolumn{6}{c}{Correction factors}\\
\cmidrule(lr){1-2}\cmidrule(lr){3-3}\cmidrule(lr){4-16}\cmidrule(lr){17-22}
 $_\text{min}$ &  $_\text{max}$ & $\frac{\d\sigma}{\d\tb}$ & $_\text{stat}$ & $_\text{RCES}$ & $_\text{JES}$ & $_\text{HadTh}$ & $_\text{ElEn}$ & $_\text{ElTh}$ & $_\text{Model}$ & $_\text{MCstat}$ & $_\text{Unfold}$ &
 $_\text{ElecID}$ & $_\text{Lumi}$ & $_\text{Uncor}$ & $_\text{QED}$ & $c_\text{QED}$ & $c_\text{NoZ}$ & $c_\text{Born}$ & $c_\text{e+p}$ & $c_\text{Had}$ & $\delta_\text{HAD}$ \\
          &  &  $_{[\text{pb}]}$   & $_{[\%]}$  & $_{[\%]}$ & $_{[\%]}$ & $_{[\%]}$ & $_{[\%]}$ & $_{[\%]}$ & $_{[\%]}$ & $_{[\%]}$ & $_{[\%]}$ & $_{[\%]}$ & $_{[\%]}$ & $_{[\%]}$ & $_{[\%]}$ & & & & & & $_{[\%]}$\\
\midrule
   0.00 &   0.05 &      11.91 &     35.5 & $     +0.1$ & $     +1.5$ & $     +0.8$ & $     -0.4$ & $     +0.5$ & $    -20.6$ & $      2.1$ & $     -4.8$ &       0.2 &      2.7 &      0.5 &       0.9 &    1.087 &  1.004 &  0.896 &  0.991 &    0.054 &       1.0 \\
   0.05 &   0.10 &      236.8 &      4.9 & $     +1.7$ & $     +1.3$ & $     +0.7$ & $     -1.3$ & $     +0.2$ & $     -3.9$ & $      0.9$ & $     +0.3$ &       0.2 &      2.7 &      0.5 &       0.3 &    1.066 &  0.996 &  0.917 &  0.991 &    0.769 &       0.5 \\
   0.10 &   0.15 &      569.4 &      2.8 & $     +1.4$ & $     +0.3$ & $     +0.2$ & $     -1.6$ & $     +0.6$ & $     +3.8$ & $      1.2$ & $     -0.7$ &       0.2 &      2.7 &      0.5 &       0.3 &    1.066 &  0.990 &  0.915 &  0.996 &    2.037 &      38.7 \\
\bottomrule
\end{tabular}\end{adjustbox}
\caption{ 
Differential 1-jettiness cross sections for $e^-p$ ($\mathcal{L}=351.1\,\pb^{-1}$) scattering in the range  $280 < \Qsq < 440$\,\GeVsq  and  $0.10 < y < 0.20$,  $\d\sigma/\d\tb ~[\pb/\Delta\tb]$. Further details are given in the caption of Tab.~\ref{tab:xs1D.em}.
}
\label{tab:xstaub1b.em351.280_0.10}
\end{table*}

\begin{table*}[t]
\tiny\centering\begin{adjustbox}{width=\textwidth}\begin{tabular}{cc|c|cccccccc ccccc|ccccc c}
\toprule
\multicolumn{21}{c}{Cross sections in $e^-p$ ($\mathcal{L}=351.1\,\pb^{-1}$) for $280 < \Qsq < 440$\,\GeVsq  and  $0.20 < y < 0.40$  --  $\d\sigma/\d\tb ~[\pb/\Delta\tb]$} \\
\midrule\multicolumn{2}{c|}{\tb range} & Results & \multicolumn{13}{c|}{Uncertainties} & \multicolumn{6}{c}{Correction factors}\\
\cmidrule(lr){1-2}\cmidrule(lr){3-3}\cmidrule(lr){4-16}\cmidrule(lr){17-22}
 $_\text{min}$ &  $_\text{max}$ & $\frac{\d\sigma}{\d\tb}$ & $_\text{stat}$ & $_\text{RCES}$ & $_\text{JES}$ & $_\text{HadTh}$ & $_\text{ElEn}$ & $_\text{ElTh}$ & $_\text{Model}$ & $_\text{MCstat}$ & $_\text{Unfold}$ &
 $_\text{ElecID}$ & $_\text{Lumi}$ & $_\text{Uncor}$ & $_\text{QED}$ & $c_\text{QED}$ & $c_\text{NoZ}$ & $c_\text{Born}$ & $c_\text{e+p}$ & $c_\text{Had}$ & $\delta_\text{HAD}$ \\
          &  &  $_{[\text{pb}]}$   & $_{[\%]}$  & $_{[\%]}$ & $_{[\%]}$ & $_{[\%]}$ & $_{[\%]}$ & $_{[\%]}$ & $_{[\%]}$ & $_{[\%]}$ & $_{[\%]}$ & $_{[\%]}$ & $_{[\%]}$ & $_{[\%]}$ & $_{[\%]}$ & & & & & & $_{[\%]}$\\
\midrule
   0.00 &   0.05 &      11.35 &     35.5 & $     -0.4$ & $     +1.5$ & $     +1.4$ & $     -0.0$ & $     +0.4$ & $    -25.2$ & $      1.9$ & $     +0.2$ &       0.2 &      2.7 &      0.5 &       0.8 &    1.065 &  1.006 &  0.941 &  0.998 &    0.048 &       0.8 \\
   0.05 &   0.10 &      247.2 &      4.5 & $     +0.7$ & $     +1.4$ & $     +0.6$ & $     -1.1$ & $     +0.4$ & $     -3.5$ & $      0.9$ & $     -0.1$ &       0.2 &      2.7 &      0.5 &       0.3 &    1.075 &  0.989 &  0.910 &  0.986 &    0.634 &       0.1 \\
   0.10 &   0.15 &      571.6 &      2.7 & $     +0.9$ & $     +0.5$ & $     +0.1$ & $     -1.0$ & $     +0.4$ & $     +5.5$ & $      0.9$ & $     +0.9$ &       0.2 &      2.7 &      0.5 &       0.2 &    1.068 &  0.998 &  0.916 &  0.994 &    1.645 &      23.9 \\
\bottomrule
\end{tabular}\end{adjustbox}
\caption{ 
Differential 1-jettiness cross sections for $e^-p$ ($\mathcal{L}=351.1\,\pb^{-1}$) scattering in the range  $280 < \Qsq < 440$\,\GeVsq  and  $0.20 < y < 0.40$,  $\d\sigma/\d\tb ~[\pb/\Delta\tb]$. Further details are given in the caption of Tab.~\ref{tab:xs1D.em}.
}
\label{tab:xstaub1b.em351.280_0.20}
\end{table*}

\begin{table*}[t]
\tiny\centering\begin{adjustbox}{width=\textwidth}\begin{tabular}{cc|c|cccccccc ccccc|ccccc c}
\toprule
\multicolumn{21}{c}{Cross sections in $e^-p$ ($\mathcal{L}=351.1\,\pb^{-1}$) for $280 < \Qsq < 440$\,\GeVsq  and  $0.40 < y < 0.70$  --  $\d\sigma/\d\tb ~[\pb/\Delta\tb]$} \\
\midrule\multicolumn{2}{c|}{\tb range} & Results & \multicolumn{13}{c|}{Uncertainties} & \multicolumn{6}{c}{Correction factors}\\
\cmidrule(lr){1-2}\cmidrule(lr){3-3}\cmidrule(lr){4-16}\cmidrule(lr){17-22}
 $_\text{min}$ &  $_\text{max}$ & $\frac{\d\sigma}{\d\tb}$ & $_\text{stat}$ & $_\text{RCES}$ & $_\text{JES}$ & $_\text{HadTh}$ & $_\text{ElEn}$ & $_\text{ElTh}$ & $_\text{Model}$ & $_\text{MCstat}$ & $_\text{Unfold}$ &
 $_\text{ElecID}$ & $_\text{Lumi}$ & $_\text{Uncor}$ & $_\text{QED}$ & $c_\text{QED}$ & $c_\text{NoZ}$ & $c_\text{Born}$ & $c_\text{e+p}$ & $c_\text{Had}$ & $\delta_\text{HAD}$ \\
          &  &  $_{[\text{pb}]}$   & $_{[\%]}$  & $_{[\%]}$ & $_{[\%]}$ & $_{[\%]}$ & $_{[\%]}$ & $_{[\%]}$ & $_{[\%]}$ & $_{[\%]}$ & $_{[\%]}$ & $_{[\%]}$ & $_{[\%]}$ & $_{[\%]}$ & $_{[\%]}$ & & & & & & $_{[\%]}$\\
\midrule
   0.00 &   0.05 &         17 &     23.3 & $     -0.3$ & $     +1.8$ & $     +0.4$ & $     -2.2$ & $     +0.5$ & $    -26.2$ & $      2.2$ & $     -9.6$ &       0.2 &      2.7 &      0.5 &       0.8 &    1.081 &  0.976 &  0.902 &  0.994 &    0.044 &       0.8 \\
   0.05 &   0.10 &        154 &      6.5 & $     -0.2$ & $     +1.9$ & $     +0.3$ & $     -0.4$ & $     +0.6$ & $     -0.1$ & $      1.1$ & $     +0.8$ &       0.2 &      2.7 &      0.5 &       0.3 &    1.073 &  0.992 &  0.914 &  0.992 &    0.543 &       0.3 \\
   0.10 &   0.15 &      448.9 &      3.3 & $     -0.9$ & $     +0.6$ & $     +0.2$ & $     +0.3$ & $     +0.3$ & $     +3.0$ & $      1.0$ & $     -0.8$ &       0.2 &      2.7 &      0.5 &       0.2 &    1.076 &  0.985 &  0.908 &  0.984 &    1.398 &      13.1 \\
\bottomrule
\end{tabular}\end{adjustbox}
\caption{ 
Differential 1-jettiness cross sections for $e^-p$ ($\mathcal{L}=351.1\,\pb^{-1}$) scattering in the range  $280 < \Qsq < 440$\,\GeVsq  and  $0.40 < y < 0.70$,  $\d\sigma/\d\tb ~[\pb/\Delta\tb]$. Further details are given in the caption of Tab.~\ref{tab:xs1D.em}.
}
\label{tab:xstaub1b.em351.280_0.40}
\end{table*}

\begin{table*}[t]
\tiny\centering\begin{adjustbox}{width=\textwidth}\begin{tabular}{cc|c|cccccccc ccccc|ccccc c}
\toprule
\multicolumn{21}{c}{Cross sections in $e^-p$ ($\mathcal{L}=351.1\,\pb^{-1}$) for $440 < \Qsq < 700$\,\GeVsq  and  $0.05 < y < 0.10$  --  $\d\sigma/\d\tb ~[\pb/\Delta\tb]$} \\
\midrule\multicolumn{2}{c|}{\tb range} & Results & \multicolumn{13}{c|}{Uncertainties} & \multicolumn{6}{c}{Correction factors}\\
\cmidrule(lr){1-2}\cmidrule(lr){3-3}\cmidrule(lr){4-16}\cmidrule(lr){17-22}
 $_\text{min}$ &  $_\text{max}$ & $\frac{\d\sigma}{\d\tb}$ & $_\text{stat}$ & $_\text{RCES}$ & $_\text{JES}$ & $_\text{HadTh}$ & $_\text{ElEn}$ & $_\text{ElTh}$ & $_\text{Model}$ & $_\text{MCstat}$ & $_\text{Unfold}$ &
 $_\text{ElecID}$ & $_\text{Lumi}$ & $_\text{Uncor}$ & $_\text{QED}$ & $c_\text{QED}$ & $c_\text{NoZ}$ & $c_\text{Born}$ & $c_\text{e+p}$ & $c_\text{Had}$ & $\delta_\text{HAD}$ \\
          &  &  $_{[\text{pb}]}$   & $_{[\%]}$  & $_{[\%]}$ & $_{[\%]}$ & $_{[\%]}$ & $_{[\%]}$ & $_{[\%]}$ & $_{[\%]}$ & $_{[\%]}$ & $_{[\%]}$ & $_{[\%]}$ & $_{[\%]}$ & $_{[\%]}$ & $_{[\%]}$ & & & & & & $_{[\%]}$\\
\midrule
   0.00 &   0.10 &      128.7 &      3.1 & $     +1.0$ & $     +0.8$ & $     +0.0$ & $     -1.4$ & $     +0.2$ & $     -0.5$ & $      0.4$ & $     +2.0$ &       0.2 &      2.7 &      0.5 &       0.3 &    1.073 &  0.993 &  0.911 &  0.998 &    0.584 &       0.1 \\
   0.10 &   0.15 &      317.4 &      2.6 & $     +1.2$ & $     -0.0$ & $     +0.2$ & $     -1.4$ & $     +0.7$ & $     +2.7$ & $      1.0$ & $     -3.2$ &       0.2 &      2.7 &      0.5 &       0.4 &    1.067 &  0.983 &  0.905 &  0.979 &    2.321 &      16.2 \\
   0.15 &   0.22 &      204.5 &      3.5 & $     -0.4$ & $     -0.1$ & $     -1.5$ & $     -1.4$ & $     +0.0$ & $     +4.3$ & $      1.8$ & $     -2.7$ &       0.2 &      2.7 &      0.5 &       0.4 &    1.061 &  0.993 &  0.910 &  0.987 &    1.556 &       5.6 \\
\bottomrule
\end{tabular}\end{adjustbox}
\caption{ 
Differential 1-jettiness cross sections for $e^-p$ ($\mathcal{L}=351.1\,\pb^{-1}$) scattering in the range  $440 < \Qsq < 700$\,\GeVsq  and  $0.05 < y < 0.10$,  $\d\sigma/\d\tb ~[\pb/\Delta\tb]$. Further details are given in the caption of Tab.~\ref{tab:xs1D.em}.
}
\label{tab:xstaub1b.em351.440_0.05}
\end{table*}

\begin{table*}[t]
\tiny\centering\begin{adjustbox}{width=\textwidth}\begin{tabular}{cc|c|cccccccc ccccc|ccccc c}
\toprule
\multicolumn{21}{c}{Cross sections in $e^-p$ ($\mathcal{L}=351.1\,\pb^{-1}$) for $440 < \Qsq < 700$\,\GeVsq  and  $0.10 < y < 0.20$  --  $\d\sigma/\d\tb ~[\pb/\Delta\tb]$} \\
\midrule\multicolumn{2}{c|}{\tb range} & Results & \multicolumn{13}{c|}{Uncertainties} & \multicolumn{6}{c}{Correction factors}\\
\cmidrule(lr){1-2}\cmidrule(lr){3-3}\cmidrule(lr){4-16}\cmidrule(lr){17-22}
 $_\text{min}$ &  $_\text{max}$ & $\frac{\d\sigma}{\d\tb}$ & $_\text{stat}$ & $_\text{RCES}$ & $_\text{JES}$ & $_\text{HadTh}$ & $_\text{ElEn}$ & $_\text{ElTh}$ & $_\text{Model}$ & $_\text{MCstat}$ & $_\text{Unfold}$ &
 $_\text{ElecID}$ & $_\text{Lumi}$ & $_\text{Uncor}$ & $_\text{QED}$ & $c_\text{QED}$ & $c_\text{NoZ}$ & $c_\text{Born}$ & $c_\text{e+p}$ & $c_\text{Had}$ & $\delta_\text{HAD}$ \\
          &  &  $_{[\text{pb}]}$   & $_{[\%]}$  & $_{[\%]}$ & $_{[\%]}$ & $_{[\%]}$ & $_{[\%]}$ & $_{[\%]}$ & $_{[\%]}$ & $_{[\%]}$ & $_{[\%]}$ & $_{[\%]}$ & $_{[\%]}$ & $_{[\%]}$ & $_{[\%]}$ & & & & & & $_{[\%]}$\\
\midrule
   0.00 &   0.05 &      11.17 &     39.1 & $     +0.7$ & $     +1.3$ & $     +1.2$ & $     -0.6$ & $     +0.1$ & $    -15.9$ & $      1.3$ & $     -2.8$ &       0.2 &      2.7 &      0.5 &       0.6 &    1.087 &  0.989 &  0.907 &  0.982 &    0.107 &       1.8 \\
   0.05 &   0.10 &      268.4 &      4.3 & $     +1.5$ & $     +0.8$ & $     +0.3$ & $     -1.6$ & $     +0.4$ & $     -0.4$ & $      0.7$ & $     +0.2$ &       0.2 &      2.7 &      0.5 &       0.3 &    1.083 &  0.986 &  0.906 &  0.987 &    1.679 &      38.7 \\
   0.10 &   0.15 &      389.6 &      3.7 & $     +0.7$ & $     -0.0$ & $     -0.3$ & $     -1.2$ & $     +0.2$ & $     +7.6$ & $      1.6$ & $     +0.7$ &       0.2 &      2.7 &      0.5 &       0.3 &    1.063 &  0.994 &  0.913 &  0.991 &    1.918 &       8.2 \\
\bottomrule
\end{tabular}\end{adjustbox}
\caption{ 
Differential 1-jettiness cross sections for $e^-p$ ($\mathcal{L}=351.1\,\pb^{-1}$) scattering in the range  $440 < \Qsq < 700$\,\GeVsq  and  $0.10 < y < 0.20$,  $\d\sigma/\d\tb ~[\pb/\Delta\tb]$. Further details are given in the caption of Tab.~\ref{tab:xs1D.em}.
}
\label{tab:xstaub1b.em351.440_0.10}
\end{table*}

\begin{table*}[t]
\tiny\centering\begin{adjustbox}{width=\textwidth}\begin{tabular}{cc|c|cccccccc ccccc|ccccc c}
\toprule
\multicolumn{21}{c}{Cross sections in $e^-p$ ($\mathcal{L}=351.1\,\pb^{-1}$) for $440 < \Qsq < 700$\,\GeVsq  and  $0.20 < y < 0.40$  --  $\d\sigma/\d\tb ~[\pb/\Delta\tb]$} \\
\midrule\multicolumn{2}{c|}{\tb range} & Results & \multicolumn{13}{c|}{Uncertainties} & \multicolumn{6}{c}{Correction factors}\\
\cmidrule(lr){1-2}\cmidrule(lr){3-3}\cmidrule(lr){4-16}\cmidrule(lr){17-22}
 $_\text{min}$ &  $_\text{max}$ & $\frac{\d\sigma}{\d\tb}$ & $_\text{stat}$ & $_\text{RCES}$ & $_\text{JES}$ & $_\text{HadTh}$ & $_\text{ElEn}$ & $_\text{ElTh}$ & $_\text{Model}$ & $_\text{MCstat}$ & $_\text{Unfold}$ &
 $_\text{ElecID}$ & $_\text{Lumi}$ & $_\text{Uncor}$ & $_\text{QED}$ & $c_\text{QED}$ & $c_\text{NoZ}$ & $c_\text{Born}$ & $c_\text{e+p}$ & $c_\text{Had}$ & $\delta_\text{HAD}$ \\
          &  &  $_{[\text{pb}]}$   & $_{[\%]}$  & $_{[\%]}$ & $_{[\%]}$ & $_{[\%]}$ & $_{[\%]}$ & $_{[\%]}$ & $_{[\%]}$ & $_{[\%]}$ & $_{[\%]}$ & $_{[\%]}$ & $_{[\%]}$ & $_{[\%]}$ & $_{[\%]}$ & & & & & & $_{[\%]}$\\
\midrule
   0.00 &   0.05 &      18.95 &     24.2 & $     +0.2$ & $     +1.7$ & $     +1.1$ & $     -1.3$ & $     +0.3$ & $    -17.9$ & $      1.2$ & $     -2.2$ &       0.2 &      2.7 &      0.5 &       0.5 &    1.086 &  0.980 &  0.913 &  0.960 &    0.097 &       1.6 \\
   0.05 &   0.10 &      279.5 &      4.1 & $     +1.2$ & $     +0.9$ & $     +0.5$ & $     -1.4$ & $     +0.6$ & $     +1.1$ & $      0.7$ & $     +1.2$ &       0.2 &      2.7 &      0.5 &       0.2 &    1.077 &  0.993 &  0.912 &  0.989 &    1.328 &      26.0 \\
   0.10 &   0.15 &      440.9 &      3.2 & $     +0.2$ & $     +0.2$ & $     -0.3$ & $     -1.2$ & $     +0.4$ & $     +6.7$ & $      1.1$ & $     -1.6$ &       0.2 &      2.7 &      0.5 &       0.2 &    1.073 &  0.987 &  0.908 &  0.984 &    1.631 &       0.2 \\
\bottomrule
\end{tabular}\end{adjustbox}
\caption{ 
Differential 1-jettiness cross sections for $e^-p$ ($\mathcal{L}=351.1\,\pb^{-1}$) scattering in the range  $440 < \Qsq < 700$\,\GeVsq  and  $0.20 < y < 0.40$,  $\d\sigma/\d\tb ~[\pb/\Delta\tb]$. Further details are given in the caption of Tab.~\ref{tab:xs1D.em}.
}
\label{tab:xstaub1b.em351.440_0.20}
\end{table*}

\begin{table*}[t]
\tiny\centering\begin{adjustbox}{width=\textwidth}\begin{tabular}{cc|c|cccccccc ccccc|ccccc c}
\toprule
\multicolumn{21}{c}{Cross sections in $e^-p$ ($\mathcal{L}=351.1\,\pb^{-1}$) for $440 < \Qsq < 700$\,\GeVsq  and  $0.40 < y < 0.70$  --  $\d\sigma/\d\tb ~[\pb/\Delta\tb]$} \\
\midrule\multicolumn{2}{c|}{\tb range} & Results & \multicolumn{13}{c|}{Uncertainties} & \multicolumn{6}{c}{Correction factors}\\
\cmidrule(lr){1-2}\cmidrule(lr){3-3}\cmidrule(lr){4-16}\cmidrule(lr){17-22}
 $_\text{min}$ &  $_\text{max}$ & $\frac{\d\sigma}{\d\tb}$ & $_\text{stat}$ & $_\text{RCES}$ & $_\text{JES}$ & $_\text{HadTh}$ & $_\text{ElEn}$ & $_\text{ElTh}$ & $_\text{Model}$ & $_\text{MCstat}$ & $_\text{Unfold}$ &
 $_\text{ElecID}$ & $_\text{Lumi}$ & $_\text{Uncor}$ & $_\text{QED}$ & $c_\text{QED}$ & $c_\text{NoZ}$ & $c_\text{Born}$ & $c_\text{e+p}$ & $c_\text{Had}$ & $\delta_\text{HAD}$ \\
          &  &  $_{[\text{pb}]}$   & $_{[\%]}$  & $_{[\%]}$ & $_{[\%]}$ & $_{[\%]}$ & $_{[\%]}$ & $_{[\%]}$ & $_{[\%]}$ & $_{[\%]}$ & $_{[\%]}$ & $_{[\%]}$ & $_{[\%]}$ & $_{[\%]}$ & $_{[\%]}$ & & & & & & $_{[\%]}$\\
\midrule
   0.00 &   0.05 &      2.629 &    131.4 & $     +0.0$ & $     +2.0$ & $     +0.8$ & $     +0.1$ & $     +0.8$ & $    -21.0$ & $      2.3$ & $     -3.0$ &       0.2 &      2.7 &      0.5 &       0.7 &    1.085 &  0.997 &  0.916 &  0.989 &    0.088 &       1.4 \\
   0.05 &   0.10 &      205.5 &      4.6 & $     +0.5$ & $     +1.4$ & $     +0.2$ & $     -0.5$ & $     +0.3$ & $     +2.2$ & $      0.9$ & $     +1.4$ &       0.2 &      2.7 &      0.5 &       0.2 &    1.084 &  0.984 &  0.910 &  0.973 &    1.098 &      17.7 \\
   0.10 &   0.15 &      344.8 &      3.7 & $     -0.1$ & $     +0.2$ & $     +0.1$ & $     +0.1$ & $     +0.5$ & $     -0.7$ & $      1.1$ & $     -2.5$ &       0.2 &      2.7 &      0.5 &       0.2 &    1.075 &  0.987 &  0.909 &  0.987 &    1.431 &       6.1 \\
\bottomrule
\end{tabular}\end{adjustbox}
\caption{ 
Differential 1-jettiness cross sections for $e^-p$ ($\mathcal{L}=351.1\,\pb^{-1}$) scattering in the range  $440 < \Qsq < 700$\,\GeVsq  and  $0.40 < y < 0.70$,  $\d\sigma/\d\tb ~[\pb/\Delta\tb]$. Further details are given in the caption of Tab.~\ref{tab:xs1D.em}.
}
\label{tab:xstaub1b.em351.440_0.40}
\end{table*}

\begin{table*}[t]
\tiny\centering\begin{adjustbox}{width=\textwidth}\begin{tabular}{cc|c|cccccccc ccccc|ccccc c}
\toprule
\multicolumn{21}{c}{Cross sections in $e^-p$ ($\mathcal{L}=351.1\,\pb^{-1}$) for $700 < \Qsq < 1100$\,\GeVsq  and  $0.05 < y < 0.10$  --  $\d\sigma/\d\tb ~[\pb/\Delta\tb]$} \\
\midrule\multicolumn{2}{c|}{\tb range} & Results & \multicolumn{13}{c|}{Uncertainties} & \multicolumn{6}{c}{Correction factors}\\
\cmidrule(lr){1-2}\cmidrule(lr){3-3}\cmidrule(lr){4-16}\cmidrule(lr){17-22}
 $_\text{min}$ &  $_\text{max}$ & $\frac{\d\sigma}{\d\tb}$ & $_\text{stat}$ & $_\text{RCES}$ & $_\text{JES}$ & $_\text{HadTh}$ & $_\text{ElEn}$ & $_\text{ElTh}$ & $_\text{Model}$ & $_\text{MCstat}$ & $_\text{Unfold}$ &
 $_\text{ElecID}$ & $_\text{Lumi}$ & $_\text{Uncor}$ & $_\text{QED}$ & $c_\text{QED}$ & $c_\text{NoZ}$ & $c_\text{Born}$ & $c_\text{e+p}$ & $c_\text{Had}$ & $\delta_\text{HAD}$ \\
          &  &  $_{[\text{pb}]}$   & $_{[\%]}$  & $_{[\%]}$ & $_{[\%]}$ & $_{[\%]}$ & $_{[\%]}$ & $_{[\%]}$ & $_{[\%]}$ & $_{[\%]}$ & $_{[\%]}$ & $_{[\%]}$ & $_{[\%]}$ & $_{[\%]}$ & $_{[\%]}$ & & & & & & $_{[\%]}$\\
\midrule
   0.00 &   0.10 &      125.3 &      3.2 & $     +1.1$ & $     +0.6$ & $     +0.0$ & $     -1.3$ & $     +0.2$ & $     +1.1$ & $      0.4$ & $     -0.1$ &       0.2 &      2.7 &      0.5 &       0.3 &    1.096 &  0.974 &  0.899 &  0.978 &    0.765 &       0.3 \\
   0.10 &   0.15 &      178.5 &      4.1 & $     +0.7$ & $     -0.5$ & $     -0.8$ & $     -1.0$ & $     +1.1$ & $     +1.2$ & $      1.9$ & $     -1.9$ &       0.2 &      2.7 &      0.5 &       0.4 &    1.050 &  0.996 &  0.909 &  1.001 &    1.821 &       5.8 \\
   0.15 &   0.22 &         93 &      6.7 & $     +0.7$ & $     +0.6$ & $     -0.0$ & $     -2.2$ & $     -0.6$ & $     +5.7$ & $      3.3$ & $     +1.9$ &       0.2 &      2.7 &      0.5 &       0.5 &    1.061 &  0.984 &  0.898 &  0.983 &    1.462 &       2.5 \\
\bottomrule
\end{tabular}\end{adjustbox}
\caption{ 
Differential 1-jettiness cross sections for $e^-p$ ($\mathcal{L}=351.1\,\pb^{-1}$) scattering in the range  $700 < \Qsq < 1100$\,\GeVsq  and  $0.05 < y < 0.10$,  $\d\sigma/\d\tb ~[\pb/\Delta\tb]$. Further details are given in the caption of Tab.~\ref{tab:xs1D.em}.
}
\label{tab:xstaub1b.em351.700_0.05}
\end{table*}

\begin{table*}[t]
\tiny\centering\begin{adjustbox}{width=\textwidth}\begin{tabular}{cc|c|cccccccc ccccc|ccccc c}
\toprule
\multicolumn{21}{c}{Cross sections in $e^-p$ ($\mathcal{L}=351.1\,\pb^{-1}$) for $700 < \Qsq < 1100$\,\GeVsq  and  $0.10 < y < 0.20$  --  $\d\sigma/\d\tb ~[\pb/\Delta\tb]$} \\
\midrule\multicolumn{2}{c|}{\tb range} & Results & \multicolumn{13}{c|}{Uncertainties} & \multicolumn{6}{c}{Correction factors}\\
\cmidrule(lr){1-2}\cmidrule(lr){3-3}\cmidrule(lr){4-16}\cmidrule(lr){17-22}
 $_\text{min}$ &  $_\text{max}$ & $\frac{\d\sigma}{\d\tb}$ & $_\text{stat}$ & $_\text{RCES}$ & $_\text{JES}$ & $_\text{HadTh}$ & $_\text{ElEn}$ & $_\text{ElTh}$ & $_\text{Model}$ & $_\text{MCstat}$ & $_\text{Unfold}$ &
 $_\text{ElecID}$ & $_\text{Lumi}$ & $_\text{Uncor}$ & $_\text{QED}$ & $c_\text{QED}$ & $c_\text{NoZ}$ & $c_\text{Born}$ & $c_\text{e+p}$ & $c_\text{Had}$ & $\delta_\text{HAD}$ \\
          &  &  $_{[\text{pb}]}$   & $_{[\%]}$  & $_{[\%]}$ & $_{[\%]}$ & $_{[\%]}$ & $_{[\%]}$ & $_{[\%]}$ & $_{[\%]}$ & $_{[\%]}$ & $_{[\%]}$ & $_{[\%]}$ & $_{[\%]}$ & $_{[\%]}$ & $_{[\%]}$ & & & & & & $_{[\%]}$\\
\midrule
   0.00 &   0.05 &       15.7 &     30.6 & $     -0.2$ & $     +1.3$ & $     +0.7$ & $     -0.4$ & $     +0.0$ & $    -13.2$ & $      0.9$ & $     -3.1$ &       0.2 &      2.7 &      0.5 &       0.4 &    1.095 &  0.970 &  0.911 &  0.975 &    0.214 &       1.8 \\
   0.05 &   0.10 &        222 &      4.9 & $     +1.5$ & $     +0.2$ & $     -0.1$ & $     -1.0$ & $     +0.4$ & $     +2.6$ & $      0.8$ & $     +1.1$ &       0.2 &      2.7 &      0.5 &       0.2 &    1.077 &  0.981 &  0.905 &  0.979 &    2.237 &      43.4 \\
   0.10 &   0.15 &      211.6 &      6.1 & $     +2.2$ & $     +0.2$ & $     -0.2$ & $     -1.6$ & $     +0.6$ & $     +5.9$ & $      2.3$ & $     -0.3$ &       0.2 &      2.7 &      0.5 &       0.3 &    1.067 &  0.975 &  0.904 &  0.972 &    1.622 &      11.3 \\
\bottomrule
\end{tabular}\end{adjustbox}
\caption{ 
Differential 1-jettiness cross sections for $e^-p$ ($\mathcal{L}=351.1\,\pb^{-1}$) scattering in the range  $700 < \Qsq < 1100$\,\GeVsq  and  $0.10 < y < 0.20$,  $\d\sigma/\d\tb ~[\pb/\Delta\tb]$. Further details are given in the caption of Tab.~\ref{tab:xs1D.em}.
}
\label{tab:xstaub1b.em351.700_0.10}
\end{table*}

\begin{table*}[t]
\tiny\centering\begin{adjustbox}{width=\textwidth}\begin{tabular}{cc|c|cccccccc ccccc|ccccc c}
\toprule
\multicolumn{21}{c}{Cross sections in $e^-p$ ($\mathcal{L}=351.1\,\pb^{-1}$) for $700 < \Qsq < 1100$\,\GeVsq  and  $0.20 < y < 0.40$  --  $\d\sigma/\d\tb ~[\pb/\Delta\tb]$} \\
\midrule\multicolumn{2}{c|}{\tb range} & Results & \multicolumn{13}{c|}{Uncertainties} & \multicolumn{6}{c}{Correction factors}\\
\cmidrule(lr){1-2}\cmidrule(lr){3-3}\cmidrule(lr){4-16}\cmidrule(lr){17-22}
 $_\text{min}$ &  $_\text{max}$ & $\frac{\d\sigma}{\d\tb}$ & $_\text{stat}$ & $_\text{RCES}$ & $_\text{JES}$ & $_\text{HadTh}$ & $_\text{ElEn}$ & $_\text{ElTh}$ & $_\text{Model}$ & $_\text{MCstat}$ & $_\text{Unfold}$ &
 $_\text{ElecID}$ & $_\text{Lumi}$ & $_\text{Uncor}$ & $_\text{QED}$ & $c_\text{QED}$ & $c_\text{NoZ}$ & $c_\text{Born}$ & $c_\text{e+p}$ & $c_\text{Had}$ & $\delta_\text{HAD}$ \\
          &  &  $_{[\text{pb}]}$   & $_{[\%]}$  & $_{[\%]}$ & $_{[\%]}$ & $_{[\%]}$ & $_{[\%]}$ & $_{[\%]}$ & $_{[\%]}$ & $_{[\%]}$ & $_{[\%]}$ & $_{[\%]}$ & $_{[\%]}$ & $_{[\%]}$ & $_{[\%]}$ & & & & & & $_{[\%]}$\\
\midrule
   0.00 &   0.05 &      15.95 &     27.4 & $     +0.3$ & $     +1.7$ & $     +1.0$ & $     -0.6$ & $     -0.0$ & $    -14.7$ & $      0.8$ & $     -0.1$ &       0.2 &      2.7 &      0.5 &       0.4 &    1.103 &  0.973 &  0.909 &  0.973 &    0.195 &       1.5 \\
   0.05 &   0.10 &      269.2 &      3.7 & $     +1.0$ & $     +0.4$ & $     -0.0$ & $     -1.3$ & $     +0.2$ & $     +4.0$ & $      0.7$ & $     +0.6$ &       0.2 &      2.7 &      0.5 &       0.2 &    1.090 &  0.975 &  0.904 &  0.964 &    1.803 &      26.2 \\
   0.10 &   0.15 &      219.1 &      5.2 & $     +0.5$ & $     -0.0$ & $     -0.3$ & $     -1.1$ & $     +0.3$ & $     +6.0$ & $      1.4$ & $     -2.4$ &       0.2 &      2.7 &      0.5 &       0.3 &    1.077 &  0.980 &  0.910 &  0.973 &    1.456 &      12.3 \\
\bottomrule
\end{tabular}\end{adjustbox}
\caption{ 
Differential 1-jettiness cross sections for $e^-p$ ($\mathcal{L}=351.1\,\pb^{-1}$) scattering in the range  $700 < \Qsq < 1100$\,\GeVsq  and  $0.20 < y < 0.40$,  $\d\sigma/\d\tb ~[\pb/\Delta\tb]$. Further details are given in the caption of Tab.~\ref{tab:xs1D.em}.
}
\label{tab:xstaub1b.em351.700_0.20}
\end{table*}

\begin{table*}[t]
\tiny\centering\begin{adjustbox}{width=\textwidth}\begin{tabular}{cc|c|cccccccc ccccc|ccccc c}
\toprule
\multicolumn{21}{c}{Cross sections in $e^-p$ ($\mathcal{L}=351.1\,\pb^{-1}$) for $700 < \Qsq < 1100$\,\GeVsq  and  $0.40 < y < 0.70$  --  $\d\sigma/\d\tb ~[\pb/\Delta\tb]$} \\
\midrule\multicolumn{2}{c|}{\tb range} & Results & \multicolumn{13}{c|}{Uncertainties} & \multicolumn{6}{c}{Correction factors}\\
\cmidrule(lr){1-2}\cmidrule(lr){3-3}\cmidrule(lr){4-16}\cmidrule(lr){17-22}
 $_\text{min}$ &  $_\text{max}$ & $\frac{\d\sigma}{\d\tb}$ & $_\text{stat}$ & $_\text{RCES}$ & $_\text{JES}$ & $_\text{HadTh}$ & $_\text{ElEn}$ & $_\text{ElTh}$ & $_\text{Model}$ & $_\text{MCstat}$ & $_\text{Unfold}$ &
 $_\text{ElecID}$ & $_\text{Lumi}$ & $_\text{Uncor}$ & $_\text{QED}$ & $c_\text{QED}$ & $c_\text{NoZ}$ & $c_\text{Born}$ & $c_\text{e+p}$ & $c_\text{Had}$ & $\delta_\text{HAD}$ \\
          &  &  $_{[\text{pb}]}$   & $_{[\%]}$  & $_{[\%]}$ & $_{[\%]}$ & $_{[\%]}$ & $_{[\%]}$ & $_{[\%]}$ & $_{[\%]}$ & $_{[\%]}$ & $_{[\%]}$ & $_{[\%]}$ & $_{[\%]}$ & $_{[\%]}$ & $_{[\%]}$ & & & & & & $_{[\%]}$\\
\midrule
   0.00 &   0.05 &      13.16 &     28.1 & $     -0.5$ & $     +1.8$ & $     +0.8$ & $     -0.3$ & $     +0.1$ & $    -13.5$ & $      1.0$ & $     +0.8$ &       0.2 &      2.7 &      0.5 &       0.6 &    1.123 &  0.977 &  0.905 &  0.940 &    0.180 &       1.3 \\
   0.05 &   0.10 &      182.1 &      4.5 & $     +0.5$ & $     +0.6$ & $     +0.1$ & $     -0.5$ & $     +0.3$ & $     +2.5$ & $      0.7$ & $     +0.2$ &       0.2 &      2.7 &      0.5 &       0.3 &    1.095 &  0.970 &  0.906 &  0.958 &    1.503 &      16.0 \\
   0.10 &   0.15 &      175.6 &      5.6 & $     +0.5$ & $     -0.0$ & $     -0.1$ & $     -1.3$ & $     +0.1$ & $     +5.7$ & $      1.1$ & $     -0.7$ &       0.2 &      2.7 &      0.5 &       0.3 &    1.073 &  0.979 &  0.905 &  0.964 &    1.339 &      14.4 \\
\bottomrule
\end{tabular}\end{adjustbox}
\caption{ 
Differential 1-jettiness cross sections for $e^-p$ ($\mathcal{L}=351.1\,\pb^{-1}$) scattering in the range  $700 < \Qsq < 1100$\,\GeVsq  and  $0.40 < y < 0.70$,  $\d\sigma/\d\tb ~[\pb/\Delta\tb]$. Further details are given in the caption of Tab.~\ref{tab:xs1D.em}.
}
\label{tab:xstaub1b.em351.700_0.40}
\end{table*}

\begin{table*}[t]
\tiny\centering\begin{adjustbox}{width=\textwidth}\begin{tabular}{cc|c|cccccccc ccccc|ccccc c}
\toprule
\multicolumn{21}{c}{Cross sections in $e^-p$ ($\mathcal{L}=351.1\,\pb^{-1}$) for $700 < \Qsq < 1100$\,\GeVsq  and  $0.70 < y < 0.94$  --  $\d\sigma/\d\tb ~[\pb/\Delta\tb]$} \\
\midrule\multicolumn{2}{c|}{\tb range} & Results & \multicolumn{13}{c|}{Uncertainties} & \multicolumn{6}{c}{Correction factors}\\
\cmidrule(lr){1-2}\cmidrule(lr){3-3}\cmidrule(lr){4-16}\cmidrule(lr){17-22}
 $_\text{min}$ &  $_\text{max}$ & $\frac{\d\sigma}{\d\tb}$ & $_\text{stat}$ & $_\text{RCES}$ & $_\text{JES}$ & $_\text{HadTh}$ & $_\text{ElEn}$ & $_\text{ElTh}$ & $_\text{Model}$ & $_\text{MCstat}$ & $_\text{Unfold}$ &
 $_\text{ElecID}$ & $_\text{Lumi}$ & $_\text{Uncor}$ & $_\text{QED}$ & $c_\text{QED}$ & $c_\text{NoZ}$ & $c_\text{Born}$ & $c_\text{e+p}$ & $c_\text{Had}$ & $\delta_\text{HAD}$ \\
          &  &  $_{[\text{pb}]}$   & $_{[\%]}$  & $_{[\%]}$ & $_{[\%]}$ & $_{[\%]}$ & $_{[\%]}$ & $_{[\%]}$ & $_{[\%]}$ & $_{[\%]}$ & $_{[\%]}$ & $_{[\%]}$ & $_{[\%]}$ & $_{[\%]}$ & $_{[\%]}$ & & & & & & $_{[\%]}$\\
\midrule
   0.00 &   0.05 &      1.341 &    319.0 & $     -1.4$ & $     +2.8$ & $     +0.4$ & $     +0.4$ & $     +0.2$ & $    -20.8$ & $      3.1$ & $     +0.2$ &       0.2 &      2.7 &      0.5 &       0.8 &    1.119 &  0.977 &  0.917 &  0.948 &    0.171 &       0.9 \\
   0.05 &   0.10 &      96.52 &     10.3 & $     -1.9$ & $     +1.1$ & $     -0.6$ & $     +0.6$ & $     +0.2$ & $     +3.8$ & $      1.6$ & $     -1.1$ &       0.2 &      2.7 &      0.5 &       0.4 &    1.102 &  0.981 &  0.906 &  0.958 &    1.418 &      18.8 \\
   0.10 &   0.15 &      87.85 &     12.6 & $     -3.0$ & $     -0.5$ & $     +0.1$ & $     +1.3$ & $     -0.6$ & $     +3.9$ & $      2.1$ & $     +2.2$ &       0.2 &      2.7 &      0.5 &       0.4 &    1.091 &  0.966 &  0.898 &  0.966 &    1.307 &      15.9 \\
\bottomrule
\end{tabular}%
\end{adjustbox}
\caption{ 
Differential 1-jettiness cross sections for $e^-p$ ($\mathcal{L}=351.1\,\pb^{-1}$) scattering in the range  $8000 < \Qsq < 20000$\,\GeVsq  and  $0.20 < y < 0.94$,  $\d\sigma/\d\tb ~[\pb/\Delta\tb]$. Further details are given in the caption of Tab.~\ref{tab:xs1D.em}.
}
\label{tab:xstaub1b.em351.8000_0.20}
\end{table*}



\begin{table*}[t]
\tiny\centering\begin{adjustbox}{width=\textwidth}\begin{tabular}{cc|c|cccccccc ccccc|ccccc c}
\toprule
\multicolumn{22}{c}{2D neutral current DIS in $e^-p$  cross sections  $\d^2\sigma/\d\Qsq\d y ~[\pb/\GeVsq]$} \\
\midrule $y$ & \Qsq & Results & \multicolumn{13}{|c|}{Uncertainties} & \multicolumn{6}{c}{Correction factors}\\
\cmidrule(lr){1-2}\cmidrule(lr){3-3}\cmidrule(lr){4-16}\cmidrule(lr){17-22}
 $_\text{min}$ &  $_\text{min}$ & $\frac{\d^2\sigma}{\d\Qsq\d y}$ & $_\text{stat}$ & $_\text{RCES}$ & $_\text{JES}$ & $_\text{HadTh}$ & $_\text{ElEn}$ & $_\text{ElTh}$ & $_\text{Model}$ & $_\text{MCstat}$ & $_\text{Unfold}$ &
 $_\text{ElecID}$ & $_\text{Lumi}$ & $_\text{Uncor}$ & $_\text{QED}$ & $c_\text{QED}$ & $c_\text{NoZ}$ & $c_\text{Born}$ & $c_\text{e+p}$ & $c_\text{Had}$ & $\delta_\text{HAD}$ \\
          & $_{[\text{GeV}]}$  &  $_{[\text{pb}/\GeVsq]}$   & $_{[\%]}$  & $_{[\%]}$ & $_{[\%]}$ & $_{[\%]}$ & $_{[\%]}$ & $_{[\%]}$ & $_{[\%]}$ & $_{[\%]}$ & $_{[\%]}$ & $_{[\%]}$ & $_{[\%]}$ & $_{[\%]}$ & $_{[\%]}$ & & & & & & $_{[\%]}$\\
\midrule
   0.05 &    150 &      174.3 &      2.9 & $     +2.1$ & $     +0.7$ & $     +0.8$ & $     +3.8$ & $     +1.3$ & $     +0.9$ & $      0.5$ & $     -0.0$ &       0.2 &      2.7 &      0.5 &       0.2 &    1.067 &  0.999 &  0.922 &  0.999 &    1.000 &       0.0 \\
   0.05 &    200 &      139.3 &      1.6 & $     +1.6$ & $     +0.3$ & $     +0.3$ & $     -1.6$ & $     +0.7$ & $     -0.3$ & $      0.2$ & $     -0.1$ &       0.2 &      2.7 &      0.5 &       0.2 &    1.071 &  0.992 &  0.911 &  0.995 &    1.000 &       0.0 \\
   0.05 &    280 &      113.6 &      1.7 & $     +1.6$ & $     +0.2$ & $     +0.3$ & $     -1.7$ & $     +0.7$ & $     -0.2$ & $      0.2$ & $     +0.0$ &       0.2 &      2.7 &      0.5 &       0.2 &    1.066 &  0.993 &  0.912 &  0.992 &    1.000 &       0.0 \\
   0.05 &    440 &      64.84 &      2.2 & $     +1.3$ & $     +0.2$ & $     +0.1$ & $     -2.0$ & $     +0.4$ & $     +0.3$ & $      0.2$ & $     -0.1$ &       0.2 &      2.7 &      0.5 &       0.2 &    1.064 &  0.991 &  0.910 &  0.991 &    1.000 &       0.0 \\
   0.05 &    700 &      35.93 &      2.4 & $     +0.7$ & $     +0.1$ & $     -0.3$ & $     -1.3$ & $     +0.3$ & $     -1.1$ & $      0.2$ & $     -0.0$ &       0.2 &      2.7 &      0.5 &       0.3 &    1.069 &  0.981 &  0.902 &  0.983 &    1.000 &       0.0 \\
   0.10 &    150 &      217.3 &      1.5 & $     +1.4$ & $     +0.4$ & $     +0.2$ & $     +1.8$ & $     +0.8$ & $     +0.5$ & $      0.2$ & $     -0.0$ &       0.2 &      2.7 &      0.5 &       0.2 &    1.073 &  0.994 &  0.919 &  0.997 &    1.000 &       0.0 \\
   0.10 &    200 &      175.6 &      1.2 & $     +0.8$ & $     +0.1$ & $     -0.0$ & $     -1.3$ & $     +0.7$ & $     +0.2$ & $      0.1$ & $     -0.0$ &       0.2 &      2.7 &      0.5 &       0.2 &    1.073 &  0.993 &  0.915 &  0.993 &    1.000 &       0.0 \\
   0.10 &    280 &      141.5 &      1.2 & $     +1.0$ & $     +0.1$ & $     +0.0$ & $     -1.1$ & $     +0.6$ & $     +0.1$ & $      0.1$ & $     -0.0$ &       0.2 &      2.7 &      0.5 &       0.1 &    1.067 &  0.991 &  0.911 &  0.992 &    1.000 &       0.0 \\
   0.10 &    440 &      83.26 &      1.6 & $     +1.2$ & $     +0.1$ & $     +0.1$ & $     -1.6$ & $     +0.5$ & $     -0.0$ & $      0.1$ & $     -0.0$ &       0.2 &      2.7 &      0.5 &       0.2 &    1.070 &  0.987 &  0.906 &  0.987 &    1.000 &       0.0 \\
   0.10 &    700 &      41.88 &      2.1 & $     +1.2$ & $     -0.0$ & $     +0.0$ & $     -1.1$ & $     +0.4$ & $     +0.2$ & $      0.1$ & $     -0.0$ &       0.2 &      2.7 &      0.5 &       0.2 &    1.063 &  0.978 &  0.907 &  0.979 &    1.000 &       0.0 \\
   0.10 &   1100 &      21.64 &      2.8 & $     +1.5$ & $     +0.1$ & $     +0.0$ & $     -1.4$ & $     +0.2$ & $     -0.8$ & $      0.1$ & $     -0.1$ &       0.2 &      2.7 &      0.5 &       0.2 &    1.068 &  0.968 &  0.902 &  0.970 &    1.000 &       0.0 \\
   0.20 &    150 &      247.6 &      1.0 & $     +0.8$ & $     +0.3$ & $     +0.0$ & $     -0.3$ & $     +0.7$ & $     +0.9$ & $      0.1$ & $     +0.0$ &       0.2 &      2.7 &      0.5 &       0.1 &    1.072 &  0.996 &  0.918 &  0.996 &    1.000 &       0.0 \\
   0.20 &    200 &      193.9 &      1.1 & $     +0.6$ & $     +0.0$ & $     -0.0$ & $     -1.0$ & $     +0.7$ & $     +1.0$ & $      0.1$ & $     -0.0$ &       0.2 &      2.7 &      0.5 &       0.1 &    1.071 &  0.994 &  0.916 &  0.994 &    1.000 &       0.0 \\
   0.20 &    280 &      164.4 &      1.1 & $     +0.7$ & $     +0.1$ & $     -0.0$ & $     -1.1$ & $     +0.5$ & $     +0.6$ & $      0.1$ & $     -0.0$ &       0.2 &      2.7 &      0.5 &       0.1 &    1.069 &  0.993 &  0.913 &  0.992 &    1.000 &       0.0 \\
   0.20 &    440 &      94.95 &      1.5 & $     +0.9$ & $     +0.0$ & $     -0.1$ & $     -1.3$ & $     +0.4$ & $     +0.3$ & $      0.1$ & $     -0.0$ &       0.2 &      2.7 &      0.5 &       0.1 &    1.069 &  0.987 &  0.908 &  0.986 &    1.000 &       0.0 \\
   0.20 &    700 &      51.29 &      1.8 & $     +0.8$ & $     -0.0$ & $     -0.1$ & $     -0.9$ & $     +0.3$ & $     +0.3$ & $      0.1$ & $     -0.0$ &       0.2 &      2.7 &      0.5 &       0.2 &    1.068 &  0.975 &  0.906 &  0.974 &    1.000 &       0.0 \\
   0.20 &   1100 &      26.03 &      2.5 & $     +1.2$ & $     -0.0$ & $     -0.0$ & $     -1.3$ & $     +0.3$ & $     +0.1$ & $      0.1$ & $     -0.0$ &       0.2 &      2.7 &      0.5 &       0.2 &    1.066 &  0.956 &  0.903 &  0.953 &    1.000 &       0.0 \\
   0.20 &   1700 &       20.2 &      3.0 & $     +2.3$ & $     +0.0$ & $     +0.1$ & $     -1.9$ & $     +0.0$ & $     -1.1$ & $      0.2$ & $     -0.2$ &       0.2 &      2.7 &      0.5 &       0.2 &    1.066 &  0.918 &  0.900 &  0.908 &    1.000 &       0.0 \\
   0.20 &   3500 &      7.902 &      4.4 & $     +3.8$ & $     +0.0$ & $     +0.6$ & $     -2.3$ & $     -0.0$ & $     -0.3$ & $      0.2$ & $     -0.3$ &       1.0 &      2.7 &      0.5 &       0.4 &    1.060 &  0.838 &  0.908 &  0.812 &    1.000 &       0.0 \\
   0.20 &   8000 &      1.266 &     11.4 & $     +4.2$ & $     -0.0$ & $     +1.1$ & $     -3.6$ & $     -0.3$ & $     +0.2$ & $      0.5$ & $     -0.7$ &       1.0 &      2.7 &      0.5 &       0.8 &    1.059 &  0.741 &  0.917 &  0.689 &    1.000 &       0.0 \\
   0.40 &    150 &      196.5 &      1.3 & $     -0.6$ & $     -1.2$ & $     -0.1$ & $     -0.6$ & $     +0.5$ & $     +2.4$ & $      0.1$ & $     -0.0$ &       0.2 &      2.7 &      0.5 &       0.1 &    1.074 &  0.993 &  0.918 &  0.994 &    1.000 &       0.0 \\
   0.40 &    200 &      157.7 &      1.3 & $     -0.4$ & $     -0.7$ & $     -0.1$ & $     -0.5$ & $     +0.4$ & $     +2.2$ & $      0.1$ & $     -0.0$ &       0.2 &      2.7 &      0.5 &       0.1 &    1.072 &  0.992 &  0.914 &  0.993 &    1.000 &       0.0 \\
   0.40 &    280 &      135.7 &      1.4 & $     -0.3$ & $     -0.4$ & $     -0.1$ & $     -0.7$ & $     +0.4$ & $     +1.8$ & $      0.1$ & $     -0.0$ &       0.2 &      2.7 &      0.5 &       0.1 &    1.071 &  0.989 &  0.910 &  0.990 &    1.000 &       0.0 \\
   0.40 &    440 &      75.34 &      1.8 & $     +0.2$ & $     -0.2$ & $     -0.1$ & $     -0.8$ & $     +0.3$ & $     +1.1$ & $      0.1$ & $     -0.0$ &       0.2 &      2.7 &      0.5 &       0.1 &    1.069 &  0.986 &  0.909 &  0.986 &    1.000 &       0.0 \\
   0.40 &    700 &      41.08 &      2.1 & $     +0.3$ & $     -0.1$ & $     -0.1$ & $     -0.7$ & $     +0.3$ & $     +0.6$ & $      0.1$ & $     -0.0$ &       0.2 &      2.7 &      0.5 &       0.2 &    1.068 &  0.975 &  0.905 &  0.971 &    1.000 &       0.0 \\
   0.40 &   1100 &      22.59 &      2.9 & $     +0.7$ & $     -0.1$ & $     -0.1$ & $     -0.9$ & $     +0.2$ & $     +0.5$ & $      0.2$ & $     -0.0$ &       0.2 &      2.7 &      0.5 &       0.2 &    1.071 &  0.953 &  0.900 &  0.943 &    1.000 &       0.0 \\
   0.40 &   1700 &      17.86 &      3.1 & $     +1.4$ & $     -0.0$ & $     -0.0$ & $     -1.0$ & $     +0.1$ & $     +0.6$ & $      0.2$ & $     -0.2$ &       0.2 &      2.7 &      0.5 &       0.3 &    1.065 &  0.905 &  0.902 &  0.889 &    1.000 &       0.0 \\
   0.40 &   3500 &        7.9 &      4.1 & $     +2.1$ & $     -0.0$ & $     +0.0$ & $     -1.6$ & $     -0.0$ & $     -0.1$ & $      0.2$ & $     -0.0$ &       1.0 &      2.7 &      0.5 &       0.4 &    1.066 &  0.782 &  0.903 &  0.725 &    1.000 &       0.0 \\
   0.40 &   8000 &      2.612 &      7.5 & $     +4.4$ & $     +0.0$ & $     +0.5$ & $     -3.2$ & $     -0.6$ & $     -1.2$ & $      0.3$ & $     -0.1$ &       1.0 &      2.7 &      0.5 &       0.7 &    1.069 &  0.633 &  0.915 &  0.523 &    1.000 &       0.0 \\
   0.70 &    700 &       21.4 &      4.8 & $     -1.7$ & $     -0.5$ & $     -0.1$ & $     +0.3$ & $     +0.1$ & $     +0.9$ & $      0.3$ & $     -0.1$ &       0.2 &      2.7 &      0.5 &       0.2 &    1.072 &  0.972 &  0.904 &  0.970 &    1.000 &       0.0 \\
   0.70 &   1100 &      10.68 &      5.1 & $     -0.3$ & $     -0.2$ & $     +0.1$ & $     +0.1$ & $     +0.2$ & $     +1.6$ & $      0.3$ & $     -0.2$ &       0.2 &      2.7 &      0.5 &       0.3 &    1.067 &  0.952 &  0.906 &  0.949 &    1.000 &       0.0 \\
   0.70 &   1700 &      9.873 &      4.3 & $     +0.3$ & $     -0.1$ & $     -0.1$ & $     -0.2$ & $     +0.2$ & $     +1.6$ & $      0.3$ & $     -0.1$ &       0.2 &      2.7 &      0.5 &       0.4 &    1.061 &  0.905 &  0.903 &  0.881 &    1.000 &       0.0 \\
   0.70 &   3500 &      4.002 &      6.4 & $     +1.3$ & $     +0.0$ & $     -0.1$ & $     -0.9$ & $     -0.0$ & $     +4.5$ & $      0.4$ & $     -0.2$ &       1.0 &      2.7 &      0.5 &       0.5 &    1.061 &  0.764 &  0.910 &  0.690 &    1.000 &       0.0 \\
   0.70 &   8000 &      2.249 &      8.0 & $     +1.9$ & $     +0.1$ & $     -0.0$ & $     -1.3$ & $     -0.3$ & $     +3.7$ & $      0.5$ & $     -0.8$ &       1.0 &      2.7 &      0.5 &       0.8 &    1.057 &  0.600 &  0.911 &  0.440 &    1.000 &       0.0 \\
\bottomrule
\end{tabular}\end{adjustbox}
\caption{ 
Double-differential neutral current DIS  cross sections  for $e^-p$ scattering, $\d^2\sigma/\d\Qsq\d y ~[\pb]$. The lower bin edges in $y$ and $Q^2$ are indicated. Further details are given in the caption of Tab.~\ref{tab:xs1D.em}.
}
\label{tab:NCDIS2D.em} 
\end{table*}

\begin{table*}[t]
\tiny\centering\begin{adjustbox}{width=\textwidth}\begin{tabular}{cc|c|cccccccc ccccc|ccccc c}
\toprule
\multicolumn{22}{c}{2D neutral current DIS in $e^+p$  cross sections  $\d^2\sigma/\d\Qsq\d y ~[\pb/\GeVsq]$} \\
\midrule $y$ & \Qsq & Results & \multicolumn{13}{|c|}{Uncertainties} & \multicolumn{6}{c}{Correction factors}\\
\cmidrule(lr){1-2}\cmidrule(lr){3-3}\cmidrule(lr){4-16}\cmidrule(lr){17-22}
 $_\text{min}$ &  $_\text{min}$ & $\frac{\d^2\sigma}{\d\Qsq\d y}$ & $_\text{stat}$ & $_\text{RCES}$ & $_\text{JES}$ & $_\text{HadTh}$ & $_\text{ElEn}$ & $_\text{ElTh}$ & $_\text{Model}$ & $_\text{MCstat}$ & $_\text{SysTauData}$ &
 $_\text{ElecID}$ & $_\text{Lumi}$ & $_\text{Uncor}$ & $_\text{QED}$ & $c_\text{QED}$ & $c_\text{NoZ}$ & $c_\text{Born}$ & $c_\text{e+p}$ & $c_\text{Had}$ & $\delta_\text{HAD}$ \\
          & $_{[\text{GeV}]}$  &  $_{[\text{pb}/\GeVsq]}$   & $_{[\%]}$  & $_{[\%]}$ & $_{[\%]}$ & $_{[\%]}$ & $_{[\%]}$ & $_{[\%]}$ & $_{[\%]}$ & $_{[\%]}$ & $_{[\%]}$ & $_{[\%]}$ & $_{[\%]}$ & $_{[\%]}$ & $_{[\%]}$ & & & & & & $_{[\%]}$\\
\midrule
   0.05 &    150 &      180.5 &      2.7 & $     +2.2$ & $     +0.6$ & $     +0.9$ & $     +3.9$ & $     +1.3$ & $     -0.3$ & $      0.4$ & $     +0.0$ &       0.2 &      2.7 &      0.5 &       0.2 &    1.070 &  0.996 &  0.924 &  1.000 &    1.000 &       0.0 \\
   0.05 &    200 &      144.3 &      1.5 & $     +1.7$ & $     +0.4$ & $     +0.3$ & $     -1.6$ & $     +0.7$ & $     +0.1$ & $      0.2$ & $     -0.1$ &       0.2 &      2.7 &      0.5 &       0.2 &    1.070 &  0.994 &  0.923 &  1.000 &    1.000 &       0.0 \\
   0.05 &    280 &      115.7 &      1.5 & $     +1.6$ & $     +0.2$ & $     +0.3$ & $     -1.5$ & $     +0.7$ & $     +0.1$ & $      0.2$ & $     -0.0$ &       0.2 &      2.7 &      0.5 &       0.2 &    1.064 &  0.998 &  0.919 &  1.000 &    1.000 &       0.0 \\
   0.05 &    440 &      66.09 &      1.9 & $     +1.3$ & $     +0.2$ & $     +0.1$ & $     -2.1$ & $     +0.4$ & $     -0.4$ & $      0.2$ & $     +0.1$ &       0.2 &      2.7 &      0.5 &       0.2 &    1.063 &  0.997 &  0.919 &  1.000 &    1.000 &       0.0 \\
   0.05 &    700 &      35.45 &      2.1 & $     +0.5$ & $     +0.1$ & $     -0.4$ & $     -1.2$ & $     +0.3$ & $     -1.0$ & $      0.2$ & $     -0.1$ &       0.2 &      2.7 &      0.5 &       0.3 &    1.069 &  0.995 &  0.909 &  1.000 &    1.000 &       0.0 \\
   0.10 &    150 &        222 &      1.4 & $     +1.5$ & $     +0.4$ & $     +0.2$ & $     +1.8$ & $     +0.8$ & $     +0.8$ & $      0.2$ & $     -0.0$ &       0.2 &      2.7 &      0.5 &       0.1 &    1.073 &  0.994 &  0.923 &  1.000 &    1.000 &       0.0 \\
   0.10 &    200 &        175 &      1.1 & $     +0.9$ & $     +0.1$ & $     -0.0$ & $     -1.4$ & $     +0.8$ & $     -0.0$ & $      0.1$ & $     +0.0$ &       0.2 &      2.7 &      0.5 &       0.1 &    1.069 &  0.997 &  0.920 &  1.000 &    1.000 &       0.0 \\
   0.10 &    280 &      140.6 &      1.2 & $     +1.1$ & $     +0.1$ & $     +0.1$ & $     -1.2$ & $     +0.5$ & $     +0.1$ & $      0.1$ & $     -0.0$ &       0.2 &      2.7 &      0.5 &       0.1 &    1.064 &  0.996 &  0.918 &  1.000 &    1.000 &       0.0 \\
   0.10 &    440 &      79.87 &      1.5 & $     +1.3$ & $     +0.0$ & $     +0.0$ & $     -1.6$ & $     +0.5$ & $     -0.1$ & $      0.1$ & $     -0.0$ &       0.2 &      2.7 &      0.5 &       0.2 &    1.066 &  0.997 &  0.916 &  1.000 &    1.000 &       0.0 \\
   0.10 &    700 &      42.95 &      1.9 & $     +1.3$ & $     -0.0$ & $     +0.0$ & $     -1.1$ & $     +0.4$ & $     +0.0$ & $      0.1$ & $     +0.0$ &       0.2 &      2.7 &      0.5 &       0.2 &    1.062 &  0.996 &  0.915 &  1.000 &    1.000 &       0.0 \\
   0.10 &   1100 &         22 &      2.5 & $     +1.5$ & $     +0.1$ & $     -0.0$ & $     -1.6$ & $     +0.2$ & $     -0.9$ & $      0.1$ & $     -0.1$ &       0.2 &      2.7 &      0.5 &       0.2 &    1.070 &  0.995 &  0.904 &  1.000 &    1.000 &       0.0 \\
   0.20 &    150 &        248 &      0.9 & $     +0.8$ & $     +0.3$ & $     +0.0$ & $     -0.3$ & $     +0.7$ & $     +0.8$ & $      0.1$ & $     -0.0$ &       0.2 &      2.7 &      0.5 &       0.1 &    1.074 &  0.997 &  0.923 &  1.000 &    1.000 &       0.0 \\
   0.20 &    200 &      196.6 &      1.0 & $     +0.7$ & $     +0.0$ & $     -0.0$ & $     -1.0$ & $     +0.6$ & $     +0.7$ & $      0.1$ & $     -0.0$ &       0.2 &      2.7 &      0.5 &       0.1 &    1.070 &  0.997 &  0.921 &  1.000 &    1.000 &       0.0 \\
   0.20 &    280 &      165.9 &      1.0 & $     +0.8$ & $     +0.0$ & $     -0.0$ & $     -1.1$ & $     +0.5$ & $     +0.6$ & $      0.1$ & $     -0.0$ &       0.2 &      2.7 &      0.5 &       0.1 &    1.068 &  0.998 &  0.919 &  1.000 &    1.000 &       0.0 \\
   0.20 &    440 &      96.09 &      1.3 & $     +1.0$ & $     +0.0$ & $     -0.1$ & $     -1.4$ & $     +0.5$ & $     +0.4$ & $      0.1$ & $     -0.0$ &       0.2 &      2.7 &      0.5 &       0.1 &    1.067 &  0.998 &  0.917 &  1.000 &    1.000 &       0.0 \\
   0.20 &    700 &      51.72 &      1.6 & $     +1.0$ & $     +0.0$ & $     -0.0$ & $     -0.9$ & $     +0.3$ & $     +0.6$ & $      0.1$ & $     -0.0$ &       0.2 &      2.7 &      0.5 &       0.2 &    1.066 &  0.999 &  0.911 &  1.000 &    1.000 &       0.0 \\
   0.20 &   1100 &      24.99 &      2.3 & $     +1.4$ & $     -0.0$ & $     -0.0$ & $     -1.3$ & $     +0.3$ & $     -0.1$ & $      0.1$ & $     -0.0$ &       0.2 &      2.7 &      0.5 &       0.2 &    1.062 &  1.000 &  0.912 &  1.000 &    1.000 &       0.0 \\
   0.20 &   1700 &      19.34 &      2.7 & $     +2.4$ & $     +0.0$ & $     +0.1$ & $     -2.1$ & $     +0.1$ & $     -0.9$ & $      0.1$ & $     -0.3$ &       0.2 &      2.7 &      0.5 &       0.2 &    1.066 &  1.009 &  0.902 &  1.000 &    1.000 &       0.0 \\
   0.20 &   3500 &      6.077 &      4.7 & $     +4.0$ & $     -0.0$ & $     +0.6$ & $     -2.5$ & $     -0.0$ & $     -0.7$ & $      0.2$ & $     -0.3$ &       1.0 &      2.7 &      0.5 &       0.4 &    1.068 &  1.029 &  0.905 &  1.000 &    1.000 &       0.0 \\
   0.20 &   8000 &     0.8582 &     13.1 & $     +4.7$ & $     +0.1$ & $     +1.1$ & $     -4.1$ & $     -0.5$ & $     -0.5$ & $      0.5$ & $     -0.1$ &       1.0 &      2.7 &      0.5 &       1.0 &    1.076 &  1.072 &  0.889 &  1.000 &    1.000 &       0.0 \\
   0.40 &    150 &      203.2 &      1.2 & $     -0.6$ & $     -1.2$ & $     -0.1$ & $     -0.6$ & $     +0.5$ & $     +2.6$ & $      0.1$ & $     -0.0$ &       0.2 &      2.7 &      0.5 &       0.1 &    1.073 &  0.996 &  0.924 &  1.000 &    1.000 &       0.0 \\
   0.40 &    200 &      161.4 &      1.2 & $     -0.4$ & $     -0.7$ & $     -0.1$ & $     -0.5$ & $     +0.4$ & $     +2.1$ & $      0.1$ & $     -0.0$ &       0.2 &      2.7 &      0.5 &       0.1 &    1.069 &  0.996 &  0.922 &  1.000 &    1.000 &       0.0 \\
   0.40 &    280 &        132 &      1.3 & $     -0.2$ & $     -0.4$ & $     -0.1$ & $     -0.7$ & $     +0.4$ & $     +1.7$ & $      0.1$ & $     -0.0$ &       0.2 &      2.7 &      0.5 &       0.1 &    1.068 &  0.996 &  0.918 &  1.000 &    1.000 &       0.0 \\
   0.40 &    440 &      76.04 &      1.6 & $     +0.3$ & $     -0.2$ & $     -0.1$ & $     -0.8$ & $     +0.3$ & $     +1.3$ & $      0.1$ & $     -0.0$ &       0.2 &      2.7 &      0.5 &       0.1 &    1.071 &  0.997 &  0.914 &  1.000 &    1.000 &       0.0 \\
   0.40 &    700 &      38.58 &      1.9 & $     +0.4$ & $     -0.1$ & $     -0.1$ & $     -0.8$ & $     +0.3$ & $     +0.7$ & $      0.1$ & $     -0.0$ &       0.2 &      2.7 &      0.5 &       0.2 &    1.066 &  1.002 &  0.912 &  1.000 &    1.000 &       0.0 \\
   0.40 &   1100 &      21.12 &      2.7 & $     +0.9$ & $     -0.0$ & $     -0.1$ & $     -1.1$ & $     +0.2$ & $     +0.5$ & $      0.2$ & $     -0.1$ &       0.2 &      2.7 &      0.5 &       0.2 &    1.064 &  1.007 &  0.909 &  1.000 &    1.000 &       0.0 \\
   0.40 &   1700 &      16.83 &      3.0 & $     +1.7$ & $     -0.0$ & $     -0.0$ & $     -1.2$ & $     +0.1$ & $     +0.4$ & $      0.2$ & $     -0.1$ &       0.2 &      2.7 &      0.5 &       0.3 &    1.072 &  1.016 &  0.899 &  1.000 &    1.000 &       0.0 \\
   0.40 &   3500 &      6.128 &      4.3 & $     +2.5$ & $     -0.0$ & $     +0.0$ & $     -1.9$ & $     -0.1$ & $     -0.9$ & $      0.2$ & $     -0.5$ &       1.0 &      2.7 &      0.5 &       0.4 &    1.062 &  1.076 &  0.898 &  1.000 &    1.000 &       0.0 \\
   0.40 &   8000 &      1.538 &      9.2 & $     +4.7$ & $     -0.1$ & $     +0.4$ & $     -3.2$ & $     -0.3$ & $     -1.2$ & $      0.5$ & $     -0.8$ &       1.0 &      2.7 &      0.5 &       0.9 &    1.066 &  1.206 &  0.898 &  1.000 &    1.000 &       0.0 \\
   0.70 &    700 &      19.83 &      4.5 & $     -1.8$ & $     -0.5$ & $     -0.1$ & $     +0.3$ & $     +0.1$ & $     +1.1$ & $      0.3$ & $     -0.0$ &       0.2 &      2.7 &      0.5 &       0.2 &    1.071 &  0.999 &  0.913 &  1.000 &    1.000 &       0.0 \\
   0.70 &   1100 &      10.73 &      4.7 & $     -0.3$ & $     -0.2$ & $     -0.0$ & $     +0.1$ & $     +0.1$ & $     +2.3$ & $      0.3$ & $     -0.0$ &       0.2 &      2.7 &      0.5 &       0.3 &    1.070 &  1.000 &  0.904 &  1.000 &    1.000 &       0.0 \\
   0.70 &   1700 &      7.528 &      4.6 & $     +0.3$ & $     -0.2$ & $     -0.1$ & $     -0.3$ & $     +0.2$ & $     +1.3$ & $      0.3$ & $     -0.3$ &       0.2 &      2.7 &      0.5 &       0.4 &    1.063 &  1.024 &  0.900 &  1.000 &    1.000 &       0.0 \\
   0.70 &   3500 &      3.205 &      6.6 & $     +1.4$ & $     +0.3$ & $     -0.0$ & $     -1.1$ & $     +0.1$ & $     +4.8$ & $      0.5$ & $     -0.3$ &       1.0 &      2.7 &      0.5 &       0.6 &    1.050 &  1.104 &  0.901 &  1.000 &    1.000 &       0.0 \\
   0.70 &   8000 &     0.7364 &     14.5 & $     +2.8$ & $     +0.4$ & $     +0.0$ & $     -2.3$ & $     -0.7$ & $     +6.3$ & $      1.1$ & $     -1.4$ &       1.0 &      2.7 &      0.5 &       1.2 &    1.090 &  1.360 &  0.896 &  1.000 &    1.000 &       0.0 \\
\bottomrule
\end{tabular}\end{adjustbox}
\caption{ 
Double-differential neutral current DIS  cross sections  for $e^+p$ scattering, $\d^2\sigma/\d\Qsq\d y ~[\pb]$. The lower bin edges in $y$ and $Q^2$ are indicated. Further details are given in the caption of Tab.~\ref{tab:xs1D.em}.
}
\label{tab:NCDIS2D.ep} 
\end{table*}

\end{document}